\documentclass[longauth]{aa}  

\usepackage{graphicx}
\usepackage{txfonts}
\usepackage{hyperref}
\newcommand{\fwhmmufwhm}{\ensuremath{6045.2^{+23.9}_{-21.6}}}
\newcommand{\fwhmsigmawfwhm}{\ensuremath{9.9^{+2.7}_{-9.4}}}
\newcommand{\fwhmGPsigmafwhm}{\ensuremath{44.3^{+16.2}_{-9.4}}}
\newcommand{\fwhmGPalphafwhm}{\ensuremath{0.00011^{+0.00065}_{-0.00009}}}
\newcommand{\fwhmGPGammafwhm}{\ensuremath{2.9^{+3.4}_{-1.6}}}
\newcommand{\fwhmGPProtfwhm}{\ensuremath{21.44^{+0.35}_{-0.41}}}

\newcommand{\jmuESPRESSOnineteen}{\ensuremath{-4476.9^{+6.1}_{-5.1}}}
\newcommand{\jsigmawESPRESSOnineteen}{\ensuremath{0.13^{+1.21}_{-0.13}}}
\newcommand{\jPpone}{\ensuremath{11.30727^{+0.00029}_{-0.00013}}}

\newcommand{\jtzeropone}{\ensuremath{2460160.4518^{+0.0099}_{-0.0059}}}
\newcommand{\jKpone}{\ensuremath{2.8^{+2.3}_{-1.7}}}
\newcommand{\jrho}{\ensuremath{2754.0^{+1384.1}_{-1109.6}}}
\newcommand{\jppone}{\ensuremath{0.01396^{+0.00097}_{-0.00107}}}
\newcommand{\jbpone}{\ensuremath{0.30^{+0.23}_{-0.20}}}
\newcommand{\jPptwo}{\ensuremath{20.225520^{+0.000037}_{-0.000041}}}

\newcommand{\jtzeroptwo}{\ensuremath{2459037.4162^{+0.0015}_{-0.0013}}}
\newcommand{\jKptwo}{\ensuremath{4.90^{+2.99}_{-2.94}}}
\newcommand{\jpptwo}{\ensuremath{0.0269^{+0.0017}_{-0.0013}}}
\newcommand{\jbptwo}{\ensuremath{0.838^{+0.050}_{-0.058}}}
\newcommand{\jqoneTESS}{\ensuremath{0.22^{+0.22}_{-0.15}}}
\newcommand{\jqtwoTESS}{\ensuremath{0.55^{+0.31}_{-0.35}}}
\newcommand{\jGPsigmarv}{\ensuremath{14.0^{+4.0}_{-2.7}}}
\newcommand{\jGPalpharv}{\ensuremath{(2.44^{+1.41}_{-0.92})\times10^{-4}}}
\newcommand{\jGPGammarv}{\ensuremath{5.0^{+1.8}_{-1.3}}}
\newcommand{\jGPProtrv}{\ensuremath{21.47^{+0.14}_{-0.16}}}
\newcommand{\jeccpone}{\ensuremath{0}}
\newcommand{\jomegapone}{\ensuremath{90}}
\newcommand{\jeccptwo}{\ensuremath{0}}
\newcommand{\jomegaptwo}{\ensuremath{90}}
\newcommand{\TESSdetrmfluxTESStwentyseven}{\ensuremath{(1.2^{+37.0}_{-41.8})\times10^{-5}}}
\newcommand{\TESSdetrsigmawTESStwentyseven}{\ensuremath{104.1^{+80.2}_{-102.6}}}
\newcommand{\TESSdetrGPsigmaTESStwentyseven}{\ensuremath{(1.27^{+0.28}_{-0.21})\times10^{-3}}}
\newcommand{\TESSdetrGPrhoTESStwentyseven}{\ensuremath{1.00^{+0.20}_{-0.15}}}

\newcommand{\TESSdetrmfluxTESStwentyeight}{\ensuremath{(-6.1^{+20.4}_{-22.0})\times10^{-5}}}
\newcommand{\TESSdetrsigmawTESStwentyeight}{\ensuremath{3.7^{+40.3}_{-3.4}}}
\newcommand{\TESSdetrGPsigmaTESStwentyeight}{\ensuremath{(1.02^{+0.14}_{-0.10})\times10^{-3}}}
\newcommand{\TESSdetrGPrhoTESStwentyeight}{\ensuremath{0.375^{+0.052}_{-0.044}}}

\newcommand{\TESSdetrmfluxTESStwentynine}{\ensuremath{(-3.9^{+17.6}_{-17.5})\times10^{-5}}}
\newcommand{\TESSdetrsigmawTESStwentynine}{\ensuremath{235.0^{+29.1}_{-33.3}}}
\newcommand{\TESSdetrGPsigmaTESStwentynine}{\ensuremath{(1.02^{+0.10}_{-0.09})\times10^{-3}}}
\newcommand{\TESSdetrGPrhoTESStwentynine}{\ensuremath{0.277^{+0.031}_{-0.025}}}

\newcommand{\TESSdetrmfluxTESSthirty}{\ensuremath{(-13.4^{+24.0}_{-22.7})\times10^{-5}}}
\newcommand{\TESSdetrsigmawTESSthirty}{\ensuremath{319.3^{+22.2}_{-22.5}}}
\newcommand{\TESSdetrGPsigmaTESSthirty}{\ensuremath{(1.33^{+0.13}_{-0.11})\times10^{-3}}}
\newcommand{\TESSdetrGPrhoTESSthirty}{\ensuremath{0.311^{+0.030}_{-0.027}}}

\newcommand{\TESSdetrmfluxTESSthirtyone}{\ensuremath{(-17.6^{+26.6}_{-28.4})\times10^{-5}}}
\newcommand{\TESSdetrsigmawTESSthirtyone}{\ensuremath{153.3^{+46.9}_{-141.1}}}
\newcommand{\TESSdetrGPsigmaTESSthirtyone}{\ensuremath{(0.95^{+0.21}_{-0.14})\times10^{-3}}}
\newcommand{\TESSdetrGPrhoTESSthirtyone}{\ensuremath{0.87^{+0.18}_{-0.13}}}

\newcommand{\TESSdetrmfluxTESSthirtytwo}{\ensuremath{(-1.1^{+9.8}_{-9.2})\times10^{-5}}}
\newcommand{\TESSdetrsigmawTESSthirtytwo}{\ensuremath{3.5^{+35.7}_{-3.2}}}
\newcommand{\TESSdetrGPsigmaTESSthirtytwo}{\ensuremath{(0.50^{+0.06}_{-0.05})\times10^{-3}}}
\newcommand{\TESSdetrGPrhoTESSthirtytwo}{\ensuremath{0.369^{+0.067}_{-0.047}}}

\newcommand{\TESSdetrmfluxTESSthirtythree}{\ensuremath{(9.2^{+15.3}_{-14.4})\times10^{-5}}}
\newcommand{\TESSdetrsigmawTESSthirtythree}{\ensuremath{200.5^{+29.6}_{-35.9}}}
\newcommand{\TESSdetrGPsigmaTESSthirtythree}{\ensuremath{(0.83^{+0.09}_{-0.07})\times10^{-3}}}
\newcommand{\TESSdetrGPrhoTESSthirtythree}{\ensuremath{0.333^{+0.043}_{-0.037}}}

\newcommand{\TESSdetrmfluxTESSthirtyfive}{\ensuremath{(3.0^{+10.8}_{-10.0})\times10^{-5}}}
\newcommand{\TESSdetrsigmawTESSthirtyfive}{\ensuremath{4.6^{+56.8}_{-4.2}}}
\newcommand{\TESSdetrGPsigmaTESSthirtyfive}{\ensuremath{(0.49^{+0.08}_{-0.06})\times10^{-3}}}
\newcommand{\TESSdetrGPrhoTESSthirtyfive}{\ensuremath{0.409^{+0.076}_{-0.057}}}

\newcommand{\TESSdetrmfluxTESSthirtysix}{\ensuremath{(-2.7^{+9.6}_{-9.5})\times10^{-5}}}
\newcommand{\TESSdetrsigmawTESSthirtysix}{\ensuremath{29.3^{+115.0}_{-28.7}}}
\newcommand{\TESSdetrGPsigmaTESSthirtysix}{\ensuremath{0.57^{+0.05}_{-0.05})\times10^{-3}}}
\newcommand{\TESSdetrGPrhoTESSthirtysix}{\ensuremath{0.260^{+0.042}_{-0.033}}}

\newcommand{\TESSdetrmfluxTESSthirtyseven}{\ensuremath{(-3.7^{+12.7}_{-13.8})\times10^{-5}}}
\newcommand{\TESSdetrsigmawTESSthirtyseven}{\ensuremath{266.0^{+22.6}_{-25.9}}}
\newcommand{\TESSdetrGPsigmaTESSthirtyseven}{\ensuremath{(0.44^{+0.12}_{-0.07})\times10^{-3}}}
\newcommand{\TESSdetrGPrhoTESSthirtyseven}{\ensuremath{0.91^{+0.29}_{-0.19}}}

\newcommand{\TESSdetrmfluxTESSthirtyeight}{\ensuremath{(-4.9^{+20.9}_{-21.8})\times10^{-5}}}
\newcommand{\TESSdetrsigmawTESSthirtyeight}{\ensuremath{176.9^{+34.7}_{-45.8}}}
\newcommand{\TESSdetrGPsigmaTESSthirtyeight}{\ensuremath{(1.03^{+0.13}_{-0.10})\times10^{-3}}}
\newcommand{\TESSdetrGPrhoTESSthirtyeight}{\ensuremath{0.471^{+0.061}_{-0.052}}}

\newcommand{\TESSdetrmfluxTESSthirtynine}{\ensuremath{(6.1^{+30.4}_{-28.6})\times10^{-5}}}
\newcommand{\TESSdetrsigmawTESSthirtynine}{\ensuremath{261.2^{+21.7}_{-22.1}}}
\newcommand{\TESSdetrGPsigmaTESSthirtynine}{\ensuremath{(1.18^{+0.20}_{-0.15})\times10^{-3}}}
\newcommand{\TESSdetrGPrhoTESSthirtynine}{\ensuremath{0.77^{+0.13}_{-0.11}}}

\newcommand{\TESSdetrmfluxTESSsixtyone}{\ensuremath{(10.4^{+19.1}_{-19.5})\times10^{-5}}}
\newcommand{\TESSdetrsigmawTESSsixtyone}{\ensuremath{2.6^{+21.3}_{-2.3}}}
\newcommand{\TESSdetrGPsigmaTESSsixtyone}{\ensuremath{(1.20^{+0.14}_{-0.12})\times10^{-3}}}
\newcommand{\TESSdetrGPrhoTESSsixtyone}{\ensuremath{0.245^{+0.029}_{-0.023}}}

\newcommand{\TESSdetrmfluxTESSsixtytwo}{\ensuremath{-6.5^{+27.6}_{-29.5})\times10^{-5}}}
\newcommand{\TESSdetrsigmawTESSsixtytwo}{\ensuremath{2.6^{+20.3}_{-2.3}}}
\newcommand{\TESSdetrGPsigmaTESSsixtytwo}{\ensuremath{(1.34^{+0.21}_{-0.13})\times10^{-3}}}
\newcommand{\TESSdetrGPrhoTESSsixtytwo}{\ensuremath{0.489^{+0.076}_{-0.052}}}

\newcommand{\TESSdetrmfluxTESSsixtythree}{\ensuremath{(3.4^{+29.2}_{-29.2})\times10^{-5}}}
\newcommand{\TESSdetrsigmawTESSsixtythree}{\ensuremath{3.1^{+24.8}_{-2.8}}}
\newcommand{\TESSdetrGPsigmaTESSsixtythree}{\ensuremath{(1.02^{+0.20}_{-0.14})\times10^{-3}}}
\newcommand{\TESSdetrGPrhoTESSsixtythree}{\ensuremath{0.94^{+0.16}_{-0.13}}}

\newcommand{\TESSdetrmfluxTESSsixtyfive}{\ensuremath{(2.3^{+43.3}_{-42.6})\times10^{-5}}}
\newcommand{\TESSdetrsigmawTESSsixtyfive}{\ensuremath{4.3^{+57.7}_{-4.0}}}
\newcommand{\TESSdetrGPsigmaTESSsixtyfive}{\ensuremath{(1.18^{+0.54}_{-0.28})\times10^{-3}}}
\newcommand{\TESSdetrGPrhoTESSsixtyfive}{\ensuremath{1.45^{+0.58}_{-0.34}}}

\newcommand{\TESSdetrmfluxTESSsixtysix}{\ensuremath{(-0.6^{+18.0}_{-17.6})\times10^{-5}}}
\newcommand{\TESSdetrsigmawTESSsixtysix}{\ensuremath{96.2^{+77.5}_{-94.7}}}
\newcommand{\TESSdetrGPsigmaTESSsixtysix}{\ensuremath{(0.60^{+0.14}_{-0.09})\times10^{-3}}}
\newcommand{\TESSdetrGPrhoTESSsixtysix}{\ensuremath{0.92^{+0.29}_{-0.19}}}

\newcommand{\TESSdetrmfluxTESSsixtyseven}{\ensuremath{(-1.2^{+7.8}_{-6.5})\times10^{-5}}}
\newcommand{\TESSdetrsigmawTESSsixtyseven}{\ensuremath{2.4^{+17.5}_{-2.1}}}
\newcommand{\TESSdetrGPsigmaTESSsixtyseven}{\ensuremath{(0.36^{+0.05}_{-0.04})\times10^{-3}}}
\newcommand{\TESSdetrGPrhoTESSsixtyseven}{\ensuremath{0.383^{+0.081}_{-0.062}}}

\newcommand{\TESSdetrmfluxTESSsixtyeight}{\ensuremath{(-8.5^{+17.3}_{-19.2})\times10^{-5}}}
\newcommand{\TESSdetrsigmawTESSsixtyeight}{\ensuremath{3.0^{+24.1}_{-2.7}}}
\newcommand{\TESSdetrGPsigmaTESSsixtyeight}{\ensuremath{(0.79^{+0.12}_{-0.09})\times10^{-3}}}
\newcommand{\TESSdetrGPrhoTESSsixtyeight}{\ensuremath{0.521^{+0.089}_{-0.076}}}

\newcommand{\TESSdetrmfluxTESSsixtynine}{\ensuremath{(0.9^{+33.7}_{-33.1})\times10^{-5}}}
\newcommand{\TESSdetrsigmawTESSsixtynine}{\ensuremath{2.4^{+20.9}_{-2.1}}}
\newcommand{\TESSdetrGPsigmaTESSsixtynine}{\ensuremath{(1.00^{+0.24}_{-0.14})\times10^{-3}}}
\newcommand{\TESSdetrGPrhoTESSsixtynine}{\ensuremath{1.11^{+0.26}_{-0.19}}}

\newcommand{\smass}[1][$M_{\odot}$]{ $ 0.7030000 _{- 0.0170000}^{ + 0.0170000} $ } 
\newcommand{\sradius}[1][$R_{\odot}$]{ $0.6620000 _{ - 0.0039000}^{ + 0.0039000} $ }
\newcommand{\stemp}[1][$\mathrm{K}$]{ $ 4664.0000000 _{- 71.0000000}^{ + 71.0000000} $  }
\newcommand{\Tzerob}[1][days]{ $ 10160.4469_{-0.0042}^{+0.0040} $ } 
\newcommand{\Pb}[1][days]{ $ 11.307170_{-0.000079}^{+0.000085} $ } 
\newcommand{\eb}[1][ ]{ $ 0.0 \pm 0.0 $ } 
\newcommand{\wb}[1][deg]{ $ 90.0 \pm 0.0 $ } 
\newcommand{\bb}[1][ ]{ $ 0.23_{-0.16}^{+0.23} $ } 
\newcommand{\dentrheeb}[1][${\rm g^{1/3}\,cm^{-1}}$]{ $ 4.00_{-1.14}^{+0.85} $ } 
\newcommand{\rrbTESS}[1][ ]{ $ 0.01376_{-0.00081}^{+0.00078} $ } 
\newcommand{\kb}[1][${\rm m\,s^{-1}}$]{ $ 0.84_{-0.56}^{+0.97} $ } 
\newcommand{\mpb}[1][$M_{\oplus}$]{ $ 2.31_{-1.54}^{+2.67} $ } 
\newcommand{\rpbTESS}[1][$R_{\oplus}$]{ $ 0.994_{-0.059}^{+0.057} $ } 
\newcommand{\Tperib}[1][days]{ $ 10160.4469_{-0.0042}^{+0.0040} $ } 
\newcommand{\prvb}[1][${\rm km\,s^{-1}}$]{ $ nan_{-nan}^{+nan} $ } 
\newcommand{\ib}[1][deg]{ $ 89.57_{-0.52}^{+0.30} $ } 
\newcommand{\arb}[1][ ]{ $ 30.02_{-3.16}^{+1.99} $ } 
\newcommand{\ab}[1][AU]{ $ 0.0924_{-0.0097}^{+0.0061} $ } 
\newcommand{\depthbTESS}[1][ppm]{ $ 189.4_{-21.6}^{+22.1} $ } 
\newcommand{\RMbTESS}[1][${\rm m\,s^{-1}}$]{ $ 0.357_{-0.042}^{+0.049} $ } 
\newcommand{\insolationb}[1][${\rm F_{\oplus}}$]{ $ 22.04_{-3.20}^{+5.34} $ } 
\newcommand{\tsmb}[1][ ]{ $ 11.44_{-6.37}^{+22.17} $ } 
\newcommand{\denstrb}[1][${\rm g\,cm^{-3}}$]{ $ 4.00_{-1.14}^{+0.85} $ } 
\newcommand{\densspb}[1][${\rm g\,cm^{-3}}$]{ $ 3.42 \pm 0.10 $ } 
\newcommand{\Teqb}[1][K]{ $ 603.1_{-23.2}^{+33.6} $ } 
\newcommand{\ttotb}[1][hours]{ $ 2.80_{-0.15}^{+0.18} $ } 
\newcommand{\tfulb}[1][hours]{ $ 2.72_{-0.16}^{+0.17} $ } 
\newcommand{\tegb}[1][hours]{ $ 0.0407_{-0.0038}^{+0.0088} $ } 
\newcommand{\deltamagb}[1][]{ $ 0.12_{-0.11}^{+0.39} $ } 
\newcommand{\denpb}[1][${\rm g\,cm^{-3}}$]{ $ 12.75_{-8.43}^{+15.98} $ } 
\newcommand{\grapb}[1][${\rm cm\,s^{-2}}$]{ $ 2423_{-1601}^{+2959} $ } 
\newcommand{\grapparsb}[1][${\rm cm\,s^{-2}}$]{ $ 2276_{-1507}^{+2792} $ } 
\newcommand{\jspb}[1][ ]{ $ 28.8_{-19.1}^{+33.5} $ } 
\newcommand{\Tzeroc}[1][days]{ $ 9037.4161_{-0.0013}^{+0.0014} $ } 
\newcommand{\Pc}[1][days]{ $ 20.225528_{-0.000044}^{+0.000039} $ } 
\newcommand{\ec}[1][ ]{ $ 0.0 \pm 0.0 $ } 
\newcommand{\wc}[1][deg]{ $ 90.0 \pm 0.0 $ } 
\newcommand{\bc}[1][ ]{ $ 0.787_{-0.041}^{+0.046} $ } 
\newcommand{\rrcTESS}[1][ ]{ $ 0.02595_{-0.00077}^{+0.00089} $ } 
\newcommand{\kc}[1][${\rm m\,s^{-1}}$]{ $ 5.37_{-1.60}^{+1.29} $ } 
\newcommand{\mpc}[1][$M_{\oplus}$]{ $ 18.10_{-5.36}^{+4.34} $ } 
\newcommand{\rpcTESS}[1][$R_{\oplus}$]{ $ 1.874_{-0.057}^{+0.066} $ } 
\newcommand{\Tperic}[1][days]{ $ 9037.4161_{-0.0013}^{+0.0014} $ } 
\newcommand{\prvc}[1][${\rm km\,s^{-1}}$]{ $ nan_{-nan}^{+nan} $ } 
\newcommand{\ic}[1][deg]{ $ 88.98_{-0.18}^{+0.11} $ } 
\newcommand{\arc}[1][ ]{ $ 44.23_{-4.66}^{+2.93} $ } 
\newcommand{\ac}[1][AU]{ $ 0.1362_{-0.0143}^{+0.0090} $ } 
\newcommand{\depthcTESS}[1][ppm]{ $ 673.3_{-39.4}^{+46.8} $ } 
\newcommand{\RMcTESS}[1][${\rm m\,s^{-1}}$]{ $ 0.829_{-0.082}^{+0.080} $ } 
\newcommand{\insolationc}[1][${\rm F_{\oplus}}$]{ $ 10.15_{-1.48}^{+2.46} $ } 
\newcommand{\tsmc}[1][ ]{ $ 53.9_{-13.7}^{+23.4} $ } 
\newcommand{\denstrc}[1][${\rm g\,cm^{-3}}$]{ $ 4.00_{-1.14}^{+0.85} $ } 
\newcommand{\densspc}[1][${\rm g\,cm^{-3}}$]{ $ 3.42 \pm 0.10 $ } 
\newcommand{\Teqc}[1][K]{ $ 496.8_{-19.1}^{+27.7} $ } 
\newcommand{\ttotc}[1][hours]{ $ 2.307_{-0.059}^{+0.066} $ } 
\newcommand{\tfulc}[1][hours]{ $ 2.002_{-0.056}^{+0.062} $ } 
\newcommand{\tegc}[1][hours]{ $ 0.147_{-0.021}^{+0.042} $ } 
\newcommand{\deltamagc}[1][]{ $ 2.11_{-0.34}^{+0.48} $ } 
\newcommand{\denpc}[1][${\rm g\,cm^{-3}}$]{ $ 14.69_{-4.26}^{+4.60} $ } 
\newcommand{\grapc}[1][${\rm cm\,s^{-2}}$]{ $ 5240_{-1824}^{+2289} $ } 
\newcommand{\grapparsc}[1][${\rm cm\,s^{-2}}$]{ $ 4972_{-1436}^{+1401} $ } 
\newcommand{\jspc}[1][ ]{ $ 144.5_{-43.1}^{+41.1} $ } 
\newcommand{\qoneTESS}[1][]{ $ 0.20_{-0.14}^{+0.20} $ } 
\newcommand{\qtwoTESS}[1][]{ $ 0.36_{-0.28}^{+0.27} $ } 
\newcommand{\uoneTESS}[1][]{ $ 0.27_{-0.21}^{+0.33} $ } 
\newcommand{\utwoTESS}[1][]{ $ 0.10_{-0.20}^{+0.30} $ } 
\newcommand{\RVESPRESSO}[1][${\rm km\,s^{-1}}$]{ $ -4.4812_{-0.0012}^{+0.0014} $ } 
\newcommand{\RVHARPS}[1][${\rm km\,s^{-1}}$]{ $ -4.4749 \pm 0.0019 $ } 
\newcommand{\FWHMESP}[1][${\rm km\,s^{-1}}$]{ $ 6041.1_{-14.7}^{+17.9} $ } 
\newcommand{\FWHMHARPS}[1][${\rm km\,s^{-1}}$]{ $ 6117.0_{-14.8}^{+17.9} $ } 
\newcommand{\BISESP}[1][${\rm km\,s^{-1}}$]{ $ 23.76_{-0.66}^{+0.87} $ } 
\newcommand{\BISHARPS}[1][${\rm km\,s^{-1}}$]{ $ 32.04_{-1.04}^{+1.11} $ } 
\newcommand{\jRVESPRESSO}[1][${\rm m\,s^{-1}}$]{ $ 4.07_{-0.63}^{+0.84} $ } 
\newcommand{\jRVHARPS}[1][${\rm m\,s^{-1}}$]{ $ 7.41_{-1.22}^{+1.32} $ } 
\newcommand{\jFWHMESP}[1][${\rm m\,s^{-1}}$]{ $ 0.35_{-0.34}^{+4.76} $ } 
\newcommand{\jFWHMHARPS}[1][${\rm m\,s^{-1}}$]{ $ 1.53_{-1.49}^{+3.33} $ } 
\newcommand{\jBISESP}[1][${\rm m\,s^{-1}}$]{ $ 2.59_{-2.56}^{+4.06} $ } 
\newcommand{\jBISHARPS}[1][${\rm m\,s^{-1}}$]{ $ 0.53_{-0.52}^{+4.99} $ } 
\newcommand{\jtrTESS}[1][]{ $ 0.000059_{-0.000042}^{+0.000060} $ } 
\newcommand{\jAzero}[1][]{ $ 0.00197_{-0.00098}^{+0.00133} $ } 
\newcommand{\jAone}[1][]{ $ 0.0214 \pm 0.0055 $ } 
\newcommand{\jAtwo}[1][]{ $ 43.83_{-8.57}^{+10.56} $ } 
\newcommand{\jAthree}[1][]{ $ 0.0 \pm 0.0 $ } 
\newcommand{\jAfour}[1][]{ $ 1.25_{-0.52}^{+0.59} $ } 
\newcommand{\jAfive}[1][]{ $ -24.29_{-5.70}^{+5.53} $ } 
\newcommand{\jlambdae}[1][]{ $ 32.56_{-4.51}^{+16.15} $ } 
\newcommand{\jlambdap}[1][]{ $ 0.317_{-0.086}^{+0.094} $ } 
\newcommand{\jPGP}[1][]{ $ 21.71_{-0.39}^{+0.18} $ } 

\newcommand{\toitwothreetwotwoab}{\ensuremath{0.0877 \pm 0.0003}}
\newcommand{\toitwothreetwotwoincb}{\ensuremath{89.4 \pm 0.46}}
\newcommand{\toitwothreetwotwoTdurb}{\ensuremath{2.93 \pm 0.22}}
\newcommand{\toitwothreetwotwomassb}{\ensuremath{<26.66}}
\newcommand{\toitwothreetwotworadb}{\ensuremath{1.01 \pm 0.08}}
\newcommand{\toitwothreetwotworhoplb}{\ensuremath{<143}}
\newcommand{\toitwothreetwotwoteqb}{\ensuremath{608.0 \pm 3.0}}
\newcommand{\toitwothreetwotwoSb}{\ensuremath{22.7 \pm 0.0}}

\newcommand{\toitwothreetwotwoac}{\ensuremath{0.1293 \pm 0.0004}}
\newcommand{\toitwothreetwotwoincc}{\ensuremath{88.85 \pm 0.08}}
\newcommand{\toitwothreetwotwoTdurc}{\ensuremath{2.17 \pm 0.31}}
\newcommand{\toitwothreetwotwomassc}{\ensuremath{<46.87}}
\newcommand{\toitwothreetwotworadc}{\ensuremath{1.94 \pm 0.13}}
\newcommand{\toitwothreetwotworhoplc}{\ensuremath{<35}}
\newcommand{\toitwothreetwotwoteqc}{\ensuremath{501.0 \pm 2.0}}
\newcommand{\toitwothreetwotwoSc}{\ensuremath{10.5 \pm 0.0}}

\begin{document}

   \title{TOI-2322: two transiting rocky planets close to the stellar rotation period and its first harmonic}

   \author{M.\,J.\,Hobson
          \inst{\ref{unige}}
          \and
          A.\,Su\'arez\,Mascare\~{n}o
          \inst{\ref{iac}, \ref{laguna}}
          \and
          C.\,Lovis
          \inst{\ref{unige}}
          \and 
          F.\,Bouchy
          \inst{\ref{unige}}
          \and
          B.\,Lavie
          \inst{\ref{unige}}
          \and
          M.\,Cretignier
          \inst{\ref{oxford}}
          \and
          A.\,M.\,Silva
          \inst{\ref{IAporto}, \ref{uniporto}}
          \and
          S.\,G.\,Sousa
          \inst{\ref{IAporto}, \ref{uniporto}}
          \and
          H.\,M.\,Tabernero
          \inst{\ref{IEEC}, \ref{ICE}}
          \and
          V.\,Adibekyan
          \inst{\ref{IAporto}, \ref{uniporto}}
          \and
          C.\,Allende\,Prieto
          \inst{\ref{iac}, \ref{laguna}}
          \and 
          Y.\,Alibert
          \inst{\ref{unibe1}, \ref{unibe2}}
          \and
          S.\,C.\,C.\,Barros
          \inst{\ref{IAporto}, \ref{uniporto}}
          \and
          A.\,Castro-Gonz\'alez
          \inst{\ref{cab}}
          \and
          K.\,A.\,Collins
          \inst{\ref{CfA}}
          \and
          S.\,Cristiani
          \inst{\ref{inaf-trieste}, \ref{IFPU}}
          \and
          V.\,D'Odorico
          \inst{\ref{inaf-trieste}}
          \and
          M.\,Damasso
          \inst{\ref{inaf-torino}}
          \and
          D.\,Dragomir
          \inst{\ref{UNM}}
          \and
          X.\,Dumusque
          \inst{\ref{unige}}
          \and
          D.\,Ehrenreich
          \inst{\ref{unige}}
          \and
          P.\,Figueira
          \inst{\ref{IAporto}, \ref{uniporto}, \ref{ESO}}
          \and
          R.\,G\'enova\,Santos
          \inst{\ref{iac}}
          \and
          B.\,Goeke
          \inst{\ref{MIT}}
          \and
          J.\,I.\,Gonz\'alez\,Hern\'andez
          \inst{\ref{iac}, \ref{laguna}}
          \and
          K.\,Hesse
          \inst{\ref{MIT}}
          \and
          J.\,Lillo-Box
          \inst{\ref{cab}}
          \and
          G.\,Lo\,Curto
          \inst{\ref{ESO}}
          \and
          C.\,J.\,A.\,P.\,Martins
          \inst{\ref{astro-porto}, \ref{IAporto}}
          \and
          A.\,Mehner
          \inst{\ref{ESO}}
          \and
          G.\,Micela
          \inst{\ref{inaf-palermo}}
          \and
          P.\,Molaro
          \inst{\ref{inaf-trieste}}
          \and
          N.\,J.\,Nunes
          \inst{\ref{lisboa}}
          \and
          E.\,Palle
          \inst{\ref{iac}, \ref{laguna}}
          \and
          V.\,M.\,Passegger
          \inst{\ref{subaru}, \ref{iac}, \ref{laguna}, \ref{hamburg}}
          \and
          F.\,Pepe
          \inst{\ref{unige}}
          \and
          R.\,Rebolo
          \inst{\ref{iac}}
          \and
          J.\,Rodrigues
          \inst{\ref{IAporto}, \ref{uniporto}}
          \and
          N.\,Santos
          \inst{\ref{IAporto}, \ref{uniporto}}
          \and
          A.\,Sozzetti
          \inst{\ref{inaf-torino}}
          \and
          B.\,M.\,Tofflemire
          \inst{\ref{SETI}}
          \and
          S.\,Udry
          \inst{\ref{unige}}
          \and
          C.\,Watkins
          \inst{\ref{CfA}}
          \and 
          M.-R.\,Zapatero\,Osorio
          \inst{\ref{cab}}
          \and
          C.\,Ziegler
          \inst{\ref{austin}}
          }

   \institute{Observatoire de Gen\`eve, D\'epartement d'Astronomie, Universit\'e de Gen\`eve, Chemin Pegasi 51b, 1290 Versoix, Switzerland \label{unige}\\
              \email{melissa.hobson@unige.ch}
    \and
    Instituto de Astrof\'isica de Canarias, c/ V\'ia L\'actea s/n, 38205 La Laguna, Tenerife, Spain\label{iac}
    \and
    Departamento de Astrof\'isica, Universidad de La Laguna, 38206 La Laguna, Tenerife, Spain\label{laguna}
    \and
    Department of Physics, University of Oxford, OX13RH Oxford, UK\label{oxford}
    \and
    Instituto de Astrof\'isica e Ci\^encias do Espa\c{c}o, CAUP, Universidade do Porto, Rua das Estrelas, 4150-762, Porto, Portugal\label{IAporto}
    \and
    Departamento de F\'isica e Astronomia, Faculdade de Ci\^encias, Universidade do Porto, Rua do Campo Alegre, 4169-007, Porto, Portugal\label{uniporto}
    \and
    Institut d'Estudis Espacials de Catalunya (IEEC), Edifici RDIT, Campus UPC, 08860 Castelldefels (Barcelona), Spain\label{IEEC}
    \and
    Institut de Ci\`encies de l’Espai (ICE, CSIC), Campus UAB, c/ de Can Magrans s/n, 08193 Cerdanyola del Vall\`es, Barcelona, Spain\label{ICE}
    \and
    Physics Institute, University of Bern, Gesellsschaftstrasse 6, CH-3012 Bern, Switzerland\label{unibe1}
    \and
    Center for Space and Habitability, University of Bern, Gesellsschaftstrasse 6, CH-3012 Bern, Switzerland\label{unibe2}
    \and
    Centro de Astrobiolog\'{i}a, CSIC-INTA, Camino Bajo del Castillo s/n, 28692 Villanueva de la Ca\~{n}ada, Madrid, Spain\label{cab}
    \and
    Center for Astrophysics \textbar \ Harvard \& Smithsonian, 60 Garden Street, Cambridge, MA 02138, USA\label{CfA}
    \and
    INAF- Osservatorio Astronomico di Trieste, via G. B. Tiepolo 11, I-34143, Trieste, Italy\label{inaf-trieste}
    \and
    IFPU–Institute for Fundamental Physics of the Universe, via Beirut 2, I-34151 Trieste, Italy\label{IFPU}
    \and
    INAF - Osservatorio Astrofisico di Torino, Via Osservatorio 20, 10025 Pino Torinese, Italy\label{inaf-torino}
    \and
    Department of Physics and Astronomy, University of New Mexico, 210 Yale Boulevard, Albuquerque, NM 87131, USA\label{UNM}
    \and
    European Southern Observatory, Av. Alonso de Cordova, 3107, Vitacura, Santiago de Chile, Chile\label{ESO}
    \and
    Kavli Institute for Astrophysics and Space Research, Massachusetts Institute of Technology,  70 Vassar St, Cambridge, MA 02139\label{MIT}
    \and
    Centro de Astrof\'{\i}sica da Universidade do Porto, Rua das Estrelas, 4150-762 Porto, Portugal\label{astro-porto}
    \and
    INAF - Osservatorio Astronomico di Palermo, Piazza del Parlamento 1, 90134, Palermo, Italy\label{inaf-palermo}
    \and
    Instituto de Astrof\'isica e Ci\^encias do Espa\c{c}o, Faculdade de Ci\^encias da Universidade de Lisboa, 1749-016 Lisboa, Portugal\label{lisboa}
    \and
    Hamburger Sternwarte, Gojenbergsweg 112, 21029, Hamburg, Germany\label{hamburg}
    \and
    Subaru Telescope, National Astronomical Observatory of Japan (NAOJ), 650 N Aohoku Place, Hilo, HI, 96720, USA\label{subaru}
    \and
    SETI Institute, Mountain View, CA 94043 USA/NASA Ames Research Center, Moffett Field, CA 94035 USA\label{SETI}
    \and
    Department of Physics, Engineering and Astronomy, Stephen F. Austin State University, 1936 North St, Nacogdoches, TX 75962, USA\label{austin}
             }

   \date{Submitted 21 May 2025; revised 14 August 2025; accepted 25 August 2025}

 
  \abstract
   {Active regions on the stellar surface can induce quasi-periodic radial velocity (RV) variations that can mimic planets and mask true planetary signals. These spurious signals can be problematic for RV surveys such as those carried out by the ESPRESSO consortium.}
   {Using ESPRESSO and HARPS RVs and activity indicators, we aim to confirm and characterize two candidate transiting planets from TESS orbiting a K4 star with strong activity signals.}
   {From the ESPRESSO FWHM, TESS photometry, and ASAS-SN photometry, we measure a stellar rotation period of $\mathrm{21.28\pm0.08\, d}$. We jointly model the TESS photometry, ESPRESSO and HARPS RVs, and activity indicators, applying a multivariate Gaussian Process (GP) framework to the spectroscopic data.}
   {We are able to disentangle the planetary and activity components, finding that TOI-2322 b has a \Pb d period, close to the first harmonic of the rotation period, a $\leq 2.03 M_\oplus$ mass upper limit and a \rpbTESS$\mathrm{R_\oplus}$ radius. TOI-2322 c orbits close to the stellar rotation period, with a \Pc d period; it has a \mpc$\mathrm{M_\oplus}$ mass and a \rpcTESS$\mathrm{R_\oplus}$ radius.}
   {The multivariate GP framework is crucial to separating the stellar and planetary signals, significantly outperforming a one-dimensional GP. Likewise, the transit data is fundamental to constraining the periods and epochs, enabling the retrieval of the planetary signals in the RVs. The internal structure of TOI-2322 c is very similar to that of Earth, making it one of the most massive planets with an Earth-like composition known.}

   \keywords{Planets and satellites: detection -- Planets and satellites: composition -- Techniques: photometric -- Techniques: radial velocities -- Stars: rotation -- Stars: individual: TOI-2322
               }

   \maketitle
%
\section{Introduction}

From the start of radial velocity (RV) exoplanet-hunting surveys, it was clear that stellar activity would pose an obstacle to their endeavours \citep[e.g.][]{Saar1997, Santos2000}. Different stellar phenomena such as oscillations, granulation, magnetic activity, and magnetic cycles induce spurious signals on a range of timescales from minutes to years (see \citealt{Dumusque2014, Dumusque2018} and references therein). Particularly problematic to the detection of exoplanets are active regions - dark spots and bright faculae - generated by magnetic activity on the stellar surface. These distort stellar line profiles, creating spurious quasi-periodic RV signals that match the stellar rotation period, as the active regions rotate into and out of view, but also evolve over time \citep[e.g.][]{Saar1997, Lagrange2010, Lovis2011}. Persistent active regions can mimic planetary signals, as shown by numerous disproven "planets" from \cite{Queloz2001}, the first such case, to systems such as Gl 581 \citep{Forveille2011,Baluev2013} or Barnard's star \citep{Lubin2021,Artigau2022} where quasi-periodic activity signals persisted for years.

In order to correctly model planets around these active stars, this stellar activity needs to be accounted for. To do this, Gaussian Processes \citep[GPs,][]{Haywood2014, Rajpaul2015} have become the typically adopted framework. To constrain the GPs, knowledge of the stellar activity is necessary. This information can be provided by spectroscopic stellar activity indicators. Broadly speaking, these consist of two types: the measurement of specific activity-sensitive stellar lines, such as the $\mathrm{H_\alpha}$ line \citep{Cincunegui2007,Bonfils2007} or the Ca II H and K lines \citep{Vaughan1978,Noyes1984}; or the measurement of changes in the shape of the cross-correlation function (CCF) used to measure the RVs \citep{Baranne1996, Queloz2001}. Persistent active regions on the stellar surface will imprint quasi-periodic variability on these indicators at the stellar rotation period and its harmonics.

In this paper, we present the TOI-2322 system, consisting of a K4 star with clear activity signals and two transiting planets orbiting it close to the rotation period and its first harmonic respectively. The planets were first identified as candidates by the Transiting Exoplanet Survey Satellite (TESS) NASA mission \citep{Ricker2015}. We confirm and characterize these planets using RVs and activity indicators from the Echelle SPectrograph for Rocky Exoplanets and Stable Spectroscopic Observations (ESPRESSO, \citealt{Pepe2013, Pepe2021}). Located at ESO's Very Large Telescope (VLT), Paranal, Chile, ESPRESSO reaches an RV precision of better than $\mathrm{25\, cm\, s^{-1}}$ during a single night. One of the goals of the ESPRESSO Guaranteed Time Observations \citep[GTO,][]{Pepe2021} is the follow-up of small planet candidates from TESS and \textit{Kepler}, for which ESPRESSO's exquisite RV precision is vital. We also incorporate RVs and activity indicators from the High Accuracy Radial velocity Planet Searcher spectrograph \citep[HARPS,][]{Mayor2003}, which is located on the 3.6 m telescope at La Silla Observatory, Chile, to increase the temporal baseline and improve the sampling.

The paper is structured as follows: we describe the data in Section \ref{s:observations}. Section \ref{s:star-params} presents the stellar characterization, and Section \ref{s:analysis} the system modelling. We discuss our results in Section \ref{s:discussion} and conclude in Section \ref{s:conclusions}.

\section{Observations}\label{s:observations}

\subsection{TESS photometry}

TOI-2322 is in the southern TESS continuous viewing zone. It was observed by TESS in sectors 1-3 and 5-13 of the prime mission, from 25 July 2018 to 14 October 2018 and 15 November 2018 to 17 July 2019; sectors 27-33 and 35-39 of the first extended mission, from 5 July 2020 to 13 January 2021 and 2 February 2021 to 24 June 2021; and sectors 61-63 and 65-69 of the second extended mission, from 18 January 2023 to 6 April 2023 and 4 May 2023 to 20 September 2023. Camera 4 was used throughout. CCD 1 was used for sectors 35-37 and 61-63; CCD 2 for sectors 11-13, 27, 38-39, and 65-67; CCD 3 for sectors 1-3, 28-30, and 68-69; and CCD 4 for sectors 5-7 and 31-33. During the prime mission, it was observed only in the full frame images at 30 minute cadence. For the extended missions, it was observed at 2-minute cadence.

The TESS photometry was processed by the TESS Science Processing Operations Center \citep[SPOC,][]{Jenkins2016} at NASA Ames Research Center. Two potential transit signals at $\mathrm{\sim 20.2\, d}$ and $\mathrm{\sim 11.3\, d}$ were identified by the TESS Quick Look Pipeline \citep{Kunimoto2021}, and designated as TESS Objects of Interest TOI-2322.01 and TOI-2322.02 by the TESS Science Office \citep{Guerrero2021} on 7 October 2020 and 12 August 2021 respectively. We highlight that the longer-period signal was identified first, due to its greater depth, and is thus numbered first. For our analysis, we use both the SPOC Simple Aperture Photometry (SAP) light curves \citep{Twicken2010, Morris2020} and the SPOC Presearch Data Conditioning Simple Aperture Photometry (PDC-SAP) light curves \citep{Stumpe2012,Stumpe2014,Smith2012}. These are only available for the 2-minute cadence data from the extended missions, which is shown in Fig. \ref{fig:pdcsap-lightcurves}. Given the wealth of 2-minute cadence data, we do not include the prime mission lower-cadence photometry. We also note that the transits of these targets are of relatively short duration, spanning less than three hours from ingress to egress; at 30-minute cadence, there are only some four to six points within each transit, making it difficult to correctly fit for the transit shape.

\begin{figure}[htb]
    \centering
    \includegraphics[width=.5\textwidth]{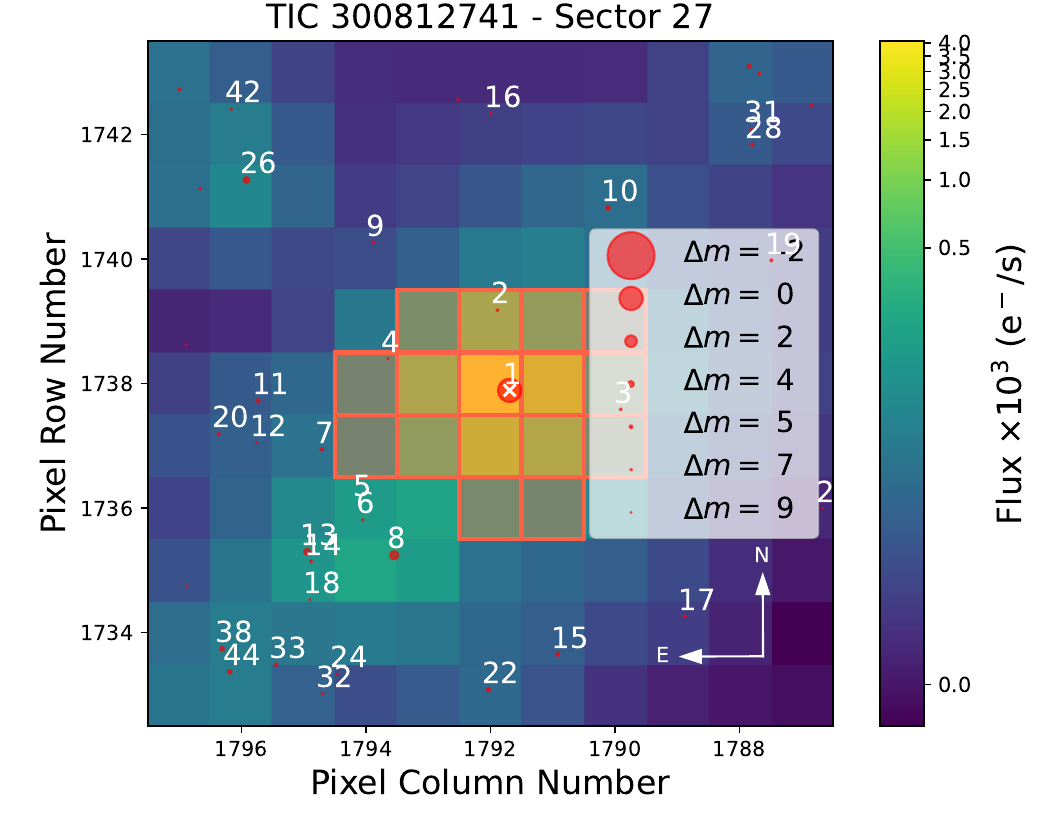}
    \caption{TESS target pixel file for TOI-2322 for sector 27. The target star is labelled as 1 and marked by a white cross. All sources from the \textit{Gaia} DR3 catalogue down to a magnitude contrast of 9 are shown as red circles, with the size proportional to the contrast. The SPOC pipeline aperture is overplotted in shaded red squares.}
    \label{fig:tpfplotter}
\end{figure}

TESS has a large pixel size of $21\arcsec$ per pixel, meaning the photometry can be contaminated by nearby stars. We used the \texttt{tpfplotter} package\footnote{Available at \url{https://github.com/jlillo/tpfplotter}} \citep{Aller2020} and the \texttt{TESS-cont} algorithm\footnote{Available at \url{https://github.com/castro-gzlz/TESS-cont}} \citep{Castro2024} to search for potential contaminants in the TESS aperture used for TOI-2322. Nearby stars with magnitude contrast down to 9 could mimic the smaller of the transit depths, that of TOI-2322.02 \citep{LilloBox2014}. Figure \ref{fig:tpfplotter} shows the TESS target pixel file and SPOC pipeline aperture for sector 27. Accounting for the relative location of the nearby sources with respect to the SPOC aperture and the TESS pixel response functions (PRFs), \texttt{TESS-cont} finds that a 6.5$\%$ eclipse in the most contaminating source (TIC 300812728; Star$\#$8 in Fig.~\ref{fig:tpfplotter}) could generate the transit signal TOI-2322.02. A similar conclusion is reached for TIC 300812734 (Star$\#$7) and TIC 764823031 (Star$\#$2), with 29.8$\%$ and 33.3$\%$ eclipses potentially generating the observed signal. We note that we discard this possibility through follow-up transit  observations (see Sect. \ref{s:ground-photom}), and that the PDCSAP data already account for flux dilution, so no additional corrections were necessary. The target pixel files for the remaining sectors with PDCSAP 2-minute cadence data are shown in Figs. \ref{fig:tpfplotter-EM1} and \ref{fig:tpfplotter-EM2}.

As a complementary measurement to the ground-based follow-up, we obtain the difference-image centroid measurement from the SPOC analysis. The Data Validation module of the SPOC pipeline \citep{Twicken2018} measures the location of the TOI-2322.01 transits to be $2.5 \pm 3.1 \arcsec$ from TOI-2322, i.e., consistent with the presumed host. This measurement also excludes the surrounding stars as sources of the transits.

\subsection{Ground-based photometry}\label{s:ground-photom}

As the TESS pixel size is large and nearby stars can thus contaminate the photometry, follow-up ground-based photometry is used to vet the nearby stars as potential background eclipsing binaries (EBs), which can cause false-positive exoplanet detections \citep{LilloBox2012}. We observed a full transit window of TOI-2322.01 continuously for 145 minutes in Sloan $i'$ band on UTC 2022 November 23 from the Las Cumbres Observatory Global Telescope (LCOGT) \citep{Brown2013} 1\,m network node at Siding Spring Observatory near Coonabarabran, Australia (SSO). The 1\,m telescope is equipped with a $4096\times4096$ SINISTRO camera having an image scale of $0\farcs389$ per pixel, resulting in a $26\arcmin\times26\arcmin$ field of view. The images were calibrated by the standard LCOGT {\tt BANZAI} pipeline \citep{McCully2018} and differential photometric data were extracted using {\tt AstroImageJ} \citep{Collins2017}. We used the {\tt TESS Transit Finder}, which is a customized version of the {\tt Tapir} software package \citep{Jensen2013}, to schedule our transit observations.

The TOI-2322.01 SPOC pipeline transit depth of 460\,ppm is generally too shallow to reliably detect with ground-based observations, so we instead checked for possible nearby eclipsing binaries (NEBs) that could be contaminating the TESS photometric aperture and causing the TESS detection. To account for possible contamination from the wings of neighbouring star PSFs, we searched for NEBs out to $2\farcm5$ from TOI-2322. If fully blended in the SPOC aperture, a neighbouring star that is fainter than the target star by 8.4 magnitudes in TESS-band could produce the SPOC-reported flux deficit at mid-transit (assuming a 100\% eclipse). To account for possible TESS magnitude uncertainties and possible delta-magnitude differences between TESS-band and Sloan $i'$ band, we included an extra 0.5 magnitudes fainter (down to TESS-band magnitude 18.9). We calculated the RMS of each of the 62 nearby star light curves (binned in 10 minute bins) that meet our search criteria and find that the values are smaller by at least a factor of 5 compared to the required NEB depth in each respective star. We then visually inspected each neighbouring star's light curve to ensure that there is no obvious eclipse-like signal. Our analysis ruled out an NEB blend as the cause of the SPOC pipeline TOI-2322.01 detection in the TESS data. All LCOGT light curve data are available on the {\tt EXOFOP-TESS} website\footnote{Located at \url{https://exofop.ipac.caltech.edu/tess/target.php?id=300812741}.}.

\subsection{ESPRESSO radial velocities}

TOI-2322 was observed by the ESPRESSO GTO 33 times between 13 November 2022 and 26 March 2023, after the 2019 fibre link replacement \citep{Pepe2021}, under program IDs 110.24CD.002, 110.24CD.003, and 110.24CD.009 (P.I. F. Pepe). The observations were done with an exposure time of 900 s, save one observation on 28 November 2022 with an increased exposure time of 1200 s due to poor weather conditions. The spectra have a median S/N of 63 at 550 nm. They were processed with the data reduction software (DRS, \citealt{Pepe2021}) v.3.2.5, in which the radial velocities are obtained through the CCF method \citep{Baranne1996}. The K6 mask was used to obtain the CCFs. The resulting RVs have a median error of 1.2 m/s.

In addition to RVs, the DRS also computes several activity indicators. Three of these are measured from the CCF: the full width at half maximum (FWHM), contrast, and bisector inverse slope \citep[BIS,][]{Queloz2001}. Others measure chromospheric emission in the cores of specific lines: the Mount-Wilson S-index ($\mathrm{S_{MW}}$, \citealt{Vaughan1978}) and $\log R'_{\rm hk}$ \citep{Noyes1984}, for the Ca II H and K lines; the $\mathrm{H_\alpha}$ index \citep{Cincunegui2007,Bonfils2007}, for the $\mathrm{H_\alpha}$ line; the Na index \citep{Diaz2007}, for the Na I D1 and D2 lines.

We downloaded the RVs and activity indicators from the Data \& Analysis Center for Exoplanets (DACE) platform\footnote{Available online at \url{https://dace.unige.ch}}. They are listed in Appendix \ref{ap:RVs}, and the RVs are shown in Fig. \ref{fig:RVs_FWHM_BIS_pyaneti}. The generalized Lomb-Scargle \citep[GLS,][]{Zechmeister2009} periodograms of the ESPRESSO RVs and activity indicators are shown in Fig. \ref{fig:spectral_periodograms_ESPRESSO}. 

\subsection{HARPS radial velocities}

TOI-2322 was observed with HARPS 19 times between 8 October 2022 and 16 January 2023, under program ID 110.242T.001 (P.I. R. Rebolo). The observations were done with an exposure time of 2700 s. The spectra have a median S/N of 52 at 550 nm. They were processed with a HARPS-adapted version of the ESPRESSO DRS v.3.2.5. As with ESPRESSO, the K6 mask was used to obtain the CCFs. The RVs have a median error of 1.8 m/s.

We again downloaded the RVs and activity indicators from the DACE platform. They are listed in Appendix \ref{ap:RVs}, and the RVs are shown in Fig. \ref{fig:RVs_FWHM_BIS_pyaneti}. The GLS periodograms of the HARPS RVs and activity indicators are shown in Fig. \ref{fig:spectral_periodograms_HARPS}.

\begin{figure}[ht!]
    \centering
    \includegraphics[width=.43\textwidth]{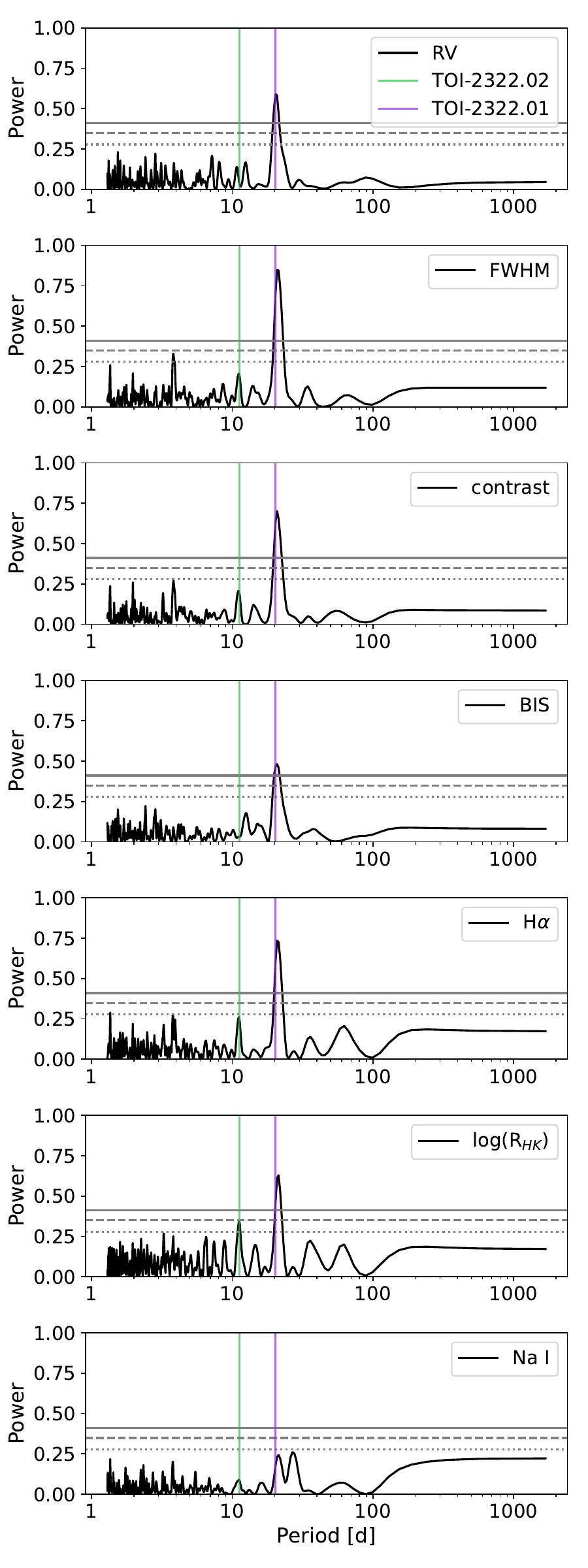}
    \caption{GLS periodograms of the combined ESPRESSO and HARPS RVs (top) and activity indicators (second to bottom: CCF FWHM, CCF contrast, CCF bisector, $\mathrm{H_\alpha}$, $\mathrm{\log R'_{HK}}$, and Na I). The vertical green and purple lines indicate the periods of TOI-2322.02 and TOI-2322.01 respectively. The dotted, dashed, and solid horizontal grey lines indicate the 10\%, 1\%, and 0.1\% FAP levels respectively.}
    \label{fig:spectral_periodograms_joint}
\end{figure} 

We show the GLS periodograms of the joint ESPRESSO and HARPs RVs and activity indicators in Fig. \ref{fig:spectral_periodograms_joint}. To compensate for any potential offset between the two instruments, we have subtracted the medians from each individual time-series prior to generating the periodograms. Significant signals appear close to the $\mathrm{\sim 20\, d}$ period of TOI-2322.01 in both the RVs and all activity indicators save the Na index. The Na index likewise does not show significant periodic signals in the individual instruments; this contrasts with the RVs and other activity indicators, which generally show clear signals in the individual HARPS and ESPRESSO datasets (see Figs. \ref{fig:spectral_periodograms_HARPS} and \ref{fig:spectral_periodograms_ESPRESSO}), although in the smaller HARPS dataset they do not reach as high significance as for ESPRESSO. In general, the Na index is a better activity tracer for M-dwarfs \citep[e.g.][]{Gomes2011}, and has been seen to be a poor tracer for K-dwarfs \citep[e.g.][]{Barragan2023}. With a $\mathrm{B-V=1.14}$, it is also close to the $\mathrm{B-V=1.1}$ threshold below which \cite{Diaz2007} found no clear correlation with activity. There are also smaller tentative peaks close to the $\mathrm{\sim 11\, d}$ period of TOI-2322.02 in several activity indicators.


\subsection{High-contrast imaging}

\begin{figure}[htb!]
    \centering
    \includegraphics[width=1\hsize]{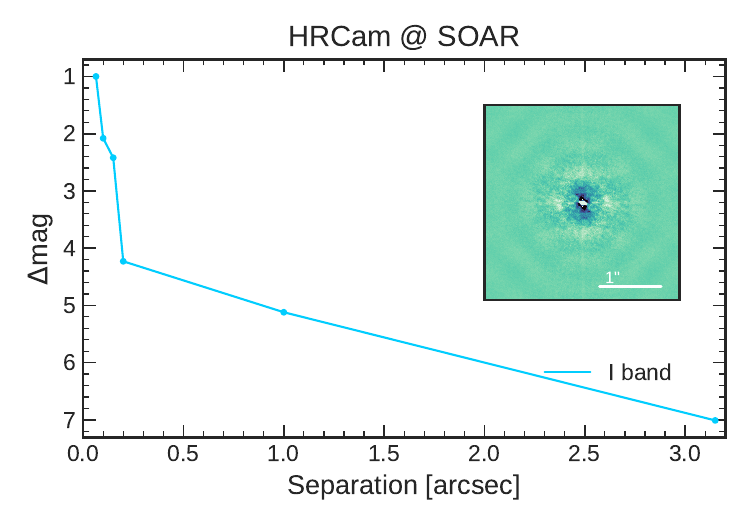}
    \caption{Contrast curve and speckle auto-correlation function from the HRCam at SOAR for TOI-2322. The cyan points and solid line indicate the $5\sigma$ contrast curve; the inset shows the speckle auto-correlation function. No nearby sources are detected.}
    \label{fig:speckle-imaging}
\end{figure}

In addition to ground-based photometry, high-contrast imaging is valuable to vet stars hosting close companions. As well as false positives from close EBs, flux from any close additional source(s) can lead to an underestimated planetary radius if not accounted for in the transit model \citep{Ciardi2015, Furlan2017, Matson2018, Castro2022}. TOI-2322 was observed with optical speckle imaging by the SOAR TESS survey \citep{Ziegler2020}, which uses the high-resolution camera (HRCam) imager on the 4.1-m Southern Astrophysical Research (SOAR) telescope at Cerro Pachón, Chile \citep{Tokovinin2018}. The observations were performed on 31 October 2020, with no nearby sources detected within $3\arcsec$. The contrast curve and speckle auto-correlation function are shown in Fig. \ref{fig:speckle-imaging}.

\section{Stellar parameters}\label{s:star-params}

The stellar parameters for TOI-2322 are listed in Table \ref{tab:starparams}. We used the \textit{Gaia} Data Release 3 \citep{GAIA2016, GaiaDR3} for its stellar coordinates, proper motions, and parallaxes. We further characterised TOI-2322 using the co-added ESPRESSO spectra, obtaining the atmospheric parameters via spectral synthesis with the {\sc SteParSyn} code\footnote{Available at \url{https://github.com/hmtabernero/SteParSyn/}} \citep{Tabernero2022}. {\sc SteParSyn} provides the effective temperature $\mathrm{T_{eff}}$, metallicity [Fe/H], surface gravity $\log{g}$, and broadening parameter $\mathrm{v_{broad}}$, which accounts for both the macroturbulence $\zeta$ and the projected rotational velocity $\mathrm{v \sin i}$. We performed an independent validation with the combined ARES+MOOG approach of \cite{Sousa2014}. The resulting parameters of $\mathrm{T_{eff} = 4499 \pm 126\,K}$, $\mathrm{[Fe/H] = -0.14 \pm 0.05}$, and $\log{g} = 4.56 \pm 0.08$
are compatible with those derived by {\sc SteParSyn}.

To determine the physical parameters of the star, we followed the procedure of \cite{Brahm2019PARSEC}, in which the broadband photometric measurements (converted to absolute magnitudes via the \textit{Gaia} DR3 \citep{GAIA2016, GaiaDR3} parallax) are compared with the stellar evolutionary models of \cite{Bressan2012}. We used the {\sc SteParSyn} $\mathrm{T_{eff}}$ and [Fe/H] as input values, taking the sum of the internal and systematic errors in quadrature. From this method we obtained the mass, radius, bolometric luminosity, and stellar age. We also obtain an independent estimation of $\log{g} = 4.64 \pm 0.01$, consistent with the spectroscopic measurement within the systematic error bar.

\begin{table}[th]
\caption{Stellar parameters for TOI-2322.}
\label{tab:starparams}
\resizebox{.5\textwidth}{!}{ 
\begin{tabular}{lcr}
\hline \hline
Parameter & Value & Reference \\
\hline
Names     &  TYC 9184-175-1  & Simbad ID          \\
          & TIC 300812741 & TESS \\
          & TOI-2322 & TESS \\
          &  	5264331179802870016 & \textit{Gaia} DR3 \\
RA \dotfill (J2000) & $\mathrm{07^h47^m53^s.51969573 \pm 0.00000071}$ & \textit{Gaia} DR3 \\
DEC \dotfill (J2000) & $-71{\degr}00{\arcmin}06{\arcsec}.2575302 \pm 0.0000096 $ &  \textit{Gaia} DR3 \\
pm$^{\rm RA}$ \hfill [mas yr$^{-1}$] & $7.866 \pm 0.014$ & \textit{Gaia} DR3 \\
pm$^{\rm DEC}$ \hfill [mas yr$^{-1}$] & $-53.146 \pm 0.011$ & \textit{Gaia} DR3 \\
$\pi$ \dotfill [mas] & $16.6964 \pm 0.0102$ & \textit{Gaia} DR3 \\
\hline
T \dotfill [mag] & 10.0534 & TESS \\
B \dotfill [mag] & 12.05 & \textit{Tycho-2} \\
V \dotfill [mag] & 10.91 & \textit{Tycho-2} \\
J \dotfill [mag] & 9.093 & 2-MASS \\
H \dotfill [mag] & 8.521 & 2-MASS \\
K \dotfill [mag] & 8.369 & 2-MASS \\
\hline 
$\mathrm{T_{eff}}$ \dotfill [K] &    $4664 \pm 14 \,(70)^\dagger$   & this work \\
Fe/H \dotfill [dex] & $-0.12 \pm 0.02 \,(0.10)$ & this work \\
$\log{g}$ \dotfill [dex] & $4.57 \pm 0.04 \,(0.10)$ & this work \\
$\mathrm{v_{broad}}$ \dotfill [$\mathrm{km \, s^{-1}}$] & $1.59	\pm 0.03$ & this work \\
R$_\star$ \dotfill [R$_\odot$] & $0.662\pm0.004\, (0.03)$ & this work \\
M$_\star$ \dotfill [M$_\odot$] & $0.703_{-0.017}^{+0.014}\, (0.04)$ & this work \\
L$_{bol,\star}$ \dotfill [L$_\odot$] & $0.181\pm0.007$ & this work \\
Age \dotfill [Gyr] & $3.9_{-2.5}^{+3.6}$ & this work \\
Spectral type \dotfill & K4 & PM13 \\
$\mathrm{\log R'_{HK}}^\ddagger$ \dotfill & $-4.587 \pm 0.053$ & this work \\
$\mathrm{P_{rot}}$ \dotfill [d] & $21.28 \pm 0.08$  & this work \\
\hline
\end{tabular}
}
Simbad: Simbad astronomical database \citep{Wenger2000}; TESS: TESS Input Catalog \citep{Stassun2019}; 2MASS: Two-micron All Sky Survey \citep{2MASS}; \textit{Gaia} DR3: \textit{Gaia} Data Release 3 \citep{GAIA2016, GaiaDR3}; \textit{Tycho-2}: the \textit{Tycho-2} Catalogue \citep{Tycho-2}; PM13: using the tables of \cite{Pecaut2013}.\\
$\dagger$: The error bars for the parameters derived in this work correspond to the internal statistical precision measurements. For the atmospheric parameters, the typical systematic error bars for late K-M stars \citep{Tabernero2022} are given in parentheses. For the mass and radius, we give the error floors of \citep{Tayar2022} in parentheses.\\
$\ddagger$: median and standard deviation of the joint median-corrected $\mathrm{\log R'_{HK}}$ values from ESPRESSO and HARPS.\\
\end{table}

\subsection{Stellar rotation}

\begin{figure}[htb!]
    \centering
    \includegraphics[width=.5\textwidth]{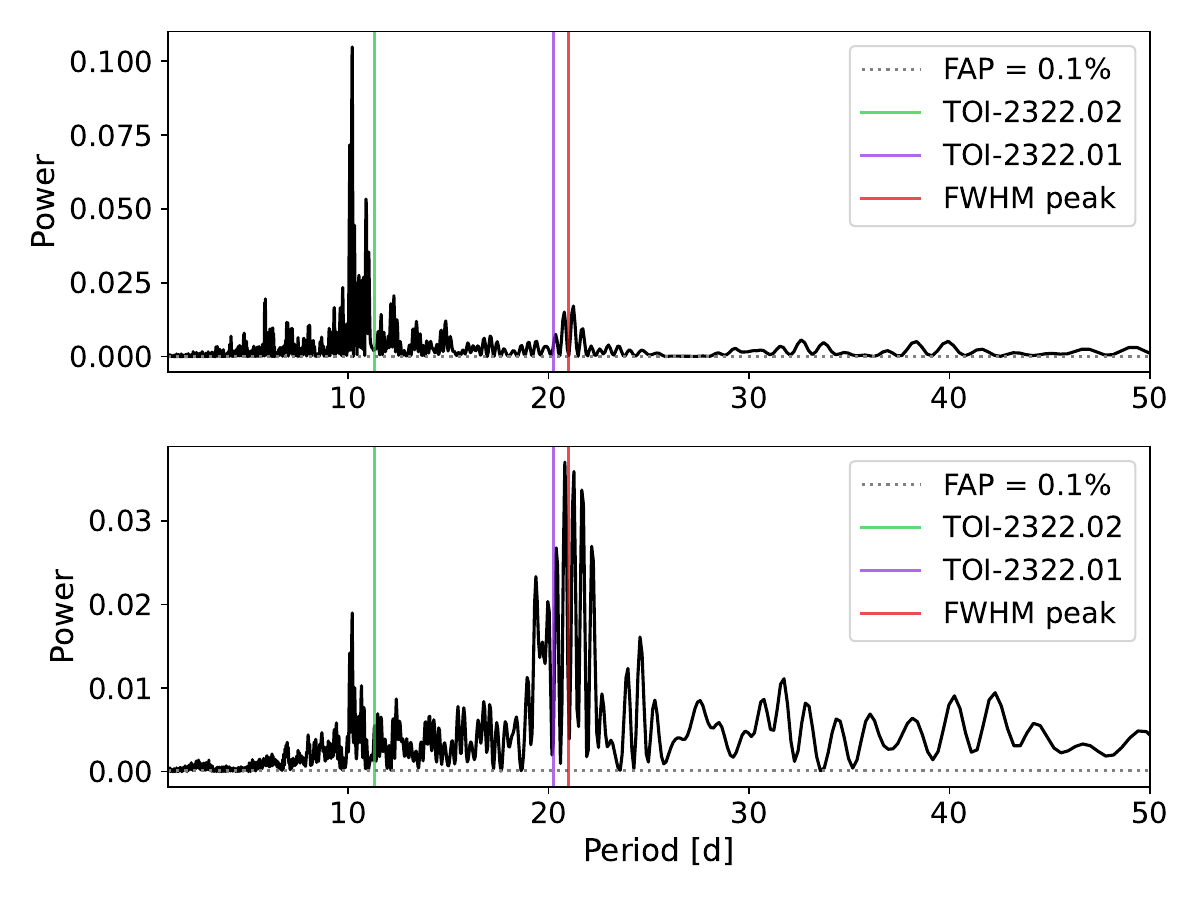}
    \caption{GLS periodogram of the TESS PDCSAP light curves (top) and SAP light curves (bottom). The vertical green and purple lines indicate the periods of TOI-2322.02 and TOI-2322.01 respectively, while the vertical red line indicates the period of the highest power seen in the FWHM periodogram. The dotted horizontal grey line indicates the 0.1\% FAP level.}
    \label{fig:tess_pdc_sap_periodogram}
\end{figure}

TOI-2322 shows strong evidence of stellar activity, with highly significant peaks at a $\mathrm{\sim 21 \,d}$ period in most of the spectral activity indicators (Fig \ref{fig:spectral_periodograms_joint}). The TESS PDCSAP light curves, which we use for our transit fits, likewise show strong variability (Fig. \ref{fig:pdcsap-lightcurves}). We computed the GLS periodogram of the PDCSAP light curves, which is shown in Fig. \ref{fig:tess_pdc_sap_periodogram}. While the highest power is obtained for a period of $\mathrm{\sim 10 \,d}$, there are also peaks at the $\mathrm{\sim 21 \,d}$ period where all the spectral activity indicators peak. Since the PDC systematic error correction algorithm is known to distort astrophysical signals with timescales longer than about half a TESS sector ($\sim$ 14 days)\footnote{See the Kepler Data Processing Handbook, \url{https://archive.stsci.edu/files/live/sites/mast/files/home/missions-and-data/kepler/_documents/KSCI-19081-003-KDPH.pdf}, for a detailed description of PDC}, we derive a more robust estimate of the stellar rotation by computing the GLS periodogram of the SPOC Simple Aperture Photometry (SAP) light curves. The light curves are shown in Fig. \ref{fig:sap-lightcurves} and the GLS periodogram in Fig. \ref{fig:tess_pdc_sap_periodogram}. The highest power is seen at $\mathrm{\sim 21 \,d}$. 

\begin{figure}[htb!]
    \centering
    \includegraphics[width=.5\textwidth]{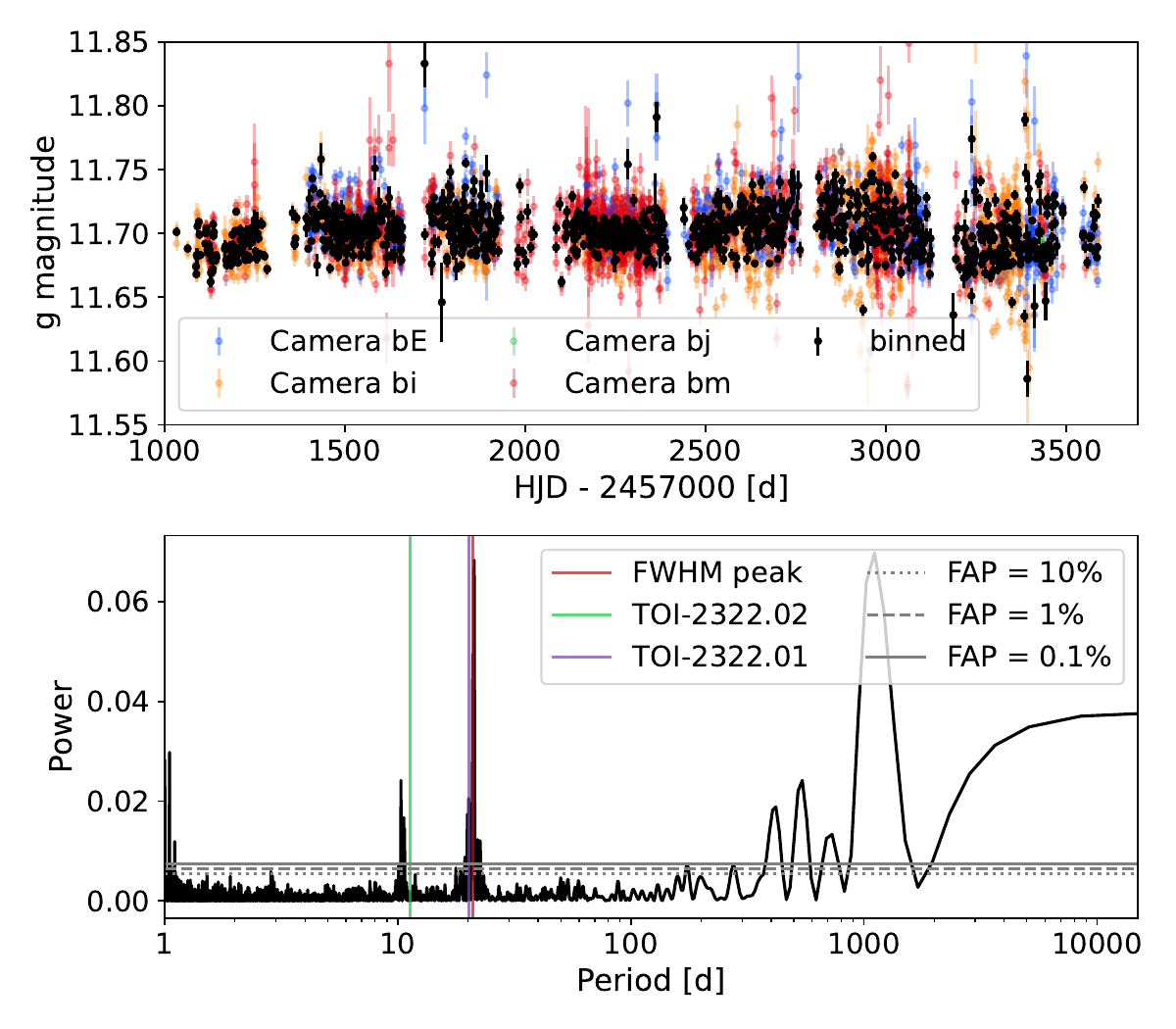}
    \caption{Light curve (top, separated by camera in colours, and 1-day binned data in black) and GLS periodogram (bottom) of the ASAS-SN g-band photometry. The vertical green and purple lines indicate the periods of TOI-2322.02 and TOI-2322.01 respectively, while the vertical red line indicates the period of highest power seen in the FWHM periodogram. The dotted, dashed, and solid horizontal grey lines indicate the 10\%, 1\%, and 0.1\% FAP levels respectively.}
    \label{fig:asas-sn_periodogram}
\end{figure}

We also obtained ground-based g-band photometry from the All-Sky Automated Survey for Supernovae \citep[ASAS-SN,][]{Shappee2014, Kochanek2017}, spanning the 7-year interval from 6 October 2017 to 5 October 2024, overlapping with the TESS and ESPRESSO observations. The light curve and periodogram are shown in Fig. \ref{fig:asas-sn_periodogram}. The strongest periodogram peaks are at $\mathrm{\sim 1111 \,d \approx 3 \,yr}$ and $\mathrm{\sim 21 \,d}$. There is also a smaller peak at $\mathrm{\sim 10 \,d}$.

Taking into account all the data, we conclude the most likely stellar rotation period is $\mathrm{P_{rot}\sim 21 \,d}$. The $\mathrm{\sim 10 \,d}$ peaks in the TESS PDSCAP periodogram would then correspond to the $\mathrm{P_{rot}/2}$ harmonic. To determine a final rotation period value, we fit Gaussians to the $\mathrm{\sim 21 \,d}$ peaks in the FWHM and ASAS-SN photometry periodograms, taking the mean and standard deviation as the $\mathrm{P_{rot}}$ value and error. For the FWHM, we obtain $\mathrm{P_{rot} = 21.3 \pm 1.4 \,d}$. For the ASAS-SN photometry, we obtain $\mathrm{P_{rot} = 21.28 \pm 0.08 \,d}$; we adopt this last, more precise value as our rotation period. 

\section{Data analysis}\label{s:analysis}

Modelling a planetary system with significant stellar activity is complex, particularly when the stellar rotation period is close to a planet period. Gaussian process regression has become a standard way to deal with quasi-periodic stellar activity signals \citep[see e.g.][]{Haywood2014, Rajpaul2015}. In this paper, we test two approaches: One incorporating a single GP with the \texttt{juliet}\footnote{Available at \url{https://github.com/nespinoza/juliet}} software \citep{Espinoza2019juliet}, detailed in Section \ref{s:juliet}, and a more sophisticated multivariate GP regression with the \texttt{pyaneti}\footnote{Available at \url{https://github.com/oscaribv/pyaneti}} software \citep{pyaneti, pyaneti2}, detailed in Section \ref{s:pyaneti}. We also test different GP kernels (Section \ref{s:kernels}) and RV extraction methods (Section \ref{s:data-red}), to ensure the robustness of our results.

\subsection{Preliminary joint fit with juliet}\label{s:juliet}

We first fit the ESPRESSO RVs and TESS PDCSAP photometry simultaneously using the \texttt{juliet} software. We tested four models: a single transiting planet on a circular orbit at the period of TOI-2322.02 (model $in$), a single transiting planet on a circular orbit at the period of TOI-2322.01 (model $out$), two transiting planets on circular orbits (model $2c$), and two transiting planets on free-eccentricity orbits (model $2e$). Given the strong stellar activity apparent in both the spectral activity indicators and the photometry, we used GPs to model it in both the RVs and photometry. 

Given the large quantities of TESS data, we use a two-step process where we first fit a GP to the out-of-transit data in each sector, then apply it to the in-transit data and use only this detrended in-transit data in the full fits. For this, we first mask the in-transit data, with a 5h padding of the transit window. We then fit the out-of-transit data using the (approximate) Matérn 3/2 kernel, as implemented in \texttt{celerite}, with broad log-uniform priors of $\mathcal{J}(1\times10^{-6}, 1\times10^6)$ for the GP amplitude $\mathrm{\sigma_{GP,sector}}$ and $\mathcal{J}(1\times10^{-3}, 1\times10^{3})$ for the GP length-scale $\mathrm{\rho_{GP,sector}}$. We also fit the instrumental parameters for each sector, taking broad priors of $\mathcal{N}(0.0,0.1)$ for the flux offset $\mathrm{m_{flux,sector}}$ and $\mathcal{J}(0.1,1000)$ for the jitter $\mathrm{\sigma_{w,sector}}$, and setting the dilution factor $\mathrm{m_{dilution,sector}}$ to 1. The priors and posteriors for these fits are listed in Tables \ref{tab:TESS_GPS_EM1} and \ref{tab:TESS_GPS_EM2}. Finally, we use these GPs to detrend the in-transit data, which is used in the full fit. This detrended in-transit data is also used for the \texttt{pyaneti} analysis. 

We present the details and full results of the \texttt{juliet} analysis in Appendix \ref{ap:juliet}. Here, we highlight that the $2c$ model is favoured by likelihood comparison, and that this single-GP approach is not sufficient to characterize this system, as we do not reach a $3\sigma$ measurement of the semi-amplitudes for either planet. With the $2c$ model, for TOI-2322.02 we find a period of $\mathrm{\jPpone\,d}$, a radius of $\mathrm{\toitwothreetwotworadb\,R_\oplus}$, and a semi-amplitude of $\mathrm{\jKpone\,m \, s^{-1}}$, leading to an upper mass limit of $\mathrm{\toitwothreetwotwomassb\,M_\oplus}$, while for TOI-2322.01 we find a period of $\mathrm{\jPptwo\,d}$, a radius of $\mathrm{\toitwothreetwotworadc\,R_\oplus}$, and a semi-amplitude of $\mathrm{\jKptwo\,m \, s^{-1}}$, leading to an upper mass limit of $\mathrm{\toitwothreetwotwomassc\,M_\oplus}$.

\subsection{Multivariate GP fit with pyaneti}\label{s:pyaneti}

While the \texttt{juliet} fit showed promising results, the uncertainties on the semi-amplitudes, and thus the masses, were large. Therefore, we decided to explore a more sophisticated modelling of the activity signal with \texttt{pyaneti}. \texttt{pyaneti} is an RV and transit modelling software, presented in \cite{pyaneti}, which allows for joint RV and transit fitting, using MCMC sampling to explore the parameter space. Its second version \citep{pyaneti2} also incorporates multidimensional GP regression, using the framework of \cite{Rajpaul2015}.

We simultaneously fit the same detrended TESS in-transit photometry and ESPRESSO RVs as used in the \texttt{juliet} fits. For this analysis, we also incorporated the HARPS RVs. We tested the same four $in$, $out$, $2c$, and $2e$ models. For all the models, we used a multivariate GP on the RVs and activity indicators to constrain the stellar activity impact on the RVs. The framework of \cite{Rajpaul2015} uses two activity indicators: one which depends only on the fractional coverage by active regions, such as $\mathrm{\log R'_{HK}}$ or FWHM; and one which, like the RV, is also affected by the stellar surface velocity at the active regions, such as BIS. \cite{Barragan2023} explored one-, two-, and three-dimensional GPs in this framework and concluded the three-dimensional GPs provided the best constraints. We thus elected to model the RVs simultaneously with two activity indicators that have the desired properties. Since our $\mathrm{\log R'_{HK}}$ time series shows some outliers and a slightly weaker periodicity at 21 d, we chose to use the FWHM as the first activity indicator. For the second, we examined the correlations between the RVs and each activity indicator. As the BIS shows a strong anti-correlation with the RVs, we selected it as our second activity indicator. We also note these indicators are the same ones that were used in \cite{Barragan2023} to model the active K3 star K2-233. Our GP model is thus expressed as:
\begin{flalign}
  \begin{gathered}
    RV = A0_{GP,rv}G(t) + A1_{GP,rv}\dot{G}(t) \hfill\\
    FWHM = A2_{GP,FWHM}G(t) \hfill\\
    BIS = A4_{GP,BIS}G(t) + A5_{GP,BIS}\dot{G}(t)
  \end{gathered}
\end{flalign}

With $G(t)$ a GP with a quasi-periodic covariance kernel:
\begin{equation}
    \gamma^{(G,G)}(t,t') = \exp \left\{-\frac{\sin^2[\pi(t-t')/P]}{2\lambda_p^2} - \frac{(t-t')^2}{2\lambda_e^2}\right\}
\end{equation}

For joint RV and transit fits, \texttt{pyaneti} takes as inputs the following parameters for each planet $\mathrm{i}$: period $\mathrm{P_{i}}$, time of mid-transit $\mathrm{t_{0,pi}}$, planet-to-star radius ratio $\mathrm{p_{i}}$, impact parameter $\mathrm{b_{i}}$, and either eccentricity $\mathrm{e_{i}}$ and angle of periastron $\mathrm{\omega_{i}}$ or a derived parametrisation such as $\mathrm{\sqrt{e_{i}} \sin \omega_{i}, \sqrt{e_{i}} \cos \omega_{i}}$, which is the parametrisation we adopt. It can fit either $\mathrm{a_{i}/R_\star}$ for each planet, or the stellar density $\rho$; we use the latter parametrisation. It also fits an offset $\mu$ and jitter $\sigma$ for each RV or activity time series, which for this analysis are fitted to the ESPRESSO and HARPS data separately, thus allowing for offsets between the two instruments. Likewise, it fits a jitter $\sigma$ and limb-darkening parameters $\mathrm{q_{1}}$ and $\mathrm{q_{2}}$ for each transit time series; here we treat all the TESS sectors as a single instrument. The parameters for the quasiperiodic kernel are expressed in slightly different formulations to those in \texttt{juliet}, such that $\lambda_e = \sqrt{1/2\alpha_{GP}}$ and $\lambda_p = \sqrt{1/2\Gamma_{GP}}$.

To set priors on the periods and times of mid-transit, we used the values from the TOI designation, as reported by the {\tt EXOFOP-TESS} website. We set broad uniform priors on the semi-amplitudes, planet-to-star radius ratios, impact parameters, stellar density, and $\mathrm{\sqrt{e_{i}} \sin \omega_{i}, \sqrt{e_{i}} \cos \omega_{i}}$ where applicable. For the RV GP hyper-parameters, we set a uniform prior around the measured stellar rotation period for the period $\mathrm{P_{GP}}$, and broad uniform priors for the length-scale $\mathrm{\lambda_{p,GP}}$ and the evolutionary time-scale $\mathrm{\lambda_{e,GP}}$ (while ensuring this last is longer than $\mathrm{P_{GP}}$). For the amplitudes $\mathrm{A0_{GP,rv}}$ - $\mathrm{A5_{GP,BIS}}$, we generally followed the prescriptions of \cite{Delisle2022} to set bounds based on the RMS of the respective data series. Preliminary fits showed positive-negative degeneracies for the $\mathrm{A2_{GP,FWHM}}$ and  $\mathrm{A5_{GP,BIS}}$ components. As the FWHM showed a positive correlation with the RVs, and the corresponding GP component for the RVs $\mathrm{A0_{GP,rv}}$ is by convention fixed positive, we additionally set $\mathrm{A2_{GP,FWHM}}$ to be positive. Likewise, as the BIS showed a strong negative correlation with the RVs, we set $\mathrm{A5_{GP,BIS}}$ to be negative. For the limb-darkening parameters we set broad uniform priors. The remaining instrumental parameters are set internally by \texttt{pyaneti} with broad uniform or log-uniform priors.

\begin{figure}[htb!]
    \centering
    \includegraphics[width=\columnwidth]{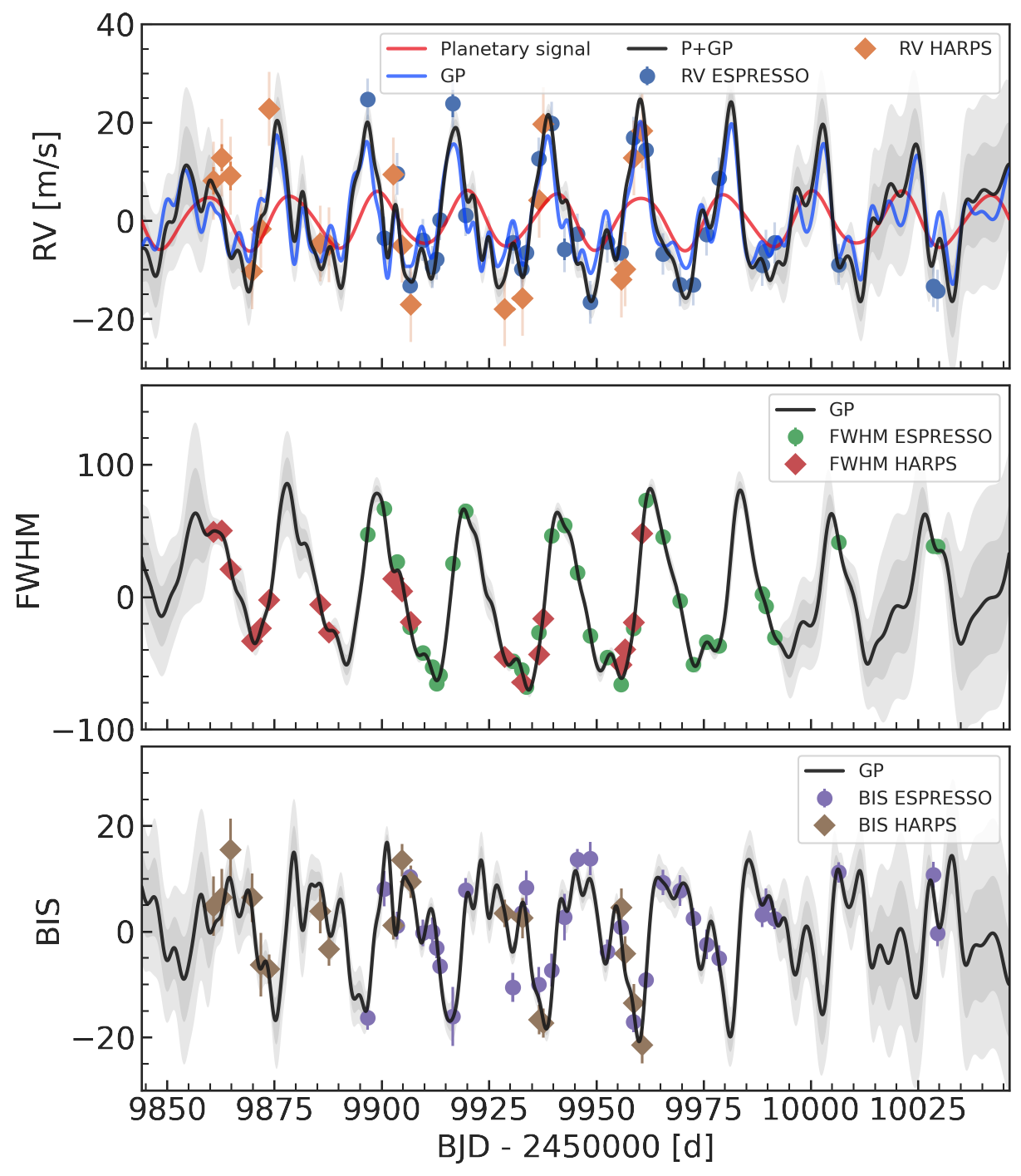}
    \caption{Top: ESPRESSO RVs (blue dots), HARPS RVs (orange diamonds), model components (GP: blue, Keplerian: red), and median model (black) for the \texttt{pyaneti} $2c$ model. Centre: ESPRESSO and HARPS FWHM values (green dots and red diamonds respectively) and joint GP model (black). Bottom: ESPRESSO and HARPS BIS values (purple dots and brown diamonds respectively) and joint GP model (black). In all panels, solid error bars show the pipeline errors, semi-transparent error bars the added jitter, and dark and light grey regions the $1\sigma$ and $2\sigma$ confidence intervals. The systemic offsets have been subtracted.}
    \label{fig:RVs_FWHM_BIS_pyaneti}
\end{figure}

\begin{figure*}[htb!]
    \centering
    \includegraphics[width=.48\textwidth]{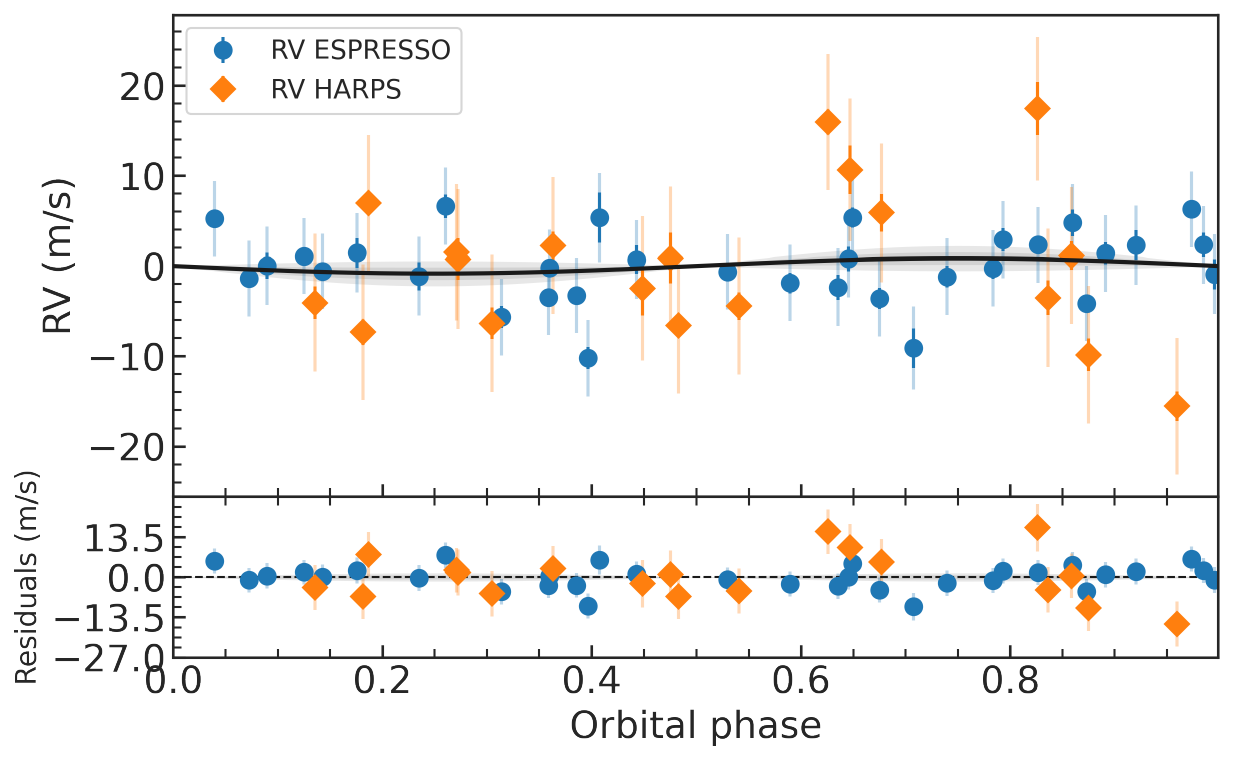}
    \includegraphics[width=.48\textwidth]{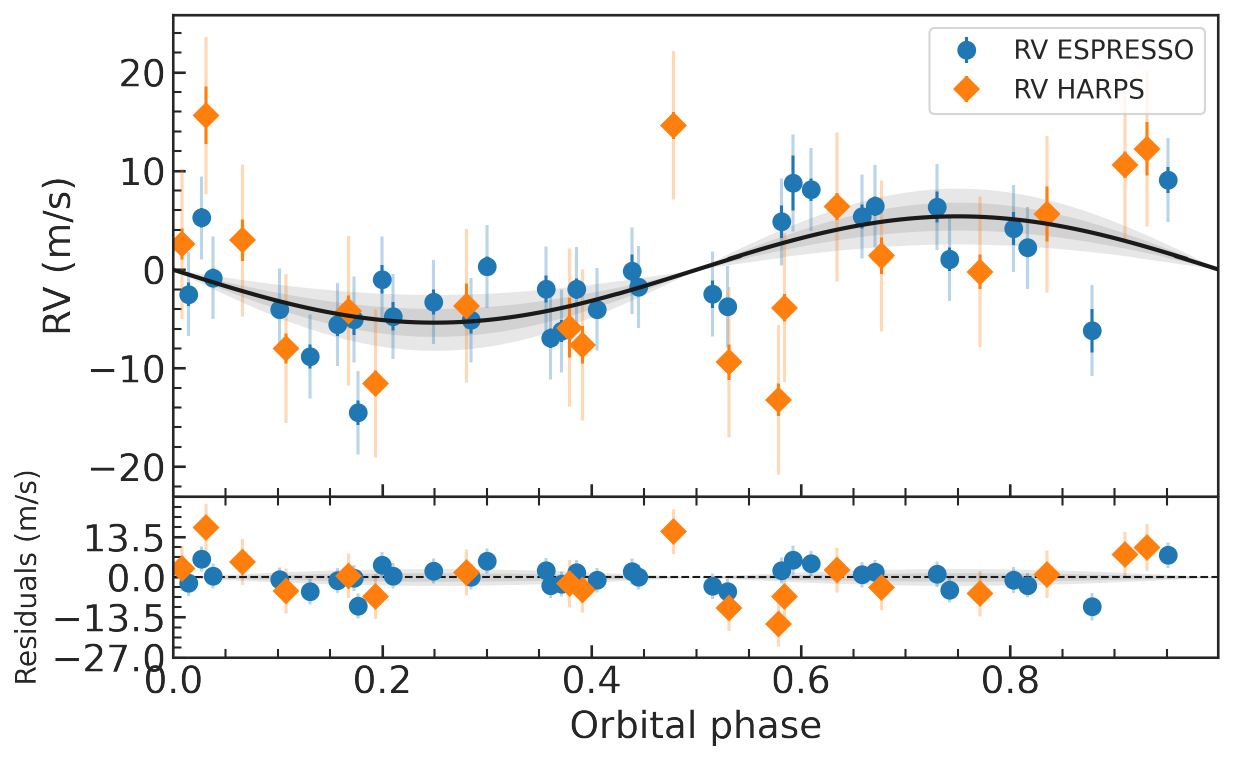}
    \caption{Top: Phase-folded ESPRESSO RVs (blue dots), HARPS RVs (orange diamonds), and median Keplerian model (black line) from the \texttt{pyaneti} $2c$ fit for the inner candidate planet TOI-2322.02 (left) and outer candidate planet TOI-2322.01 (right). The shaded grey regions show the $1\sigma$ and $2\sigma$ confidence intervals. Bottom: residuals to the fit.}
    \label{fig:RVs_2plcirc_pyaneti}
\end{figure*}

\begin{figure*}[htb!]
    \centering
    \includegraphics[width=.48\textwidth]{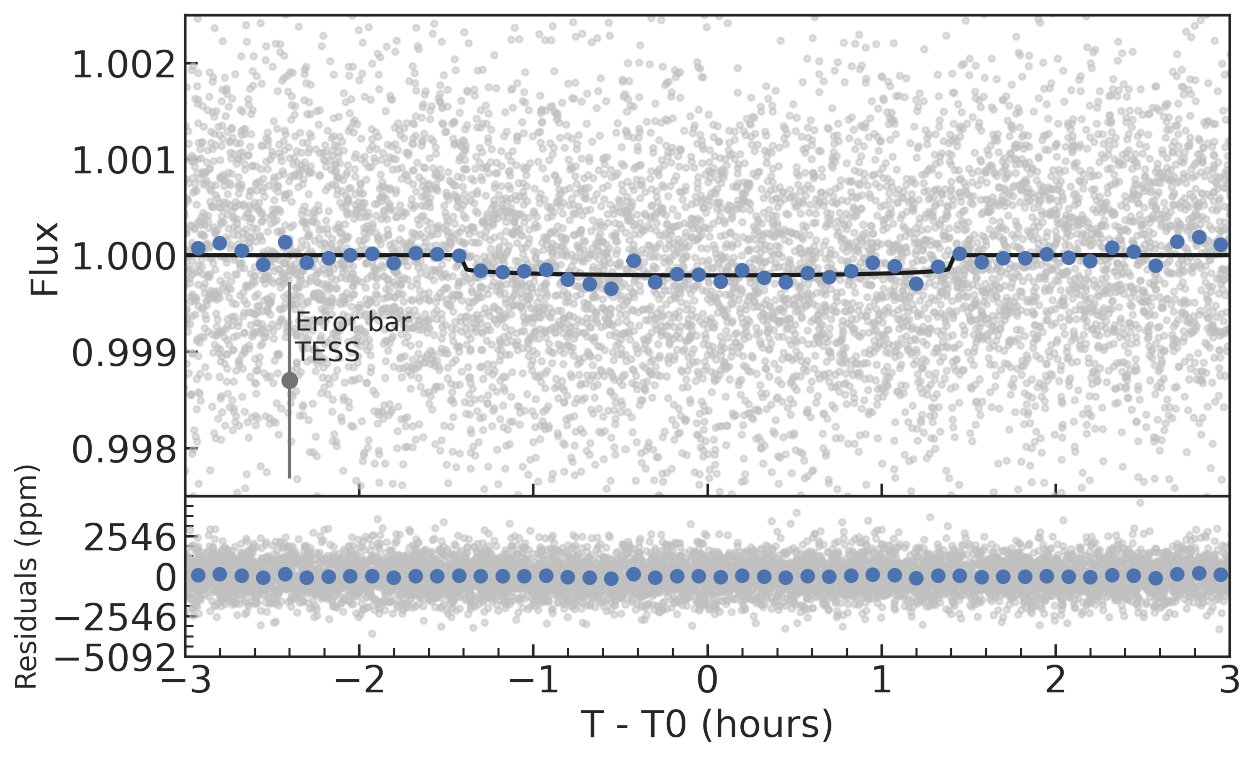}
    \includegraphics[width=.48\textwidth]{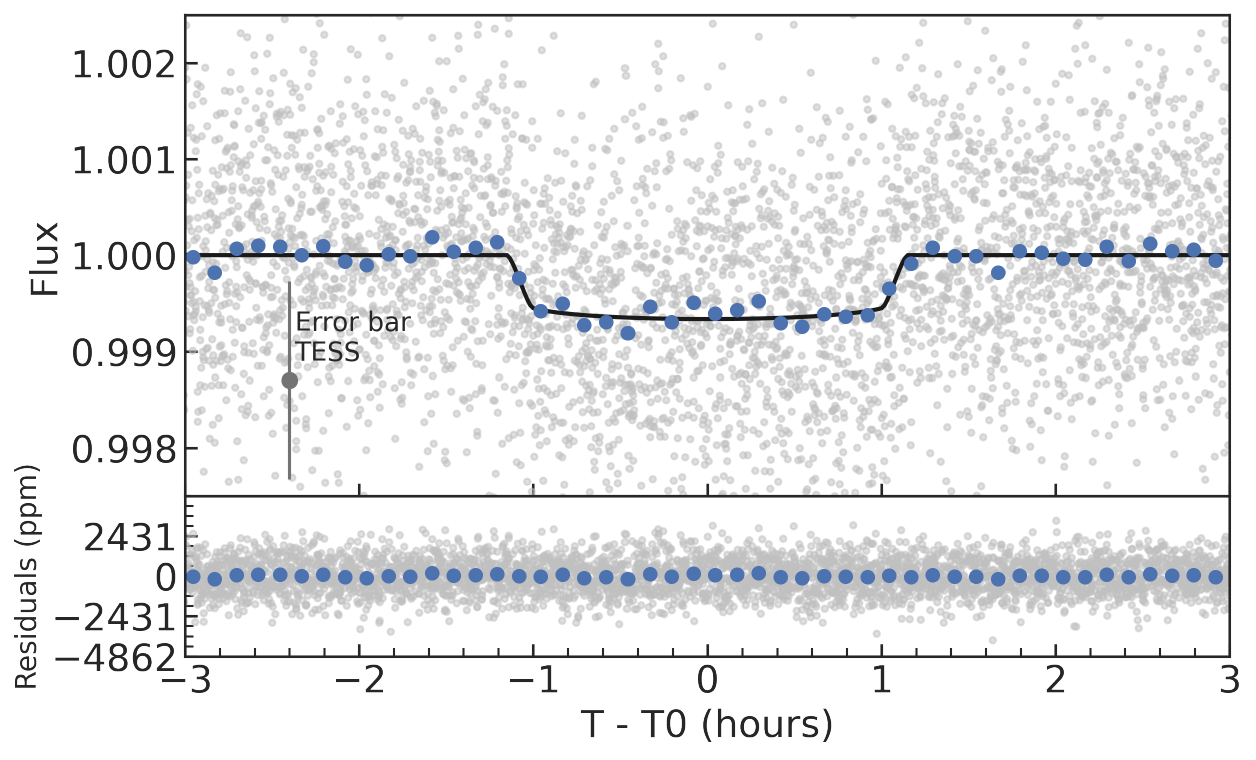}
    \caption{Top: Stacked phase-folded PDCSAP TESS data (grey dots), binned TESS data (blue dots), and the median model (black line) from the \texttt{pyaneti} $2c$ fit, for the inner candidate TOI-2322.02 (left) and outer candidate TOI-2322.01 (right). Bottom: residuals to the fit.}
    \label{fig:TESS_stacked_2plcirc_pyaneti}
\end{figure*}

As with the \texttt{juliet} analysis, the two-planet circular model is favoured by likelihood and BIC comparison, with $\Delta \ln \mathcal{L}_{2c-in} \approx 255$, $\Delta \ln \mathcal{L}_{2c-out} \approx 62$, and $\Delta \ln \mathcal{L}_{2c-2e} \approx 8$. The priors and posteriors for all the parameters are listed in Table \ref{tab:TOI-2322_2cqp_priors_posteriors_pyaneti}. Figure \ref{fig:RVs_FWHM_BIS_pyaneti} shows the GP models (and, for the RVs, two-Keplerian model) fitted to the RV, FWHM, and BIS time series. We highlight the clear agreement between the HARPS and ESPRESSO time series. The phase-folded RVs and light curves are shown in Figs. \ref{fig:RVs_2plcirc_pyaneti} and \ref{fig:TESS_stacked_2plcirc_pyaneti} respectively. The final parameters are consistent with those of the \texttt{juliet} fit, but with tighter constraints. In particular, the semi-amplitude $\mathrm{K_{c}}$ of the outer planet is measured to $3.36\sigma$, significantly better than the $1.67\sigma$ measurement obtained with \texttt{juliet}, thus providing a solid detection and mass measurement rather than a mass upper limit. For the inner planet, we continue to not reach a $3\sigma$ measurement of $\mathrm{K_{b}}$, as although the uncertainty decreases by a factor $\sim3$, the best-fit value decreases by a similar factor. The typically reported $3\sigma$ upper limit on the mass, however, would lead to an unphysically dense planet, as we will discuss in Section \ref{s:discussion}. Therefore, we report in Table \ref{tab:TOI-2322_2cqp_priors_posteriors_pyaneti} the computed mass and uncertainties resulting from the best-fit value of $\mathrm{K_{b}}$. From here onwards, we refer to TOI-2322.02 as TOI-2322 b, and to TOI-2322.01 as TOI-2322 c.

\begin{table}[ht!] 
\begin{center} 
\caption{Prior and posterior planetary parameter distributions obtained with \texttt{pyaneti} for the $2c$ model. \textit{Top}: Fitted parameters. \textit{Bottom}: derived orbital parameters and physical parameters.} 
\label{tab:TOI-2322_2cqp_priors_posteriors_pyaneti} 
\centering 
\resizebox{0.9\columnwidth}{!}{%
\begin{tabular}{lll} 
\hline  \hline 
Parameter & Prior\tablefootmark{$\star$} & Posterior \\ 
\hline 
$\mathrm{\mu_{RV, ESPRESSO}}$ \dotfill [$\mathrm{km \, s^{-1}}$] & $\mathcal{U}(-4.9978 , -3.9564)$ & \RVESPRESSO \\
$\mathrm{\sigma_{ESPRESSO}}$ \dotfill [$\mathrm{m \, s^{-1}}$] & $\mathcal{J}(0.001,10)$ & \jRVESPRESSO \\
$\mathrm{\mu_{FWHM, ESPRESSO}}$ \dotfill & $\mathcal{U}(5972.6590 , 6114.5837)$ & \FWHMESP \\
$\mathrm{\sigma_{FWHM, ESPRESSO}}$ \dotfill & $\mathcal{J}(0.001,10)$ & \jFWHMESP \\
$\mathrm{\mu_{BIS, ESPRESSO}}$ \dotfill & $\mathcal{U}(6.1894 , 38.0809)$ & \BISESP \\
$\mathrm{\sigma_{BIS, ESPRESSO}}$ \dotfill & $\mathcal{J}(0.001,10)$ & \jBISESP \\
$\mathrm{\mu_{RV, HARPS}}$ \dotfill [$\mathrm{km \, s^{-1}}$] & $\mathcal{U}(-4.9929 , -3.9522 )$ & \RVHARPS \\
$\mathrm{\sigma_{RV, HARPS}}$ \dotfill [$\mathrm{m \, s^{-1}}$] & $\mathcal{J}(0.001,10)$ & \jRVHARPS \\
$\mathrm{\mu_{FWHM, HARPS}}$ \dotfill & $\mathcal{U}(6052.0933 , 6167.5497)$ & \FWHMHARPS \\
$\mathrm{\sigma_{FWHM, HARPS}}$ \dotfill & $\mathcal{J}(0.001,10)$ & \jFWHMHARPS \\
$\mathrm{\mu_{BIS, HARPS}}$ \dotfill & $\mathcal{U}(10.0728 , 48.0468)$ & \BISHARPS \\
$\mathrm{\sigma_{BIS, HARPS}}$ \dotfill & $\mathcal{J}(0.001,10)$ & \jBISHARPS \\
$\mathrm{P_{b}}$ \dotfill [d] & $\mathcal{N}(11.307,0.001)$ & \Pb \\
$\mathrm{t_{0,b}}$ \dotfill [BJD-2450000] & $\mathcal{N}(10160.451,0.01)$ & \Tzerob \\
$\mathrm{K_{b}}$ \dotfill  [$\mathrm{m \, s^{-1}}$] & $\mathcal{U}(0,10)$ & \kb \\
$\mathrm{\rho_{}}$ \dotfill [$\mathrm{g \, cm^{-3}}$] & $\mathcal{J}(0.1,10)$ & \dentrheeb \\
$\mathrm{p_{b}}$ \dotfill & $\mathcal{U}(0.005,0.05)$ & \rrbTESS \\
$\mathrm{b_{b}}$ \dotfill & $\mathcal{U}(0,1)$ & \bb \\
$\mathrm{P_{c}}$ \dotfill [d] & $\mathcal{N}(20.225,0.001)$ & \Pc \\
$\mathrm{t_{0,b}}$ \dotfill [BJD-2450000] & $\mathcal{N}(9037.42,0.01)$ & \Tzeroc \\
$\mathrm{K_{c}}$ \dotfill [$\mathrm{m \, s^{-1}}$] & $\mathcal{U}(0,10)$ & \kc \\
$\mathrm{p_{c}}$ \dotfill & $\mathcal{U}(0.005,0.05)$ & \rrcTESS \\
$\mathrm{b_{c}}$ \dotfill & $\mathcal{U}(0,1)$ & \bc \\
$\mathrm{q_{1,TESS}}$ \dotfill & $\mathcal{U}(0,1)$ & \qoneTESS \\
$\mathrm{q_{2,TESS}}$ \dotfill & $\mathcal{U}(0,1)$ & \qtwoTESS \\
$\mathrm{\sigma_{TESS}}$ \dotfill [ppm] & $\mathcal{J}(1,100)$ & \jtrTESS \\
$\mathrm{A0_{GP,rv}}$ \dotfill & $\mathcal{U}(0,0.012)$ & \jAzero \\
$\mathrm{A1_{GP,rv}}$ \dotfill & $\mathcal{U}(-0.12,0.12)$ & \jAone \\
$\mathrm{A2_{GP,FWHM}}$ \dotfill & $\mathcal{U}(0,90)$ & \jAtwo \\
$\mathrm{A3_{GP,FWHM}}$ \dotfill & fixed & 0 \\
$\mathrm{A4_{GP,BIS}}$ \dotfill & $\mathcal{U}(-9,9)$ & \jAfour \\
$\mathrm{A5_{GP,BIS}}$ \dotfill & $\mathcal{U}(-90,0)$ & \jAfive \\
$\mathrm{\lambda_{e,GP}}$ \dotfill & $\mathcal{U}(25,125)$ & \jlambdae \\
$\mathrm{\lambda_{p,GP}}$ \dotfill & $\mathcal{U}(0.1,10)$ & \jlambdap \\
$\mathrm{P_{GP}}$ \dotfill & $\mathcal{U}(19,23)$ & \jPGP \\
\hline 
$\mathrm{e_{b}}$ \dotfill & $\mathrm{fixed}$ & \jeccpone \\
$\mathrm{\omega_{b}}$ \dotfill [$\degr$] & $\mathrm{fixed}$ & \jomegapone \\
$\mathrm{a_{b}}$ \dotfill [au] & $-$ & \ab \\
$\mathrm{i_{b}}$ \dotfill [$\degr$] & $-$ & \ib \\
$\mathrm{T_{14,b}}$ \dotfill [h] & $-$ & \ttotb \\
$\mathrm{M_{b}}$ \dotfill [$\mathrm{M_\oplus}$] & $-$ & \mpb \\
$\mathrm{M_{b, phys}}$\tablefootmark{$\dagger$} \dotfill [$\mathrm{M_\oplus}$] & $-$ & $\leq2.03$ \\
$\mathrm{R_{b}}$ \dotfill [$\mathrm{R_\oplus}$] & $-$ & \rpbTESS \\
$\mathrm{\rho_{b}}$ \dotfill [$\mathrm{g \, cm^{-3}}$] & $-$ & \denpb \\
$\mathrm{\rho_{b, phys}}$\tablefootmark{$\dagger$} \dotfill [$\mathrm{g \, cm^{-3}}$] & $-$ & $\leq11.39$ \\
$\mathrm{T_{eq,b}}$ \dotfill [K] & $-$ & \Teqb \\
$\mathrm{S_{ins,b}}$ \dotfill [$\mathrm{S_\oplus}$] & $-$ & \insolationb \\
$\mathrm{e_{c}}$ \dotfill & $\mathrm{fixed}$ & \jeccptwo \\
$\mathrm{\omega_{c}}$ \dotfill [$\degr$] & $\mathrm{fixed}$ & \jomegaptwo \\
$\mathrm{a_{c}}$ \dotfill [au] & $-$ & \ac \\
$\mathrm{i_{c}}$ \dotfill [$\degr$] & $-$ & \ic  \\
$\mathrm{T_{14,c}}$ \dotfill [h] & $-$ & \ttotc \\
$\mathrm{M_{c}}$ \dotfill [$\mathrm{M_\oplus}$] & $-$ & \mpc  \\
$\mathrm{R_{c}}$ \dotfill [$\mathrm{R_\oplus}$] & $-$ & \rpcTESS \\
$\mathrm{\rho_{c}}$ \dotfill [$\mathrm{g \, cm^{-3}}$] & $-$ & \denpc \\
$\mathrm{T_{eq,c}}$ \dotfill [K] & $-$ & \Teqc \\
$\mathrm{S_{ins,c}}$ \dotfill [$\mathrm{S_\oplus}$] & $-$ & \insolationc \\
$\mathrm{\ln \mathcal{L}}$ \dotfill  & $-$ & $75965.99$ \\
\hline 
\end{tabular} 
}
\end{center} 
\tablefoot{\tablefoottext{$\star$}{$\mathcal{U}(a,b)$ indicates a uniform distribution between a and b; $\mathcal{J}(a, b)$ a Jeffreys or log-uniform distribution between a and b; $\mathcal{N}(a,b)$ a normal distribution with mean $a$ and standard deviation $b$.}\\
\tablefoottext{$\dagger$}{Physical upper mass and density limits, derived from a 100\% iron composition model at the radius of TOI-2322 b.}}
\end{table} 

\subsection{Testing different GP kernels}\label{s:kernels}

The quasiperiodic GP kernel is a common choice to model stellar activity for various reasons: it is physically motivated, given we expect stellar activity signals to be modulated by the stellar rotation period; it has a low number of hyper-parameters, making it computationally attractive; and it has been shown to reproduce both simulated and real data well \citep{Haywood2014, Rajpaul2015, pyaneti2}. However, other kernel choices are possible. To verify the robustness of our results to different kernel choices, we performed additional fits of the detrended TESS photometry and HARPS and ESPRESSO RVs with alternative kernels for the RV GP using the S+LEAF\footnote{Available at \url{https://www.astro.unige.ch/~delisle/spleaf/doc/index.html}} software \citep{Delisle2020, Delisle2022, Hara2023}. Specifically, we tested the following four cases: a linear combination of two stochastically-driven harmonic oscillator \citep[SHO,][]{Foreman-Mackey2017} kernels; a Matérn 3/2 exponential periodic \citep[MEP,][]{Delisle2022} kernel; an exponential-sine periodic \citep[ESP,][]{Delisle2022} kernel; and an ESP kernel with four defined harmonic components rather than the default two components (ESP-4). The different kernels used and the RV semi-amplitudes obtained for the two planets with each kernel are listed in Table \ref{tab:gp_comp}. In all cases, the obtained semi-amplitudes are consistent both with the \texttt{pyaneti} analysis and with each other at $1\sigma$, highlighting the robustness of our results. We show the fitted models for each kernel in Appendix \ref{ap:s+leaf}. All the models strongly resemble both each other and the \texttt{pyaneti} model shown in Fig. \ref{fig:RVs_FWHM_BIS_pyaneti}. We highlight as commonalities the large amplitude of the RV GPs, which are much larger than the Keplerian model amplitudes; the sharp peaks of the FWHM model; and the broader, flatter peaks of the BIS model.

\begin{table}[htb!]
    \begin{center}
        \caption{{Results from different S+LEAF GP kernels with the \texttt{pyaneti} values for comparison} \label{tab:gp_comp}}
        \begin{tabular}[center]{l l l}
            \hline
            \hline
            Kernel & $\mathrm{K_b\,[m\,s^{-1}]}$ & $\mathrm{K_c\,[m\,s^{-1}]}$ \\
            \hline
            2x SHO & 1.56$^{+1.97}_{-0.91}$ & 6.2$^{+3.7}_{-1.9}$ \\
            MEP    & 1.56$^{+2.0}_{-0.91}$ & 6.2$^{+3.8}_{-1.9}$  \\
            ESP    & 1.6$^{+2.2}_{-1.0}$    & 5.9$^{+4.4}_{-2.2}$ \\
            ESP-4  & 1.20$^{+1.75}_{-0.75}$ & 5.2$^{+3.7}_{-1.9}$ \\        
            \hline
            \texttt{pyaneti} & \kb & \kc \\
             \hline
        \end{tabular}
    \end{center}
\end{table}

\subsection{Testing different data reductions}\label{s:data-red}

While the CCF method for computing radial velocities is a long-enduring field standard, newer techniques based on different principles such as template-matching \citep{Anglada2012, Astudillo-Defru2015thesis} and line-by-line \citep[LBL, ][]{Dumusque2018, Artigau2022} computation have been shown to improve the RV extraction, especially for active stars (e.g. \citealt{Zhao2022} and references therein). We first re-extracted the ESPRESSO RVs with the S-BART pipeline \citep{Silva2022}, which uses template matching to extract RV measurements in a semi-Bayesian framework, and tested the \texttt{juliet} $2c$ model with these RVs. Subsequently, we re-extracted both HARPS and ESPRESSO RVs, FWHM, and BIS using the YARARA pipeline \citep{Cretignier2021, Cretignier2023}, which corrects for telluric, stellar, and instrumental effects at the spectral level before extracting LBL RVs, and tested the \texttt{pyaneti} $2c$ model with these RVs and activity indicators. While in both cases our results were fully compatible with those obtained using the DRS RVs and activity indicators, in neither case did we obtain an improvement on the precision of the fitted semi-amplitudes. This is likely due to the spectra only reaching a moderate S/N, the relatively low number of measurements, and the host star being outside the optimal spectral types for either technique. YARARA, for example, is optimized for sun-like stars and TOI-2322 lies close to the lower edge of its range of validity in $\mathrm{T_{eff}}$. S-BART, meanwhile, shows the most significant gain in RV scatter and median uncertainty for M-dwarfs, while TOI-2322 is a K4 star. 

\subsection{Impact of including the HARPS data}
As the HARPS data have a lower precision, and show more scatter around the best-fit model obtained with the full dataset compared to the ESPRESSO data, we ran a comparison fit using \texttt{pyaneti} with the ESPRESSO data alone. Other than the removal of the HARPS offsets and jitters, all priors are identical to those in Table \ref{tab:TOI-2322_2cqp_priors_posteriors_pyaneti}. We find overall larger error bars for the planets' semi-amplitudes and the GP coefficients, indicating that despite their lower precision the HARPS data are contributing to the fit. We also note that their inclusion extends the temporal baseline covered by approximately a rotation period and a half, as seen in Fig. \ref{fig:RVs_FWHM_BIS_pyaneti}. Since the majority of the ESPRESSO data are clustered in only five rotation periods, this is a substantial increase.

\subsection{Search for transit timing variations}

\begin{figure}[htb!]
    \centering
    \includegraphics[width=.5\textwidth]{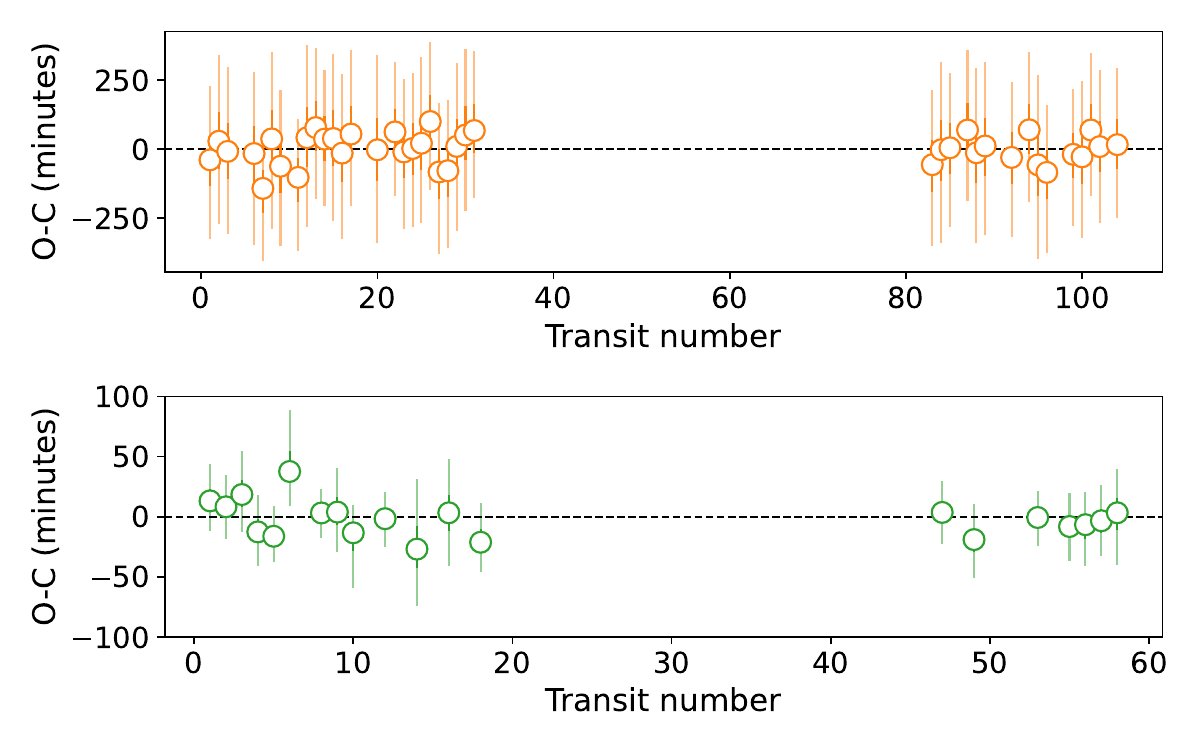}
    \caption{O-C plots for TOI-2322.02 (top) and TOI-2322.01 (bottom). Solid error bars correspond to the $1-\sigma$ error, fainter bars to the $3-\sigma$ error.}
    \label{fig:tess_TTVs}
\end{figure}

Since the periods of the planet candidates are close to the 2:1 resonance, we carried out a search for transit timing variations (TTVs) with \texttt{juliet}. The input parameters for a TTV fit are the expected time of each transit; the stellar density $\rho$; the planet-to-star radius ratio $\mathrm{p_{i}}$, impact parameter $\mathrm{b_{i}}$, and either eccentricity $\mathrm{e_{i}}$ and angle of periastron $\mathrm{\omega_{i}}$ or a derived parametrisation such as $\mathrm{\sqrt{e_{i}} \sin \omega_{i}, \sqrt{e_{i}} \cos \omega_{i}}$ for each planet $\mathrm{i}$; and the dilution factor $\mathrm{m_{dilution,instrument}}$, flux offset $\mathrm{m_{flux,instrument}}$, jitter $\mathrm{\sigma_{w,instrument}}$, and limb-darkening parameters $\mathrm{q_{1,instrument}}$ and $\mathrm{q_{2,instrument}}$ for each transit instrument. To make the fit computationally feasible, we used the sector-by-sector GP fits to the out-of-transit data described in Appendix \ref{ap:juliet} to detrend the in-transit data, then treated it all as coming from a single instrument. We find no strong evidence for TTVs for either planet, as shown in Fig. \ref{fig:tess_TTVs}. For TOI-2322 b, the best-fit O-C values have a median amplitude of $\sim38$ minutes with a median error of $\sim96$ minutes, while for TOI-2322 c the best-fit O-C values have a median amplitude of $\sim10$ minutes with a median error of $\sim10$ minutes. We note that TOI-2322 b especially has a small transit depth of $\sim210$ ppm, as can be seen in Fig. \ref{fig:TESS_stacked_2plcirc_pyaneti}, so the individual transit fits have high uncertainty, leading to the large errors on the O-C values.

\section{Discussion}\label{s:discussion}

\subsection{Planetary composition}

\begin{figure*}[htb!]
    \centering
    \includegraphics[width=\textwidth]{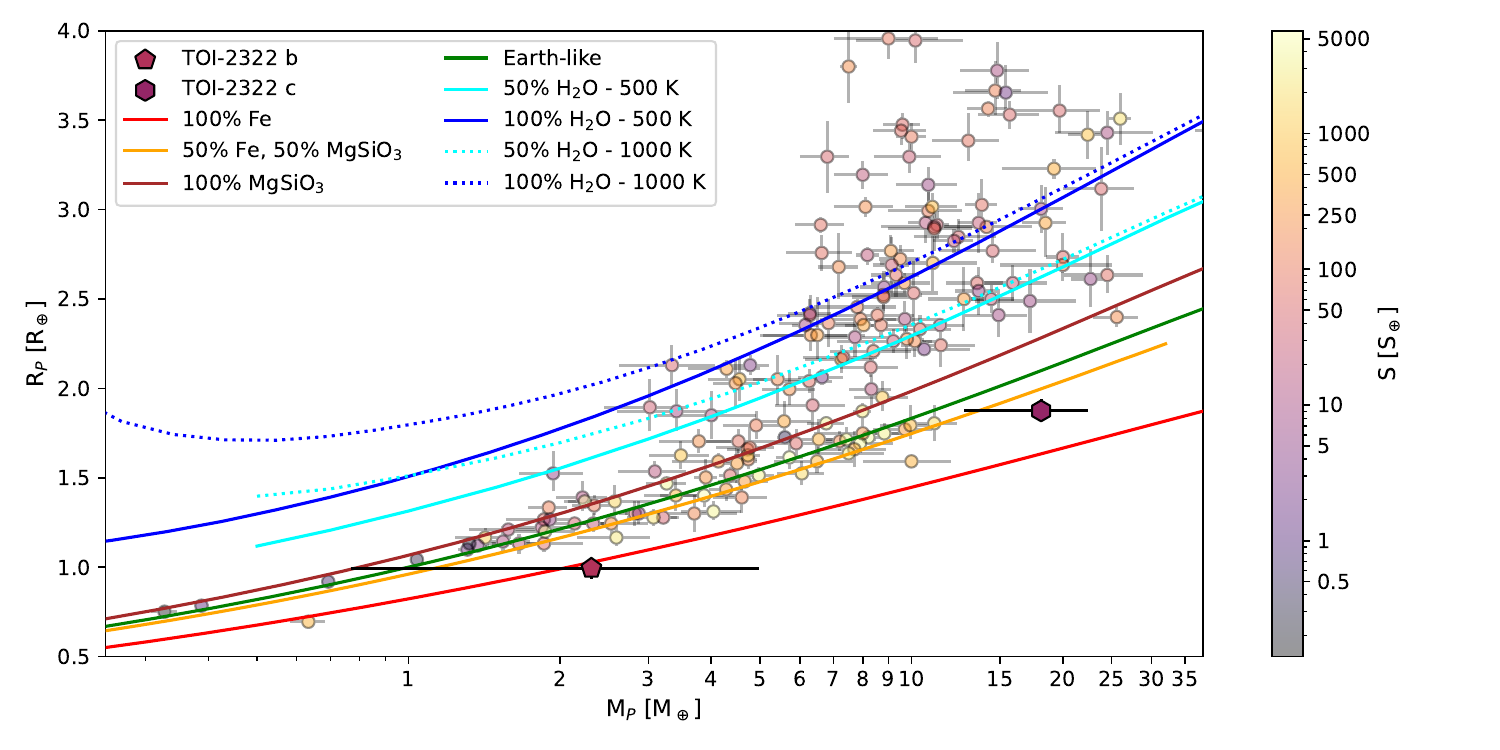}
    \caption{Mass-radius diagram of the well-characterized small planet population ($\mathrm{M_p<40M_\oplus, R_p<4R_\oplus}$, mass error better than 25\%, radius error better than 8\%) from the PlanetS catalogue \citep{Otegi2020, Parc2024} (semi-transparent circles), coloured by insolation. TOI-2322 b and c are shown as a pentagon and a hexagon respectively. The composition models of \cite{Zeng2019} are shown by the coloured curves.}
    \label{fig:PlanetS_population}
\end{figure*}

To explore the compositions of the planets of the TOI-2322 system and place them in a population context, we show the planets in Fig. \ref{fig:PlanetS_population} together with the small well-characterized planets from the PlanetS catalogue \citep{Otegi2020, Parc2024}. The mass and radius derived from the best-fit model for TOI-2322 b place it practically on the 100\% Fe model. This may indicate a formation in an extremely volatile-poor environment, or a stripping of volatiles by radiation from the active host star; however, the uncertainties on the mass are very large. Since masses larger than that corresponding to a 100\% Fe composition are unphysical, we compute an upper mass limit from the 100\% Fe model of $\mathrm{M_b \leq 2.03 M_\oplus}$, slightly lower than but fully compatible with the mass derived from the best-fit model. TOI-2322 c, meanwhile, lies close to the 50\% Fe, 50\% silicates line, and becomes the most massive planet with this composition.

\begin{figure}[htb!]
    \centering
    \includegraphics[width=\columnwidth]{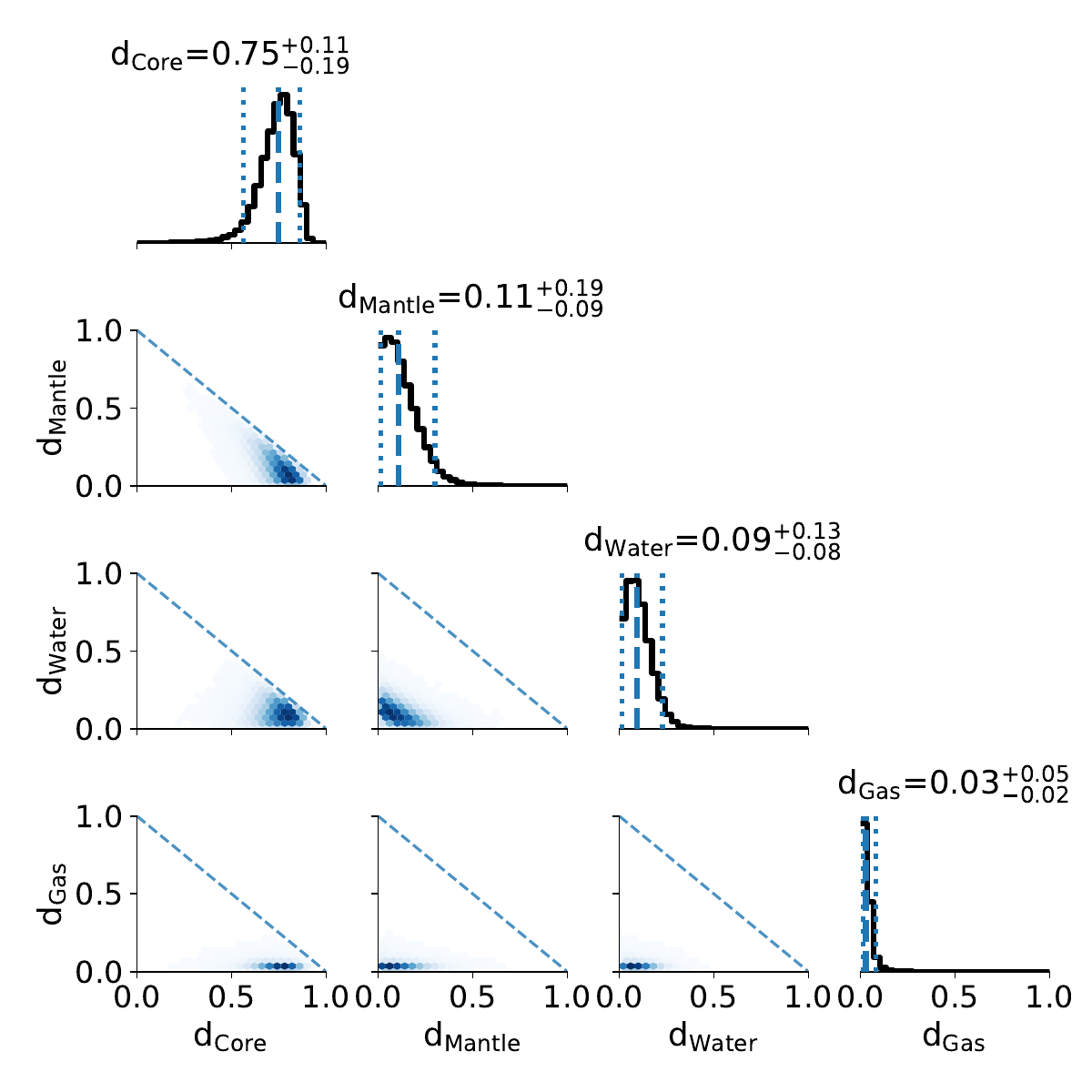}
    \caption{Posterior distributions of the ExoMDN model for TOI-2322 c, showing an internal structure similar to that of the Earth.}
    \label{fig:ExoMDN_c}
\end{figure}

To further characterize the internal structure of TOI-2322 c, we performed interior modelling via the ExoMDN code \citep{Baumeister2023}\footnote{Available at \url{https://github.com/philippbaumeister/ExoMDN}}. ExoMDN uses mixture density networks to perform interior inference modelling. From the planet's mass, radius, and equilibrium temperature, the software fits a four-layer model consisting of an iron core, a silicate mantle, a water layer, and a H/He atmosphere. 
The posterior distributions for the ExoMDN model for TOI-2322 c are shown in Fig. \ref{fig:ExoMDN_c}. We retrieve a model with very similar layer fractions to the Earth of $\mathrm{d_{Core} = 0.75^{+0.11}_{-0.19}}$, $\mathrm{d_{Mantle} = 0.11^{+0.19}_{-0.09}}$, $\mathrm{d_{Water} = 0.09^{+0.13}_{-0.08}}$, and $\mathrm{d_{Gas} = 0.03^{+0.05}_{-0.02}}$. For comparison, the fractions retrieved for the Earth are $\mathrm{d_{Core} = 0.69^{+0.06}_{-0.09}}$, $\mathrm{d_{Mantle} = 0.14^{+0.20}_{-0.13}}$, $\mathrm{d_{Water} = 0.09^{+0.12}_{-0.08}}$, and $\mathrm{d_{Gas} = 0.05^{+0.08}_{-0.04}}$. We thus characterize TOI-2322 c as one of the most massive planets with an Earth-like composition discovered to date, with a mass of \mpc $\mathrm{M_\oplus}$. 

The closest analogues to TOI-2322 c in mass and composition are Kepler-411 b \citep{Sun2019} and TOI-1347 b \citep{Rubenzahl2024}. Kepler 411 b is more massive and larger than TOI-2322 c, with a TTV-determined mass of $\mathrm{25.6 \pm 2.6\,M_\oplus}$ and a radius of $\mathrm{2.401\pm0.053\,R_\oplus}$. It likely has a larger rocky mantle and smaller iron core, with \cite{Sun2019} estimating a $21\pm21\%$ iron core mass fraction, compared to a core mass fraction from ExoMDN of $85^{+10}_{-27}\%$ for TOI-2322 c. TOI-1347 b, meanwhile, is somewhat smaller and less massive than TOI-2322 c; it has an RV-measured mass of $\mathrm{11.1 \pm 1.2\,M_\oplus}$ and a radius of $\mathrm{1.8\pm0.1\,R_\oplus}$, leading to a larger core mass fraction of $41\pm27\%$ which makes it more directly comparable to TOI-2322 c. Both planets have shorter periods than TOI-2322 c, with TOI-1347 b in particular belonging to the ultra-short-period (USP) class, and receive significantly higher insolation fluxes which may have contributed to the stripping of any primordial H/He envelopes. While TOI-2322 c receives much lower insolation, the active host star may have contributed to stripping a primordial envelope, or it may have formed in a gas-poor environment and thus never have accreted an envelope at all.

\subsection{Importance of the transit data to the planet detection}

Historically, when faced with signals which are seen at the same period (or harmonics thereof) in both RVs and activity indicators, RV blind surveys have tended to attribute these signals entirely to stellar activity \citep[e.g.][]{Butler2017, Mignon2024}. However, as the TOI-2322 system evidences, planets can and do exist at similar periods to their host star's rotation period. In order to explore whether the RV data alone could have conclusively revealed the outer planet, which has the clearest RV signal, we perform a \texttt{pyaneti} fit only on the RVs, FWHM, and BIS for the outer planet alone. The priors are analogous to those listed in Table \ref{tab:TOI-2322_2cqp_priors_posteriors_pyaneti}, with the exception of the period $\mathrm{P_{c}}$ and epoch $\mathrm{t_{0,c}}$. Lacking the information from the transit, to set a prior on the period we fit a Gaussian to the highest peak in the joint RV periodogram, which returns a median of $\mathrm{20.7\,d}$ and a standard deviation of $\mathrm{1.4\,d}$. We conservatively set the prior for the period as a slightly broader Gaussian of $\mathcal{N}(20.7,2)$. For $\mathrm{t_{0,c}}$ we set an uninformative uniform prior spanning a $\mathrm{22d}$ range centred on the $\mathrm{t_{0,c}}$ value from the transit data, to avoid convergence issues at the edge of the prior. 

The resulting fit converges to a significantly shorter and more poorly constrained period of $\mathrm{P =19.82_{-2.29}^{+1.57}\,d}$. The epoch posterior is very broad and spans a large part of the prior space, with error bars of $\mathrm{\pm 6\,d}$. The semi-amplitude of $\mathrm{K = 2.95_{-2.02}^{+2.89}\,ms^{-1}}$ barely reaches $1.5\sigma$ confidence, and is notably lower than the value obtained in the full fit. In contrast, the $\mathrm{A0_{GP,rv}}$ and $\mathrm{A1_{GP,rv}}$ coefficients are larger, suggesting that part of the planetary signal is being absorbed by the GP. We also note that the RV residuals show no significant signals at the period of TOI-2322 b (or indeed any period), so from a pure-RV survey this planet would not have been detected, and therefore is not included in this analysis. 

\subsection{TOI-2322 as a benchmark system for activity correction}

The RV-only analysis demonstrates the vital importance of transiting exoplanet systems like TOI-2322, which are the only systems in which we have as ground truth the existence of both a planet and significant stellar activity at similar periods. With the planet's period and epoch tightly constrained by the transit data, the RV signal can be properly fitted, enabling a full characterization of the planet(s). This contrasts with RV-only datasets, where a known stellar activity signal may be masking the presence of a planet at a close period.

The TOI-2322 system is thus an ideal benchmark for testing methods of disentangling these planets from stellar activity. More such planets are likely to emerge as the ongoing extended TESS missions and upcoming transit missions like PLATO begin uncovering more long-period transiting planets with periods in the range of typical stellar rotation periods for main-sequences solar-type stars. Further observations of TOI-2322 would be valuable to extend the observing baseline, enabling the exploration of the time-variation of the stellar activity impact on the RVs, refinement of the planetary masses, and the search for additional long-period companions. 
Simultaneous optical and near-infrared (nIR) spectroscopy would be particularly useful, as stellar activity is wavelength-dependent and has been shown to decrease in the nIR \citep{Carmona2023}. Likewise, co-eval RVs and photometry could enable the simultaneous characterization of the stellar activity in both datasets \citep[e.g.][]{Hara2023}.

\section{Conclusions}\label{s:conclusions}

We have presented the confirmation and characterization of TOI-2322 b and c, two planets orbiting an active K star. The inner, smaller planet TOI-2322 b has a period slightly above the first harmonic of the rotation period. As we do not reach a $3\sigma$ measurement of the semi-amplitude, we use physical considerations to establish an upper mass limit of $\mathrm{M_b \leq 2.03 M_\oplus}$, corresponding to the mass of a 100\% Fe planet with the \rpbTESS $\mathrm{R_\oplus}$ radius extracted from the TESS photometry. The outer, larger planet TOI-2322 c has a period slightly below the stellar rotation period, at which significant periodicity is seen in the photometry and spectroscopic activity indicators. It has a mass of \mpc $\mathrm{M_\oplus}$ and a radius of \rpcTESS $\mathrm{R_\oplus}$, yielding an internal structure very similar to that of the Earth, and so becoming the largest planet known with an Earth-like composition.

A fitting process involving the photometry, the RVs, and a multivariate GP applied to the RVs and the FWHM and BIS activity indicators was necessary to fully characterize the system, and in particular to separate the stellar activity effects from the planetary signals in the RVs. The TOI-2322 system is an ideal system for testing methods of disentangling planetary and activity signals in RVs. It is also an excellent candidate for follow-up nIR spectroscopy to explore how the interplay between signals of stellar and planetary origin changes across wavelength.

\begin{acknowledgements}
We thank the referee for their their careful reading and helpful comments that have improved this paper.
We thank the Swiss National Science Foundation (SNSF) and the Geneva University for their continuous support to our planet low-mass companion search programmes. This work has been carried out within the framework of the National Centre of Competence in Research PlanetS supported by the Swiss National Science Foundation. 
This publication makes use of The Data \& Analysis Center for Exoplanets (DACE), which is a facility based at the University of Geneva (CH) dedicated to extrasolar planets data visualisation, exchange and analysis. DACE is a platform of the Swiss National Centre of Competence in Research (NCCR) PlanetS, federating the Swiss expertise in Exoplanet research. The DACE platform is available at \url{https://dace.unige.ch}.
This work made use of \texttt{tpfplotter} by J. Lillo-Box (publicly available in \url{www.github.com/jlillo/tpfplotter}), which also made use of the python packages \texttt{astropy}, \texttt{lightkurve}, \texttt{matplotlib} and \texttt{numpy}.
This work made use of \texttt{TESS-cont} (\url{https://github.com/castro-gzlz/TESS-cont}), which also made use of \texttt{tpfplotter} \citep{Aller2020} and \texttt{TESS-PRF} \citep{Bell2022TESSPRF}.
This research has made use of the Exoplanet Follow-up Observation Program (ExoFOP; DOI: 10.26134/ExoFOP5) website, which is operated by the California Institute of Technology, under contract with the National Aeronautics and Space Administration under the Exoplanet Exploration Program.
Funding for the TESS mission is provided by NASA's Science Mission Directorate.
This paper made use of data collected by the TESS mission and are publicly available from the Mikulski Archive for Space Telescopes (MAST) operated by the Space Telescope Science Institute (STScI). 
We acknowledge the use of public TESS data from pipelines at the TESS Science Office and at the TESS Science Processing Operations Center. 
Resources supporting this work were provided by the NASA High-End Computing (HEC) Program through the NASA Advanced Supercomputing (NAS) Division at Ames Research Center for the production of the SPOC data products.
This work makes use of observations from the LCOGT network. Part of the LCOGT telescope time was granted by NOIRLab through the Mid-Scale Innovations Program (MSIP). MSIP is funded by NSF.
Based on observations made with ESO Telescopes at the La Silla Paranal Observatory under programme IDs 110.24CD.002, 110.24CD.003, and 110.24CD.009.
Based on observations carried out at the European Southern Observatory (ESO; La Silla, Chile) using the 3.6m telescope, under ESO programme 110.242T.001.
This work has made use of data from the European Space Agency (ESA) mission {\it Gaia} (\url{https://www.cosmos.esa.int/gaia}), processed by the {\it Gaia} Data Processing and Analysis Consortium (DPAC, \url{https://www.cosmos.esa.int/web/gaia/dpac/consortium}). Funding for the DPAC has been provided by national institutions, in particular the institutions participating in the {\it Gaia} Multilateral Agreement.
Co-funded by the European Union (ERC, FIERCE, 101052347). Views and opinions expressed are however those of the author(s) only and do not necessarily reflect those of the European Union or the European Research Council. Neither the European Union nor the granting authority can be held responsible for them. This work was also supported by FCT - Fundação para a Ciência e a Tecnologia through national funds by these grants: UIDB/04434/2020 DOI: 10.54499/UIDB/04434/2020, UIDP/04434/2020 DOI: 10.54499/UIDP/04434/2020, PTDC/FIS-AST/4862/2020, UID/04434/2025. 
S.G.S. acknowledges the support from FCT through Investigador FCT contract nr. CEECIND/00826/2018 and  POPH/FSE (EC).
S.C.C.B. acknowledges the support from Fundação para a Ciência e Tecnologia (FCT) in the form of work contract through the Scientific Employment Incentive program with reference 2023.06687.CEECIND and  DOI 10.54499/2023.06687.CEECIND/CP2839/CT0002.
A.C.-G. is funded by the Spanish Ministry of Science through MCIN/AEI/10.13039/501100011033 grant PID2019-107061GB-C61. 
K.A.C. acknowledges support from the TESS mission via subaward s3449 from MIT.
"The INAF authors acknowledge financial support of the Italian Ministry of Education, University, and Research with PRIN 201278X4FL and the ""Progetti Premiali"" funding scheme.
X.D. acknowledges the support from the European Research Council (ERC) under the European Union’s Horizon 2020 research and innovation programme (grant agreement SCORE No 851555) and from the Swiss National Science Foundation under the grant SPECTRE (No 200021\_215200).
J.I.G.H., A.S.M., R.R., and C.A.P. acknowledge financial support from the Spanish Ministry of Science, Innovation and Universities (MICIU) projects PID2020-117493GB-I00 and PID2023-149982NB-I00.
J.L.-B. is funded by the Spanish Ministry of Science and Universities (MICIU/AEI/10.13039/501100011033) and NextGenerationEU/PRTR grants PID2019-107061GB-C61 and CNS2023-144309.
This work was financed by Portuguese funds through FCT (Funda\c c\~ao para a Ci\^encia e a Tecnologia) in the framework of the project 2022.04048.PTDC (Phi in the Sky, DOI 10.54499/2022.04048.PTDC). C.J.M. also acknowledges FCT and POCH/FSE (EC) support through Investigador FCT Contract 2021.01214.CEECIND/CP1658/CT0001 (DOI 10.54499/2021.01214.CEECIND/CP1658/CT0001).
We acknowledge financial support from the Agencia Estatal de Investigaci\'on of the Ministerio de Ciencia e Innovaci\'on MCIN/AEI/10.13039/501100011033 and the ERDF “A way of making Europe” through project PID2021-125627OB-C32, and from the Centre of Excellence “Severo Ochoa” award to the Instituto de Astrofisica de Canarias.
F.P. and C.L. would like to acknowledge the Swiss National Science Foundation (SNSF) for supporting research with ESPRESSO through the SNSF grants nr. 140649, 152721, 166227, 184618 and 215190. The ESPRESSO Instrument Project was partially funded through SNSF’s FLARE Programme for large infrastructures.
V.~A. acknowledges support from FCT through a work contract funded by the FCT Scientific Employment Stimulus program (reference 2023.06055.CEECIND/CP2839/CT0005, DOI: 10.54499/2023.06055.CEECIND/CP2839/CT0005).

\end{acknowledgements}

\bibliographystyle{aa} 
\bibliography{biblio} 

\begin{thebibliography}{93}
\expandafter\ifx\csname natexlab\endcsname\relax\def\natexlab#1{#1}\fi

\bibitem[{{Aller} {et~al.}(2020){Aller}, {Lillo-Box}, {Jones}, {Miranda}, \& {Barcel{\'o} Forteza}}]{Aller2020}
{Aller}, A., {Lillo-Box}, J., {Jones}, D., {Miranda}, L.~F., \& {Barcel{\'o} Forteza}, S. 2020, \aap, 635, A128

\bibitem[{{Anglada-Escud{\'e}} \& {Butler}(2012)}]{Anglada2012}
{Anglada-Escud{\'e}}, G. \& {Butler}, R.~P. 2012, \apjs, 200, 15

\bibitem[{{Artigau} {et~al.}(2022){Artigau}, {Cadieux}, {Cook}, {Doyon}, {Vandal}, {Donati}, {Moutou}, {Delfosse}, {Fouqu{\'e}}, {Martioli}, {Bouchy}, {Parsons}, {Carmona}, {Dumusque}, {Astudillo-Defru}, {Bonfils}, \& {Mignon}}]{Artigau2022}
{Artigau}, {\'E}., {Cadieux}, C., {Cook}, N.~J., {et~al.} 2022, \aj, 164, 84

\bibitem[{Astudillo-Defru(2015)}]{Astudillo-Defru2015thesis}
Astudillo-Defru, N. 2015, PhD thesis, thèse de doctorat dirigée par Delfosse, Xavier et Bonfils, Xavier Astrophysique et milieux dilués Université Grenoble Alpes (ComUE) 2015

\bibitem[{{Baluev}(2013)}]{Baluev2013}
{Baluev}, R.~V. 2013, \mnras, 429, 2052

\bibitem[{{Baranne} {et~al.}(1996){Baranne}, {Queloz}, {Mayor}, {Adrianzyk}, {Knispel}, {Kohler}, {Lacroix}, {Meunier}, {Rimbaud}, \& {Vin}}]{Baranne1996}
{Baranne}, A., {Queloz}, D., {Mayor}, M., {et~al.} 1996, \aaps, 119, 373

\bibitem[{{Barrag{\'a}n} {et~al.}(2022){Barrag{\'a}n}, {Aigrain}, {Rajpaul}, \& {Zicher}}]{pyaneti2}
{Barrag{\'a}n}, O., {Aigrain}, S., {Rajpaul}, V.~M., \& {Zicher}, N. 2022, \mnras, 509, 866

\bibitem[{Barrag\'an {et~al.}(2019)Barrag\'an, Gandolfi, \& Antoniciello}]{pyaneti}
Barrag\'an, O., Gandolfi, D., \& Antoniciello, G. 2019, \mnras, 482, 1017

\bibitem[{{Barrag{\'a}n} {et~al.}(2023){Barrag{\'a}n}, {Gillen}, {Aigrain}, {Meech}, {Klein}, {Nielsen}, {Yu}, {O'Sullivan}, {Nicholson}, \& {Lillo-Box}}]{Barragan2023}
{Barrag{\'a}n}, O., {Gillen}, E., {Aigrain}, S., {et~al.} 2023, \mnras, 522, 3458

\bibitem[{{Baumeister} \& {Tosi}(2023)}]{Baumeister2023}
{Baumeister}, P. \& {Tosi}, N. 2023, \aap, 676, A106

\bibitem[{{Bell} \& {Higgins}(2022)}]{Bell2022TESSPRF}
{Bell}, K.~J. \& {Higgins}, M.~E. 2022, {TESS\_PRF: Display the TESS pixel response function}, Astrophysics Source Code Library, record ascl:2207.008

\bibitem[{{Bonfils} {et~al.}(2007){Bonfils}, {Mayor}, {Delfosse}, {Forveille}, {Gillon}, {Perrier}, {Udry}, {Bouchy}, {Lovis}, {Pepe}, {Queloz}, {Santos}, \& {Bertaux}}]{Bonfils2007}
{Bonfils}, X., {Mayor}, M., {Delfosse}, X., {et~al.} 2007, \aap, 474, 293

\bibitem[{{Brahm} {et~al.}(2019){Brahm}, {Espinoza}, {Jord{\'a}n}, {Henning}, {Sarkis}, {Jones}, {D{\'\i}az}, {Jenkins}, {Vanzi}, {Zapata}, {Petrovich}, {Kossakowski}, {Rabus}, {Rojas}, \& {Torres}}]{Brahm2019PARSEC}
{Brahm}, R., {Espinoza}, N., {Jord{\'a}n}, A., {et~al.} 2019, \aj, 158, 45

\bibitem[{{Bressan} {et~al.}(2012){Bressan}, {Marigo}, {Girardi}, {Salasnich}, {Dal Cero}, {Rubele}, \& {Nanni}}]{Bressan2012}
{Bressan}, A., {Marigo}, P., {Girardi}, L., {et~al.} 2012, \mnras, 427, 127

\bibitem[{{Brown} {et~al.}(2013){Brown}, {Baliber}, {Bianco}, {Bowman}, {Burleson}, {Conway}, {Crellin}, {Depagne}, {De Vera}, {Dilday}, {Dragomir}, {Dubberley}, {Eastman}, {Elphick}, {Falarski}, {Foale}, {Ford}, {Fulton}, {Garza}, {Gomez}, {Graham}, {Greene}, {Haldeman}, {Hawkins}, {Haworth}, {Haynes}, {Hidas}, {Hjelstrom}, {Howell}, {Hygelund}, {Lister}, {Lobdill}, {Martinez}, {Mullins}, {Norbury}, {Parrent}, {Paulson}, {Petry}, {Pickles}, {Posner}, {Rosing}, {Ross}, {Sand}, {Saunders}, {Shobbrook}, {Shporer}, {Street}, {Thomas}, {Tsapras}, {Tufts}, {Valenti}, {Vander Horst}, {Walker}, {White}, \& {Willis}}]{Brown2013}
{Brown}, T.~M., {Baliber}, N., {Bianco}, F.~B., {et~al.} 2013, \pasp, 125, 1031

\bibitem[{{Butler} {et~al.}(2017){Butler}, {Vogt}, {Laughlin}, {Burt}, {Rivera}, {Tuomi}, {Teske}, {Arriagada}, {Diaz}, {Holden}, \& {Keiser}}]{Butler2017}
{Butler}, R.~P., {Vogt}, S.~S., {Laughlin}, G., {et~al.} 2017, \aj, 153, 208

\bibitem[{{Carmona} {et~al.}(2023){Carmona}, {Delfosse}, {Bellotti}, {Cort{\'e}s-Zuleta}, {Ould-Elhkim}, {Heidari}, {Mignon}, {Donati}, {Moutou}, {Cook}, {Artigau}, {Fouqu{\'e}}, {Martioli}, {Cadieux}, {Morin}, {Forveille}, {Boisse}, {H{\'e}brard}, {D{\'\i}az}, {Lafreni{\`e}re}, {Kiefer}, {Petit}, {Doyon}, {Acu{\~n}a}, {Arnold}, {Bonfils}, {Bouchy}, {Bourrier}, {Dalal}, {Deleuil}, {Demangeon}, {Dumusque}, {Hara}, {Hoyer}, {Mousis}, {Santerne}, {S{\'e}grasan}, {Stalport}, \& {Udry}}]{Carmona2023}
{Carmona}, A., {Delfosse}, X., {Bellotti}, S., {et~al.} 2023, \aap, 674, A110

\bibitem[{{Castro-Gonz{\'a}lez} {et~al.}(2023){Castro-Gonz{\'a}lez}, {Demangeon}, {Lillo-Box}, {Lovis}, {Lavie}, {Adibekyan}, {Acu{\~n}a}, {Deleuil}, {Aguichine}, {Zapatero Osorio}, {Tabernero}, {Davoult}, {Alibert}, {Santos}, {Sousa}, {Antoniadis-Karnavas}, {Borsa}, {Winn}, {Allende Prieto}, {Figueira}, {Jenkins}, {Sozzetti}, {Damasso}, {Silva}, {Astudillo-Defru}, {Barros}, {Bonfils}, {Cristiani}, {Di Marcantonio}, {Gonz{\'a}lez Hern{\'a}ndez}, {Curto}, {Martins}, {Nunes}, {Palle}, {Pepe}, {Seager}, \& {Su{\'a}rez Mascare{\~n}o}}]{Castro2023}
{Castro-Gonz{\'a}lez}, A., {Demangeon}, O.~D.~S., {Lillo-Box}, J., {et~al.} 2023, \aap, 675, A52

\bibitem[{{Castro-Gonz{\'a}lez} {et~al.}(2022){Castro-Gonz{\'a}lez}, {D{\'\i}ez Alonso}, {Men{\'e}ndez Blanco}, {Livingston}, {de Leon}, {Lillo-Box}, {Korth}, {Fern{\'a}ndez Men{\'e}ndez}, {Recio}, {Izquierdo-Ruiz}, {Coya Lozano}, {Garc{\'\i}a de la Cuesta}, {G{\'o}mez Hern{\'a}ndez}, {Vidal Blanco}, {Hevia D{\'\i}az}, {Pardo Silva}, {P{\'e}rez Acevedo}, {Polancos Ruiz}, {Padilla Tijer{\'\i}n}, {V{\'a}zquez Garc{\'\i}a}, {Su{\'a}rez G{\'o}mez}, {Garc{\'\i}a Riesgo}, {Gonz{\'a}lez Guti{\'e}rrez}, {Bonavera}, {Gonz{\'a}lez-Nuevo}, {Rodr{\'\i}guez Pereira}, {S{\'a}nchez Lasheras}, {S{\'a}nchez Rodr{\'\i}guez}, {Mu{\~n}iz}, {Santos Rodr{\'\i}guez}, \& {de Cos Juez}}]{Castro2022}
{Castro-Gonz{\'a}lez}, A., {D{\'\i}ez Alonso}, E., {Men{\'e}ndez Blanco}, J., {et~al.} 2022, \mnras, 509, 1075

\bibitem[{{Castro-Gonz{\'a}lez} {et~al.}(2024){Castro-Gonz{\'a}lez}, {Lillo-Box}, {Armstrong}, {Acu{\~n}a}, {Aguichine}, {Bourrier}, {Gandhi}, {Sousa}, {Delgado-Mena}, {Moya}, {Adibekyan}, {Correia}, {Barrado}, {Damasso}, {Winn}, {Santos}, {Barkaoui}, {Barros}, {Benkhaldoun}, {Bouchy}, {Brice{\~n}o}, {Caldwell}, {Collins}, {Essack}, {Ghachoui}, {Gillon}, {Hounsell}, {Jehin}, {Jenkins}, {Keniger}, {Law}, {Mann}, {Nielsen}, {Pozuelos}, {Schanche}, {Seager}, {Tan}, {Timmermans}, {Villase{\~n}or}, {Watkins}, \& {Ziegler}}]{Castro2024}
{Castro-Gonz{\'a}lez}, A., {Lillo-Box}, J., {Armstrong}, D.~J., {et~al.} 2024, \aap, 691, A233

\bibitem[{{Ciardi} {et~al.}(2015){Ciardi}, {Beichman}, {Horch}, \& {Howell}}]{Ciardi2015}
{Ciardi}, D.~R., {Beichman}, C.~A., {Horch}, E.~P., \& {Howell}, S.~B. 2015, \apj, 805, 16

\bibitem[{{Cincunegui} {et~al.}(2007){Cincunegui}, {D{\'\i}az}, \& {Mauas}}]{Cincunegui2007}
{Cincunegui}, C., {D{\'\i}az}, R.~F., \& {Mauas}, P.~J.~D. 2007, \aap, 469, 309

\bibitem[{{Collins} {et~al.}(2017){Collins}, {Kielkopf}, {Stassun}, \& {Hessman}}]{Collins2017}
{Collins}, K.~A., {Kielkopf}, J.~F., {Stassun}, K.~G., \& {Hessman}, F.~V. 2017, \aj, 153, 77

\bibitem[{{Cretignier} {et~al.}(2023){Cretignier}, {Dumusque}, {Aigrain}, \& {Pepe}}]{Cretignier2023}
{Cretignier}, M., {Dumusque}, X., {Aigrain}, S., \& {Pepe}, F. 2023, \aap, 678, A2

\bibitem[{{Cretignier} {et~al.}(2021){Cretignier}, {Dumusque}, {Hara}, \& {Pepe}}]{Cretignier2021}
{Cretignier}, M., {Dumusque}, X., {Hara}, N.~C., \& {Pepe}, F. 2021, \aap, 653, A43

\bibitem[{{Delisle} {et~al.}(2020){Delisle}, {Hara}, \& {S{\'e}gransan}}]{Delisle2020}
{Delisle}, J.~B., {Hara}, N., \& {S{\'e}gransan}, D. 2020, \aap, 638, A95

\bibitem[{{Delisle} {et~al.}(2022){Delisle}, {Unger}, {Hara}, \& {S{\'e}gransan}}]{Delisle2022}
{Delisle}, J.~B., {Unger}, N., {Hara}, N.~C., \& {S{\'e}gransan}, D. 2022, \aap, 659, A182

\bibitem[{{D{\'\i}az} {et~al.}(2007){D{\'\i}az}, {Cincunegui}, \& {Mauas}}]{Diaz2007}
{D{\'\i}az}, R.~F., {Cincunegui}, C., \& {Mauas}, P. J.~D. 2007, \mnras, 378, 1007

\bibitem[{{Dumusque}(2018)}]{Dumusque2018}
{Dumusque}, X. 2018, \aap, 620, A47

\bibitem[{{Dumusque} {et~al.}(2014){Dumusque}, {Boisse}, \& {Santos}}]{Dumusque2014}
{Dumusque}, X., {Boisse}, I., \& {Santos}, N.~C. 2014, \apj, 796, 132

\bibitem[{{Espinoza} {et~al.}(2019){Espinoza}, {Kossakowski}, \& {Brahm}}]{Espinoza2019juliet}
{Espinoza}, N., {Kossakowski}, D., \& {Brahm}, R. 2019, \mnras, 490, 2262

\bibitem[{{Foreman-Mackey} {et~al.}(2017){Foreman-Mackey}, {Agol}, {Ambikasaran}, \& {Angus}}]{Foreman-Mackey2017}
{Foreman-Mackey}, D., {Agol}, E., {Ambikasaran}, S., \& {Angus}, R. 2017, \aj, 154, 220

\bibitem[{{Forveille} {et~al.}(2011){Forveille}, {Bonfils}, {Delfosse}, {Alonso}, {Udry}, {Bouchy}, {Gillon}, {Lovis}, {Neves}, {Mayor}, {Pepe}, {Queloz}, {Santos}, {Segransan}, {Almenara}, {Deeg}, \& {Rabus}}]{Forveille2011}
{Forveille}, T., {Bonfils}, X., {Delfosse}, X., {et~al.} 2011, arXiv e-prints, arXiv:1109.2505

\bibitem[{{Fulton} {et~al.}(2018){Fulton}, {Petigura}, {Blunt}, \& {Sinukoff}}]{Fulton2018}
{Fulton}, B.~J., {Petigura}, E.~A., {Blunt}, S., \& {Sinukoff}, E. 2018, \pasp, 130, 044504

\bibitem[{{Furlan} \& {Howell}(2017)}]{Furlan2017}
{Furlan}, E. \& {Howell}, S.~B. 2017, \aj, 154, 66

\bibitem[{{Gaia Collaboration} {et~al.}(2016){Gaia Collaboration}, {Prusti}, {de Bruijne}, {Brown}, {Vallenari}, {Babusiaux}, {Bailer-Jones}, {Bastian}, {Biermann}, {Evans}, {Eyer}, {Jansen}, {Jordi}, {Klioner}, {Lammers}, {Lindegren}, {Luri}, {Mignard}, {Milligan}, {Panem}, {Poinsignon}, {Pourbaix}, {Randich}, {Sarri}, {Sartoretti}, {Siddiqui}, {Soubiran}, {Valette}, {van Leeuwen}, {Walton}, {Aerts}, {Arenou}, {Cropper}, {Drimmel}, {H{\o}g}, {Katz}, {Lattanzi}, {O'Mullane}, {Grebel}, {Holland}, {Huc}, {Passot}, {Bramante}, {Cacciari}, {Casta{\~n}eda}, {Chaoul}, {Cheek}, {De Angeli}, {Fabricius}, {Guerra}, {Hern{\'a}ndez}, {Jean-Antoine-Piccolo}, {Masana}, {Messineo}, {Mowlavi}, {Nienartowicz}, {Ord{\'o}{\~n}ez-Blanco}, {Panuzzo}, {Portell}, {Richards}, {Riello}, {Seabroke}, {Tanga}, {Th{\'e}venin}, {Torra}, {Els}, {Gracia-Abril}, {Comoretto}, {Garcia-Reinaldos}, {Lock}, {Mercier}, {Altmann}, {Andrae}, {Astraatmadja}, {Bellas-Velidis}, {Benson}, {Berthier}, {Blomme}, {Busso}, {Carry}, {Cellino}, {Clementini},
  {Cowell}, {Creevey}, {Cuypers}, {Davidson}, {De Ridder}, {de Torres}, {Delchambre}, {Dell'Oro}, {Ducourant}, {Fr{\'e}mat}, {Garc{\'\i}a-Torres}, {Gosset}, {Halbwachs}, {Hambly}, {Harrison}, {Hauser}, {Hestroffer}, {Hodgkin}, {Huckle}, {Hutton}, {Jasniewicz}, {Jordan}, {Kontizas}, {Korn}, {Lanzafame}, {Manteiga}, {Moitinho}, {Muinonen}, {Osinde}, {Pancino}, {Pauwels}, {Petit}, {Recio-Blanco}, {Robin}, {Sarro}, {Siopis}, {Smith}, {Smith}, {Sozzetti}, {Thuillot}, {van Reeven}, {Viala}, {Abbas}, {Abreu Aramburu}, {Accart}, {Aguado}, {Allan}, {Allasia}, {Altavilla}, {{\'A}lvarez}, {Alves}, {Anderson}, {Andrei}, {Anglada Varela}, {Antiche}, {Antoja}, {Ant{\'o}n}, {Arcay}, {Atzei}, {Ayache}, {Bach}, {Baker}, {Balaguer-N{\'u}{\~n}ez}, {Barache}, {Barata}, {Barbier}, {Barblan}, {Baroni}, {Barrado y Navascu{\'e}s}, {Barros}, {Barstow}, {Becciani}, {Bellazzini}, {Bellei}, {Bello Garc{\'\i}a}, {Belokurov}, {Bendjoya}, {Berihuete}, {Bianchi}, {Bienaym{\'e}}, {Billebaud}, {Blagorodnova}, {Blanco-Cuaresma}, {Boch},
  {Bombrun}, {Borrachero}, {Bouquillon}, {Bourda}, {Bouy}, {Bragaglia}, {Breddels}, {Brouillet}, {Br{\"u}semeister}, {Bucciarelli}, {Budnik}, {Burgess}, {Burgon}, {Burlacu}, {Busonero}, {Buzzi}, {Caffau}, {Cambras}, {Campbell}, {Cancelliere}, {Cantat-Gaudin}, {Carlucci}, {Carrasco}, {Castellani}, {Charlot}, {Charnas}, {Charvet}, {Chassat}, {Chiavassa}, {Clotet}, {Cocozza}, {Collins}, {Collins}, {Costigan}, {Crifo}, {Cross}, {Crosta}, {Crowley}, {Dafonte}, {Damerdji}, {Dapergolas}, {David}, {David}, {De Cat}, {de Felice}, {de Laverny}, {De Luise}, {De March}, {de Martino}, {de Souza}, {Debosscher}, {del Pozo}, {Delbo}, {Delgado}, {Delgado}, {di Marco}, {Di Matteo}, {Diakite}, {Distefano}, {Dolding}, {Dos Anjos}, {Drazinos}, {Dur{\'a}n}, {Dzigan}, {Ecale}, {Edvardsson}, {Enke}, {Erdmann}, {Escolar}, {Espina}, {Evans}, {Eynard Bontemps}, {Fabre}, {Fabrizio}, {Faigler}, {Falc{\~a}o}, {Farr{\`a}s Casas}, {Faye}, {Federici}, {Fedorets}, {Fern{\'a}ndez-Hern{\'a}ndez}, {Fernique}, {Fienga}, {Figueras}, {Filippi},
  {Findeisen}, {Fonti}, {Fouesneau}, {Fraile}, {Fraser}, {Fuchs}, {Furnell}, {Gai}, {Galleti}, {Galluccio}, {Garabato}, {Garc{\'\i}a-Sedano}, {Gar{\'e}}, {Garofalo}, {Garralda}, {Gavras}, {Gerssen}, {Geyer}, {Gilmore}, {Girona}, {Giuffrida}, {Gomes}, {Gonz{\'a}lez-Marcos}, {Gonz{\'a}lez-N{\'u}{\~n}ez}, {Gonz{\'a}lez-Vidal}, {Granvik}, {Guerrier}, {Guillout}, {Guiraud}, {G{\'u}rpide}, {Guti{\'e}rrez-S{\'a}nchez}, {Guy}, {Haigron}, {Hatzidimitriou}, {Haywood}, {Heiter}, {Helmi}, {Hobbs}, {Hofmann}, {Holl}, {Holland }, {Hunt}, {Hypki}, {Icardi}, {Irwin}, {Jevardat de Fombelle}, {Jofr{\'e}}, {Jonker}, {Jorissen}, {Julbe}, {Karampelas}, {Kochoska}, {Kohley}, {Kolenberg}, {Kontizas}, {Koposov}, {Kordopatis}, {Koubsky}, {Kowalczyk}, {Krone-Martins}, {Kudryashova}, {Kull}, {Bachchan}, {Lacoste-Seris}, {Lanza}, {Lavigne}, {Le Poncin-Lafitte}, {Lebreton}, {Lebzelter}, {Leccia}, {Leclerc}, {Lecoeur-Taibi}, {Lemaitre}, {Lenhardt}, {Leroux}, {Liao}, {Licata}, {Lindstr{\o}m}, {Lister}, {Livanou}, {Lobel}, {L{\"o}ffler},
  {L{\'o}pez}, {Lopez-Lozano}, {Lorenz}, {Loureiro}, {MacDonald}, {Magalh{\~a}es Fernandes}, {Managau}, {Mann}, {Mantelet}, {Marchal}, {Marchant}, {Marconi}, {Marie}, {Marinoni}, {Marrese}, {Marschalk{\'o}}, {Marshall}, {Mart{\'\i}n-Fleitas}, {Martino}, {Mary}, {Matijevi{\v{c}}}, {Mazeh}, {McMillan}, {Messina}, {Mestre}, {Michalik}, {Millar}, {Miranda}, {Molina}, {Molinaro}, {Molinaro}, {Moln{\'a}r}, {Moniez}, {Montegriffo}, {Monteiro}, {Mor}, {Mora}, {Morbidelli}, {Morel}, {Morgenthaler}, {Morley}, {Morris}, {Mulone}, {Muraveva}, {Musella}, {Narbonne}, {Nelemans}, {Nicastro}, {Noval}, {Ord{\'e}novic}, {Ordieres-Mer{\'e}}, {Osborne}, {Pagani}, {Pagano}, {Pailler}, {Palacin}, {Palaversa}, {Parsons}, {Paulsen}, {Pecoraro}, {Pedrosa}, {Pentik{\"a}inen}, {Pereira}, {Pichon}, {Piersimoni}, {Pineau}, {Plachy}, {Plum}, {Poujoulet}, {Pr{\v{s}}a}, {Pulone}, {Ragaini}, {Rago}, {Rambaux}, {Ramos-Lerate}, {Ranalli}, {Rauw}, {Read}, {Regibo}, {Renk}, {Reyl{\'e}}, {Ribeiro}, {Rimoldini}, {Ripepi}, {Riva}, {Rixon},
  {Roelens}, {Romero-G{\'o}mez}, {Rowell}, {Royer}, {Rudolph}, {Ruiz-Dern}, {Sadowski}, {Sagrist{\`a} Sell{\'e}s}, {Sahlmann}, {Salgado}, {Salguero}, {Sarasso}, {Savietto}, {Schnorhk}, {Schultheis}, {Sciacca}, {Segol}, {Segovia}, {Segransan}, {Serpell}, {Shih}, {Smareglia}, {Smart}, {Smith}, {Solano}, {Solitro}, {Sordo}, {Soria Nieto}, {Souchay}, {Spagna}, {Spoto}, {Stampa}, {Steele}, {Steidelm{\"u}ller}, {Stephenson}, {Stoev}, {Suess}, {S{\"u}veges}, {Surdej}, {Szabados}, {Szegedi-Elek}, {Tapiador}, {Taris}, {Tauran}, {Taylor}, {Teixeira}, {Terrett}, {Tingley}, {Trager}, {Turon}, {Ulla}, {Utrilla}, {Valentini}, {van Elteren}, {Van Hemelryck}, {van Leeuwen}, {Varadi}, {Vecchiato}, {Veljanoski}, {Via}, {Vicente}, {Vogt}, {Voss}, {Votruba}, {Voutsinas}, {Walmsley}, {Weiler}, {Weingrill}, {Werner}, {Wevers}, {Whitehead}, {Wyrzykowski}, {Yoldas}, {{\v{Z}}erjal}, {Zucker}, {Zurbach}, {Zwitter}, {Alecu}, {Allen}, {Allende Prieto}, {Amorim}, {Anglada-Escud{\'e}}, {Arsenijevic}, {Azaz}, {Balm}, {Beck}, {Bernstein},
  {Bigot}, {Bijaoui}, {Blasco}, {Bonfigli}, {Bono}, {Boudreault}, {Bressan}, {Brown}, {Brunet}, {Bunclark}, {Buonanno}, {Butkevich}, {Carret}, {Carrion}, {Chemin}, {Ch{\'e}reau}, {Corcione}, {Darmigny}, {de Boer}, {de Teodoro}, {de Zeeuw}, {Delle Luche}, {Domingues}, {Dubath}, {Fodor}, {Fr{\'e}zouls}, {Fries}, {Fustes}, {Fyfe}, {Gallardo}, {Gallegos}, {Gardiol}, {Gebran}, {Gomboc}, {G{\'o}mez}, {Grux}, {Gueguen}, {Heyrovsky}, {Hoar}, {Iannicola}, {Isasi Parache}, {Janotto}, {Joliet}, {Jonckheere}, {Keil}, {Kim}, {Klagyivik}, {Klar}, {Knude}, {Kochukhov}, {Kolka}, {Kos}, {Kutka}, {Lainey}, {LeBouquin}, {Liu}, {Loreggia}, {Makarov}, {Marseille}, {Martayan}, {Martinez-Rubi}, {Massart}, {Meynadier}, {Mignot}, {Munari}, {Nguyen}, {Nordlander}, {Ocvirk}, {O'Flaherty}, {Olias Sanz}, {Ortiz}, {Osorio}, {Oszkiewicz}, {Ouzounis}, {Palmer}, {Park}, {Pasquato}, {Peltzer}, {Peralta}, {P{\'e}turaud}, {Pieniluoma}, {Pigozzi}, {Poels}, {Prat}, {Prod'homme}, {Raison}, {Rebordao}, {Risquez}, {Rocca-Volmerange}, {Rosen},
  {Ruiz-Fuertes}, {Russo}, {Sembay}, {Serraller Vizcaino}, {Short}, {Siebert}, {Silva}, {Sinachopoulos}, {Slezak}, {Soffel}, {Sosnowska}, {Strai{\v{z}}ys}, {ter Linden}, {Terrell}, {Theil}, {Tiede}, {Troisi}, {Tsalmantza}, {Tur}, {Vaccari}, {Vachier}, {Valles}, {Van Hamme}, {Veltz}, {Virtanen}, {Wallut}, {Wichmann}, {Wilkinson}, {Ziaeepour}, \& {Zschocke}}]{GAIA2016}
{Gaia Collaboration}, {Prusti}, T., {de Bruijne}, J.~H.~J., {et~al.} 2016, \aap, 595, A1

\bibitem[{{Gaia Collaboration} {et~al.}(2023){Gaia Collaboration}, {Vallenari}, {Brown}, {Prusti}, {de Bruijne}, {Arenou}, {Babusiaux}, {Biermann}, {Creevey}, {Ducourant}, {Evans}, {Eyer}, {Guerra}, {Hutton}, {Jordi}, {Klioner}, {Lammers}, {Lindegren}, {Luri}, {Mignard}, {Panem}, {Pourbaix}, {Randich}, {Sartoretti}, {Soubiran}, {Tanga}, {Walton}, {Bailer-Jones}, {Bastian}, {Drimmel}, {Jansen}, {Katz}, {Lattanzi}, {van Leeuwen}, {Bakker}, {Cacciari}, {Casta{\~n}eda}, {De Angeli}, {Fabricius}, {Fouesneau}, {Fr{\'e}mat}, {Galluccio}, {Guerrier}, {Heiter}, {Masana}, {Messineo}, {Mowlavi}, {Nicolas}, {Nienartowicz}, {Pailler}, {Panuzzo}, {Riclet}, {Roux}, {Seabroke}, {Sordo}, {Th{\'e}venin}, {Gracia-Abril}, {Portell}, {Teyssier}, {Altmann}, {Andrae}, {Audard}, {Bellas-Velidis}, {Benson}, {Berthier}, {Blomme}, {Burgess}, {Busonero}, {Busso}, {C{\'a}novas}, {Carry}, {Cellino}, {Cheek}, {Clementini}, {Damerdji}, {Davidson}, {de Teodoro}, {Nu{\~n}ez Campos}, {Delchambre}, {Dell'Oro}, {Esquej},
  {Fern{\'a}ndez-Hern{\'a}ndez}, {Fraile}, {Garabato}, {Garc{\'\i}a-Lario}, {Gosset}, {Haigron}, {Halbwachs}, {Hambly}, {Harrison}, {Hern{\'a}ndez}, {Hestroffer}, {Hodgkin}, {Holl}, {Jan{\ss}en}, {Jevardat de Fombelle}, {Jordan}, {Krone-Martins}, {Lanzafame}, {L{\"o}ffler}, {Marchal}, {Marrese}, {Moitinho}, {Muinonen}, {Osborne}, {Pancino}, {Pauwels}, {Recio-Blanco}, {Reyl{\'e}}, {Riello}, {Rimoldini}, {Roegiers}, {Rybizki}, {Sarro}, {Siopis}, {Smith}, {Sozzetti}, {Utrilla}, {van Leeuwen}, {Abbas}, {{\'A}brah{\'a}m}, {Abreu Aramburu}, {Aerts}, {Aguado}, {Ajaj}, {Aldea-Montero}, {Altavilla}, {{\'A}lvarez}, {Alves}, {Anders}, {Anderson}, {Anglada Varela}, {Antoja}, {Baines}, {Baker}, {Balaguer-N{\'u}{\~n}ez}, {Balbinot}, {Balog}, {Barache}, {Barbato}, {Barros}, {Barstow}, {Bartolom{\'e}}, {Bassilana}, {Bauchet}, {Becciani}, {Bellazzini}, {Berihuete}, {Bernet}, {Bertone}, {Bianchi}, {Binnenfeld}, {Blanco-Cuaresma}, {Blazere}, {Boch}, {Bombrun}, {Bossini}, {Bouquillon}, {Bragaglia}, {Bramante}, {Breedt},
  {Bressan}, {Brouillet}, {Brugaletta}, {Bucciarelli}, {Burlacu}, {Butkevich}, {Buzzi}, {Caffau}, {Cancelliere}, {Cantat-Gaudin}, {Carballo}, {Carlucci}, {Carnerero}, {Carrasco}, {Casamiquela}, {Castellani}, {Castro-Ginard}, {Chaoul}, {Charlot}, {Chemin}, {Chiaramida}, {Chiavassa}, {Chornay}, {Comoretto}, {Contursi}, {Cooper}, {Cornez}, {Cowell}, {Crifo}, {Cropper}, {Crosta}, {Crowley}, {Dafonte}, {Dapergolas}, {David}, {David}, {de Laverny}, {De Luise}, {De March}, {De Ridder}, {de Souza}, {de Torres}, {del Peloso}, {del Pozo}, {Delbo}, {Delgado}, {Delisle}, {Demouchy}, {Dharmawardena}, {Di Matteo}, {Diakite}, {Diener}, {Distefano}, {Dolding}, {Edvardsson}, {Enke}, {Fabre}, {Fabrizio}, {Faigler}, {Fedorets}, {Fernique}, {Fienga}, {Figueras}, {Fournier}, {Fouron}, {Fragkoudi}, {Gai}, {Garcia-Gutierrez}, {Garcia-Reinaldos}, {Garc{\'\i}a-Torres}, {Garofalo}, {Gavel}, {Gavras}, {Gerlach}, {Geyer}, {Giacobbe}, {Gilmore}, {Girona}, {Giuffrida}, {Gomel}, {Gomez}, {Gonz{\'a}lez-N{\'u}{\~n}ez},
  {Gonz{\'a}lez-Santamar{\'\i}a}, {Gonz{\'a}lez-Vidal}, {Granvik}, {Guillout}, {Guiraud}, {Guti{\'e}rrez-S{\'a}nchez}, {Guy}, {Hatzidimitriou}, {Hauser}, {Haywood}, {Helmer}, {Helmi}, {Sarmiento}, {Hidalgo}, {Hilger}, {H{\l}adczuk}, {Hobbs}, {Holland}, {Huckle}, {Jardine}, {Jasniewicz}, {Jean-Antoine Piccolo}, {Jim{\'e}nez-Arranz}, {Jorissen}, {Juaristi Campillo}, {Julbe}, {Karbevska}, {Kervella}, {Khanna}, {Kontizas}, {Kordopatis}, {Korn}, {K{\'o}sp{\'a}l}, {Kostrzewa-Rutkowska}, {Kruszy{\'n}ska}, {Kun}, {Laizeau}, {Lambert}, {Lanza}, {Lasne}, {Le Campion}, {Lebreton}, {Lebzelter}, {Leccia}, {Leclerc}, {Lecoeur-Taibi}, {Liao}, {Licata}, {Lindstr{\o}m}, {Lister}, {Livanou}, {Lobel}, {Lorca}, {Loup}, {Madrero Pardo}, {Magdaleno Romeo}, {Managau}, {Mann}, {Manteiga}, {Marchant}, {Marconi}, {Marcos}, {Marcos Santos}, {Mar{\'\i}n Pina}, {Marinoni}, {Marocco}, {Marshall}, {Martin Polo}, {Mart{\'\i}n-Fleitas}, {Marton}, {Mary}, {Masip}, {Massari}, {Mastrobuono-Battisti}, {Mazeh}, {McMillan}, {Messina}, {Michalik},
  {Millar}, {Mints}, {Molina}, {Molinaro}, {Moln{\'a}r}, {Monari}, {Mongui{\'o}}, {Montegriffo}, {Montero}, {Mor}, {Mora}, {Morbidelli}, {Morel}, {Morris}, {Muraveva}, {Murphy}, {Musella}, {Nagy}, {Noval}, {Oca{\~n}a}, {Ogden}, {Ordenovic}, {Osinde}, {Pagani}, {Pagano}, {Palaversa}, {Palicio}, {Pallas-Quintela}, {Panahi}, {Payne-Wardenaar}, {Pe{\~n}alosa Esteller}, {Penttil{\"a}}, {Pichon}, {Piersimoni}, {Pineau}, {Plachy}, {Plum}, {Poggio}, {Pr{\v{s}}a}, {Pulone}, {Racero}, {Ragaini}, {Rainer}, {Raiteri}, {Rambaux}, {Ramos}, {Ramos-Lerate}, {Re Fiorentin}, {Regibo}, {Richards}, {Rios Diaz}, {Ripepi}, {Riva}, {Rix}, {Rixon}, {Robichon}, {Robin}, {Robin}, {Roelens}, {Rogues}, {Rohrbasser}, {Romero-G{\'o}mez}, {Rowell}, {Royer}, {Ruz Mieres}, {Rybicki}, {Sadowski}, {S{\'a}ez N{\'u}{\~n}ez}, {Sagrist{\`a} Sell{\'e}s}, {Sahlmann}, {Salguero}, {Samaras}, {Sanchez Gimenez}, {Sanna}, {Santove{\~n}a}, {Sarasso}, {Schultheis}, {Sciacca}, {Segol}, {Segovia}, {S{\'e}gransan}, {Semeux}, {Shahaf}, {Siddiqui}, {Siebert},
  {Siltala}, {Silvelo}, {Slezak}, {Slezak}, {Smart}, {Snaith}, {Solano}, {Solitro}, {Souami}, {Souchay}, {Spagna}, {Spina}, {Spoto}, {Steele}, {Steidelm{\"u}ller}, {Stephenson}, {S{\"u}veges}, {Surdej}, {Szabados}, {Szegedi-Elek}, {Taris}, {Taylor}, {Teixeira}, {Tolomei}, {Tonello}, {Torra}, {Torra}, {Torralba Elipe}, {Trabucchi}, {Tsounis}, {Turon}, {Ulla}, {Unger}, {Vaillant}, {van Dillen}, {van Reeven}, {Vanel}, {Vecchiato}, {Viala}, {Vicente}, {Voutsinas}, {Weiler}, {Wevers}, {Wyrzykowski}, {Yoldas}, {Yvard}, {Zhao}, {Zorec}, {Zucker}, \& {Zwitter}}]{GaiaDR3}
{Gaia Collaboration}, {Vallenari}, A., {Brown}, A.~G.~A., {et~al.} 2023, \aap, 674, A1

\bibitem[{{Gomes da Silva} {et~al.}(2011){Gomes da Silva}, {Santos}, {Bonfils}, {Delfosse}, {Forveille}, \& {Udry}}]{Gomes2011}
{Gomes da Silva}, J., {Santos}, N.~C., {Bonfils}, X., {et~al.} 2011, \aap, 534, A30

\bibitem[{{Guerrero} {et~al.}(2021){Guerrero}, {Seager}, {Huang}, {Vanderburg}, {Garcia Soto}, {Mireles}, {Hesse}, {Fong}, {Glidden}, {Shporer}, {Latham}, {Collins}, {Quinn}, {Burt}, {Dragomir}, {Crossfield}, {Vanderspek}, {Fausnaugh}, {Burke}, {Ricker}, {Daylan}, {Essack}, {G{\"u}nther}, {Osborn}, {Pepper}, {Rowden}, {Sha}, {Villanueva}, {Yahalomi}, {Yu}, {Ballard}, {Batalha}, {Berardo}, {Chontos}, {Dittmann}, {Esquerdo}, {Mikal-Evans}, {Jayaraman}, {Krishnamurthy}, {Louie}, {Mehrle}, {Niraula}, {Rackham}, {Rodriguez}, {Rowden}, {Sousa-Silva}, {Watanabe}, {Wong}, {Zhan}, {Zivanovic}, {Christiansen}, {Ciardi}, {Swain}, {Lund}, {Mullally}, {Fleming}, {Rodriguez}, {Boyd}, {Quintana}, {Barclay}, {Col{\'o}n}, {Rinehart}, {Schlieder}, {Clampin}, {Jenkins}, {Twicken}, {Caldwell}, {Coughlin}, {Henze}, {Lissauer}, {Morris}, {Rose}, {Smith}, {Tenenbaum}, {Ting}, {Wohler}, {Bakos}, {Bean}, {Berta-Thompson}, {Bieryla}, {Bouma}, {Buchhave}, {Butler}, {Charbonneau}, {Doty}, {Ge}, {Holman}, {Howard}, {Kaltenegger}, {Kane},
  {Kjeldsen}, {Kreidberg}, {Lin}, {Minsky}, {Narita}, {Paegert}, {P{\'a}l}, {Palle}, {Sasselov}, {Spencer}, {Sozzetti}, {Stassun}, {Torres}, {Udry}, \& {Winn}}]{Guerrero2021}
{Guerrero}, N.~M., {Seager}, S., {Huang}, C.~X., {et~al.} 2021, \apjs, 254, 39

\bibitem[{{Hara} \& {Delisle}(2023)}]{Hara2023}
{Hara}, N.~C. \& {Delisle}, J.-B. 2023, arXiv e-prints, arXiv:2304.08489

\bibitem[{{Haywood} {et~al.}(2014){Haywood}, {Collier Cameron}, {Queloz}, {Barros}, {Deleuil}, {Fares}, {Gillon}, {Lanza}, {Lovis}, {Moutou}, {Pepe}, {Pollacco}, {Santerne}, {S{\'e}gransan}, \& {Unruh}}]{Haywood2014}
{Haywood}, R.~D., {Collier Cameron}, A., {Queloz}, D., {et~al.} 2014, \mnras, 443, 2517

\bibitem[{{H{\o}g} {et~al.}(2000){H{\o}g}, {Fabricius}, {Makarov}, {Urban}, {Corbin}, {Wycoff}, {Bastian}, {Schwekendiek}, \& {Wicenec}}]{Tycho-2}
{H{\o}g}, E., {Fabricius}, C., {Makarov}, V.~V., {et~al.} 2000, \aap, 355, L27

\bibitem[{{Jenkins} {et~al.}(2016){Jenkins}, {Twicken}, {McCauliff}, {Campbell}, {Sanderfer}, {Lung}, {Mansouri-Samani}, {Girouard}, {Tenenbaum}, {Klaus}, {Smith}, {Caldwell}, {Chacon}, {Henze}, {Heiges}, {Latham}, {Morgan}, {Swade}, {Rinehart}, \& {Vanderspek}}]{Jenkins2016}
{Jenkins}, J.~M., {Twicken}, J.~D., {McCauliff}, S., {et~al.} 2016, in Society of Photo-Optical Instrumentation Engineers (SPIE) Conference Series, Vol. 9913, Software and Cyberinfrastructure for Astronomy IV, ed. G.~{Chiozzi} \& J.~C. {Guzman}, 99133E

\bibitem[{{Jensen}(2013)}]{Jensen2013}
{Jensen}, E. 2013, {Tapir: A web interface for transit/eclipse observability}, Astrophysics Source Code Library, record ascl:1306.007

\bibitem[{{Kipping}(2013)}]{Kipping2013}
{Kipping}, D.~M. 2013, \mnras, 435, 2152

\bibitem[{{Kochanek} {et~al.}(2017){Kochanek}, {Shappee}, {Stanek}, {Holoien}, {Thompson}, {Prieto}, {Dong}, {Shields}, {Will}, {Britt}, {Perzanowski}, \& {Pojma{\'n}ski}}]{Kochanek2017}
{Kochanek}, C.~S., {Shappee}, B.~J., {Stanek}, K.~Z., {et~al.} 2017, \pasp, 129, 104502

\bibitem[{{Kreidberg}(2015)}]{Kreidberg2015}
{Kreidberg}, L. 2015, \pasp, 127, 1161

\bibitem[{{Kunimoto} {et~al.}(2021){Kunimoto}, {Huang}, {Tey}, {Fong}, {Hesse}, {Shporer}, {Guerrero}, {Fausnaugh}, {Vanderspek}, \& {Ricker}}]{Kunimoto2021}
{Kunimoto}, M., {Huang}, C., {Tey}, E., {et~al.} 2021, Research Notes of the American Astronomical Society, 5, 234

\bibitem[{{Lagrange} {et~al.}(2010){Lagrange}, {Desort}, \& {Meunier}}]{Lagrange2010}
{Lagrange}, A.~M., {Desort}, M., \& {Meunier}, N. 2010, \aap, 512, A38

\bibitem[{{Lavie} {et~al.}(2023){Lavie}, {Bouchy}, {Lovis}, {Zapatero Osorio}, {Deline}, {Barros}, {Figueira}, {Sozzetti}, {Gonz{\'a}lez Hern{\'a}ndez}, {Lillo-Box}, {Rodrigues}, {Mehner}, {Damasso}, {Adibekyan}, {Alibert}, {Allende Prieto}, {Cristiani}, {D'Odorico}, {Di Marcantonio}, {Ehrenreich}, {G{\'e}nova Santos}, {Lo Curto}, {Martins}, {Micela}, {Molaro}, {Nunes}, {Palle}, {Pepe}, {Poretti}, {Rebolo}, {Santos}, {Sousa}, {Su{\'a}rez Mascare{\~n}o}, {Tabrenero}, \& {Udry}}]{Lavie2023}
{Lavie}, B., {Bouchy}, F., {Lovis}, C., {et~al.} 2023, \aap, 673, A69

\bibitem[{{Lillo-Box} {et~al.}(2012){Lillo-Box}, {Barrado}, \& {Bouy}}]{LilloBox2012}
{Lillo-Box}, J., {Barrado}, D., \& {Bouy}, H. 2012, \aap, 546, A10

\bibitem[{{Lillo-Box} {et~al.}(2014){Lillo-Box}, {Barrado}, \& {Bouy}}]{LilloBox2014}
{Lillo-Box}, J., {Barrado}, D., \& {Bouy}, H. 2014, \aap, 566, A103

\bibitem[{{Lillo-Box} {et~al.}(2020){Lillo-Box}, {Figueira}, {Leleu}, {Acu{\~n}a}, {Faria}, {Hara}, {Santos}, {Correia}, {Robutel}, {Deleuil}, {Barrado}, {Sousa}, {Bonfils}, {Mousis}, {Almenara}, {Astudillo-Defru}, {Marcq}, {Udry}, {Lovis}, \& {Pepe}}]{LilloBox2020}
{Lillo-Box}, J., {Figueira}, P., {Leleu}, A., {et~al.} 2020, \aap, 642, A121

\bibitem[{{Lovis} {et~al.}(2011){Lovis}, {Dumusque}, {Santos}, {Bouchy}, {Mayor}, {Pepe}, {Queloz}, {S{\'e}gransan}, \& {Udry}}]{Lovis2011}
{Lovis}, C., {Dumusque}, X., {Santos}, N.~C., {et~al.} 2011, arXiv e-prints, arXiv:1107.5325

\bibitem[{{Lubin} {et~al.}(2021){Lubin}, {Robertson}, {Stefansson}, {Ninan}, {Mahadevan}, {Endl}, {Ford}, {Wright}, {Beard}, {Bender}, {Cochran}, {Diddams}, {Fredrick}, {Halverson}, {Kanodia}, {Metcalf}, {Ramsey}, {Roy}, {Schwab}, \& {Terrien}}]{Lubin2021}
{Lubin}, J., {Robertson}, P., {Stefansson}, G., {et~al.} 2021, \aj, 162, 61

\bibitem[{{Matson} {et~al.}(2018){Matson}, {Howell}, {Horch}, \& {Everett}}]{Matson2018}
{Matson}, R.~A., {Howell}, S.~B., {Horch}, E.~P., \& {Everett}, M.~E. 2018, \aj, 156, 31

\bibitem[{{Mayor} {et~al.}(2003){Mayor}, {Pepe}, {Queloz}, {Bouchy}, {Rupprecht}, {Lo Curto}, {Avila}, {Benz}, {Bertaux}, {Bonfils}, {Dall}, {Dekker}, {Delabre}, {Eckert}, {Fleury}, {Gilliotte}, {Gojak}, {Guzman}, {Kohler}, {Lizon}, {Longinotti}, {Lovis}, {Megevand}, {Pasquini}, {Reyes}, {Sivan}, {Sosnowska}, {Soto}, {Udry}, {van Kesteren}, {Weber}, \& {Weilenmann}}]{Mayor2003}
{Mayor}, M., {Pepe}, F., {Queloz}, D., {et~al.} 2003, The Messenger, 114, 20

\bibitem[{{McCully} {et~al.}(2018){McCully}, {Volgenau}, {Harbeck}, {Lister}, {Saunders}, {Turner}, {Siiverd}, \& {Bowman}}]{McCully2018}
{McCully}, C., {Volgenau}, N.~H., {Harbeck}, D.-R., {et~al.} 2018, in Society of Photo-Optical Instrumentation Engineers (SPIE) Conference Series, Vol. 10707, \procspie, 107070K

\bibitem[{{Mignon} {et~al.}(2024){Mignon}, {Delfosse}, {Bonfils}, {Meunier}, {Astudillo-Defru}, {Gaisne}, {Forveille}, {Bouchy}, {Lo Curto}, {Udry}, {Segransan}, {Unger}, {Lovis}, {Santos}, \& {Mayor}}]{Mignon2024}
{Mignon}, L., {Delfosse}, X., {Bonfils}, X., {et~al.} 2024, \aap, 689, A32

\bibitem[{{Morris} {et~al.}(2020){Morris}, {Twicken}, {Smith}, {Clarke}, {Jenkins}, {Bryson}, {Girouard}, \& {Klaus}}]{Morris2020}
{Morris}, R.~L., {Twicken}, J.~D., {Smith}, J.~C., {et~al.} 2020, {Kepler Data Processing Handbook: Photometric Analysis}, Kepler Science Document KSCI-19081-003

\bibitem[{{Noyes} {et~al.}(1984){Noyes}, {Hartmann}, {Baliunas}, {Duncan}, \& {Vaughan}}]{Noyes1984}
{Noyes}, R.~W., {Hartmann}, L.~W., {Baliunas}, S.~L., {Duncan}, D.~K., \& {Vaughan}, A.~H. 1984, \apj, 279, 763

\bibitem[{{Otegi} {et~al.}(2020){Otegi}, {Bouchy}, \& {Helled}}]{Otegi2020}
{Otegi}, J.~F., {Bouchy}, F., \& {Helled}, R. 2020, \aap, 634, A43

\bibitem[{{Parc} {et~al.}(2024){Parc}, {Bouchy}, {Venturini}, {Dorn}, \& {Helled}}]{Parc2024}
{Parc}, L., {Bouchy}, F., {Venturini}, J., {Dorn}, C., \& {Helled}, R. 2024, \aap, 688, A59

\bibitem[{{Pecaut} \& {Mamajek}(2013)}]{Pecaut2013}
{Pecaut}, M.~J. \& {Mamajek}, E.~E. 2013, \apjs, 208, 9

\bibitem[{{Pepe} {et~al.}(2021){Pepe}, {Cristiani}, {Rebolo}, {Santos}, {Dekker}, {Cabral}, {Di Marcantonio}, {Figueira}, {Lo Curto}, {Lovis}, {Mayor}, {M{\'e}gevand}, {Molaro}, {Riva}, {Zapatero Osorio}, {Amate}, {Manescau}, {Pasquini}, {Zerbi}, {Adibekyan}, {Abreu}, {Affolter}, {Alibert}, {Aliverti}, {Allart}, {Allende Prieto}, {{\'A}lvarez}, {Alves}, {Avila}, {Baldini}, {Bandy}, {Barros}, {Benz}, {Bianco}, {Borsa}, {Bourrier}, {Bouchy}, {Broeg}, {Calderone}, {Cirami}, {Coelho}, {Conconi}, {Coretti}, {Cumani}, {Cupani}, {D'Odorico}, {Damasso}, {Deiries}, {Delabre}, {Demangeon}, {Dumusque}, {Ehrenreich}, {Faria}, {Fragoso}, {Genolet}, {Genoni}, {G{\'e}nova Santos}, {Gonz{\'a}lez Hern{\'a}ndez}, {Hughes}, {Iwert}, {Kerber}, {Knudstrup}, {Landoni}, {Lavie}, {Lillo-Box}, {Lizon}, {Maire}, {Martins}, {Mehner}, {Micela}, {Modigliani}, {Monteiro}, {Monteiro}, {Moschetti}, {Murphy}, {Nunes}, {Oggioni}, {Oliveira}, {Oshagh}, {Pall{\'e}}, {Pariani}, {Poretti}, {Rasilla}, {Rebord{\~a}o}, {Redaelli}, {Santana Tschudi},
  {Santin}, {Santos}, {S{\'e}gransan}, {Schmidt}, {Segovia}, {Sosnowska}, {Sozzetti}, {Sousa}, {Span{\`o}}, {Su{\'a}rez Mascare{\~n}o}, {Tabernero}, {Tenegi}, {Udry}, \& {Zanutta}}]{Pepe2021}
{Pepe}, F., {Cristiani}, S., {Rebolo}, R., {et~al.} 2021, \aap, 645, A96

\bibitem[{{Pepe} {et~al.}(2013){Pepe}, {Cristiani}, {Rebolo}, {Santos}, {Dekker}, {M{\'e}gevand}, {Zerbi}, {Cabral}, {Molaro}, {Di Marcantonio}, {Abreu}, {Affolter}, {Aliverti}, {Allende Prieto}, {Amate}, {Avila}, {Baldini}, {Bristow}, {Broeg}, {Cirami}, {Coelho}, {Conconi}, {Coretti}, {Cupani}, {D'Odorico}, {De Caprio}, {Delabre}, {Dorn}, {Figueira}, {Fragoso}, {Galeotta}, {Genolet}, {Gomes}, {Gonz{\'a}lez Hern{\'a}ndez}, {Hughes}, {Iwert}, {Kerber}, {Landoni}, {Lizon}, {Lovis}, {Maire}, {Mannetta}, {Martins}, {Monteiro}, {Oliveira}, {Poretti}, {Rasilla}, {Riva}, {Santana Tschudi}, {Santos}, {Sosnowska}, {Sousa}, {Span{\`o}}, {Tenegi}, {Toso}, {Vanzella}, {Viel}, \& {Zapatero Osorio}}]{Pepe2013}
{Pepe}, F., {Cristiani}, S., {Rebolo}, R., {et~al.} 2013, The Messenger, 153, 6

\bibitem[{{Queloz} {et~al.}(2001){Queloz}, {Henry}, {Sivan}, {Baliunas}, {Beuzit}, {Donahue}, {Mayor}, {Naef}, {Perrier}, \& {Udry}}]{Queloz2001}
{Queloz}, D., {Henry}, G.~W., {Sivan}, J.~P., {et~al.} 2001, \aap, 379, 279

\bibitem[{{Rajpaul} {et~al.}(2015){Rajpaul}, {Aigrain}, {Osborne}, {Reece}, \& {Roberts}}]{Rajpaul2015}
{Rajpaul}, V., {Aigrain}, S., {Osborne}, M.~A., {Reece}, S., \& {Roberts}, S. 2015, \mnras, 452, 2269

\bibitem[{{Ricker} {et~al.}(2015){Ricker}, {Winn}, {Vanderspek}, {Latham}, {Bakos}, {Bean}, {Berta-Thompson}, {Brown}, {Buchhave}, {Butler}, {Butler}, {Chaplin}, {Charbonneau}, {Christensen-Dalsgaard}, {Clampin}, {Deming}, {Doty}, {De Lee}, {Dressing}, {Dunham}, {Endl}, {Fressin}, {Ge}, {Henning}, {Holman}, {Howard}, {Ida}, {Jenkins}, {Jernigan}, {Johnson}, {Kaltenegger}, {Kawai}, {Kjeldsen}, {Laughlin}, {Levine}, {Lin}, {Lissauer}, {MacQueen}, {Marcy}, {McCullough}, {Morton}, {Narita}, {Paegert}, {Palle}, {Pepe}, {Pepper}, {Quirrenbach}, {Rinehart}, {Sasselov}, {Sato}, {Seager}, {Sozzetti}, {Stassun}, {Sullivan}, {Szentgyorgyi}, {Torres}, {Udry}, \& {Villasenor}}]{Ricker2015}
{Ricker}, G.~R., {Winn}, J.~N., {Vanderspek}, R., {et~al.} 2015, Journal of Astronomical Telescopes, Instruments, and Systems, 1, 014003

\bibitem[{{Rubenzahl} {et~al.}(2024){Rubenzahl}, {Dai}, {Howard}, {Lissauer}, {Van Zandt}, {Beard}, {Giacalone}, {Murphy}, {Chontos}, {Lubin}, {Brinkman}, {Tyler}, {MacDougall}, {Rice}, {Dalba}, {Mayo}, {Weiss}, {Polanski}, {Blunt}, {Yee}, {Hill}, {Angelo}, {Turtelboom}, {Holcomb}, {Behmard}, {Pidhorodetska}, {Batalha}, {Crossfield}, {Dressing}, {Fulton}, {Huber}, {Isaacson}, {Kane}, {Petigura}, {Robertson}, {Scarsdale}, {Mocnik}, {Fetherolf}, {Malavolta}, {Mortier}, {Fiorenzano}, \& {Pedani}}]{Rubenzahl2024}
{Rubenzahl}, R.~A., {Dai}, F., {Howard}, A.~W., {et~al.} 2024, \aj, 167, 153

\bibitem[{{Saar} \& {Donahue}(1997)}]{Saar1997}
{Saar}, S.~H. \& {Donahue}, R.~A. 1997, \apj, 485, 319

\bibitem[{{Santos} {et~al.}(2000){Santos}, {Mayor}, {Naef}, {Pepe}, {Queloz}, {Udry}, \& {Blecha}}]{Santos2000}
{Santos}, N.~C., {Mayor}, M., {Naef}, D., {et~al.} 2000, \aap, 361, 265

\bibitem[{{Shappee} {et~al.}(2014){Shappee}, {Prieto}, {Grupe}, {Kochanek}, {Stanek}, {De Rosa}, {Mathur}, {Zu}, {Peterson}, {Pogge}, {Komossa}, {Im}, {Jencson}, {Holoien}, {Basu}, {Beacom}, {Szczygie{\l}}, {Brimacombe}, {Adams}, {Campillay}, {Choi}, {Contreras}, {Dietrich}, {Dubberley}, {Elphick}, {Foale}, {Giustini}, {Gonzalez}, {Hawkins}, {Howell}, {Hsiao}, {Koss}, {Leighly}, {Morrell}, {Mudd}, {Mullins}, {Nugent}, {Parrent}, {Phillips}, {Pojmanski}, {Rosing}, {Ross}, {Sand}, {Terndrup}, {Valenti}, {Walker}, \& {Yoon}}]{Shappee2014}
{Shappee}, B.~J., {Prieto}, J.~L., {Grupe}, D., {et~al.} 2014, \apj, 788, 48

\bibitem[{{Silva} {et~al.}(2022){Silva}, {Faria}, {Santos}, {Sousa}, {Viana}, {Martins}, {Figueira}, {Lovis}, {Pepe}, {Cristiani}, {Rebolo}, {Allart}, {Cabral}, {Mehner}, {Sozzetti}, {Su{\'a}rez Mascare{\~n}o}, {Martins}, {Ehrenreich}, {M{\'e}gevand}, {Palle}, {Lo Curto}, {Tabernero}, {Lillo-Box}, {Gonz{\'a}lez Hern{\'a}ndez}, {Zapatero Osorio}, {Hara}, {Nunes}, {Di Marcantonio}, {Udry}, {Adibekyan}, \& {Dumusque}}]{Silva2022}
{Silva}, A.~M., {Faria}, J.~P., {Santos}, N.~C., {et~al.} 2022, \aap, 663, A143

\bibitem[{{Skrutskie} {et~al.}(2006){Skrutskie}, {Cutri}, {Stiening}, {Weinberg}, {Schneider}, {Carpenter}, {Beichman}, {Capps}, {Chester}, {Elias}, {Huchra}, {Liebert}, {Lonsdale}, {Monet}, {Price}, {Seitzer}, {Jarrett}, {Kirkpatrick}, {Gizis}, {Howard}, {Evans}, {Fowler}, {Fullmer}, {Hurt}, {Light}, {Kopan}, {Marsh}, {McCallon}, {Tam}, {Van Dyk}, \& {Wheelock}}]{2MASS}
{Skrutskie}, M.~F., {Cutri}, R.~M., {Stiening}, R., {et~al.} 2006, \aj, 131, 1163

\bibitem[{{Smith} {et~al.}(2012){Smith}, {Stumpe}, {Van Cleve}, {Jenkins}, {Barclay}, {Fanelli}, {Girouard}, {Kolodziejczak}, {McCauliff}, {Morris}, \& {Twicken}}]{Smith2012}
{Smith}, J.~C., {Stumpe}, M.~C., {Van Cleve}, J.~E., {et~al.} 2012, \pasp, 124, 1000

\bibitem[{{Sousa}(2014)}]{Sousa2014}
{Sousa}, S.~G. 2014, in Determination of Atmospheric Parameters of B, 297--310

\bibitem[{{Speagle}(2020)}]{Speagle2020}
{Speagle}, J.~S. 2020, \mnras, 493, 3132

\bibitem[{{Stassun}(2019)}]{Stassun2019}
{Stassun}, K.~G. 2019, VizieR Online Data Catalog, IV/38

\bibitem[{{Stumpe} {et~al.}(2014){Stumpe}, {Smith}, {Catanzarite}, {Van Cleve}, {Jenkins}, {Twicken}, \& {Girouard}}]{Stumpe2014}
{Stumpe}, M.~C., {Smith}, J.~C., {Catanzarite}, J.~H., {et~al.} 2014, \pasp, 126, 100

\bibitem[{{Stumpe} {et~al.}(2012){Stumpe}, {Smith}, {Van Cleve}, {Twicken}, {Barclay}, {Fanelli}, {Girouard}, {Jenkins}, {Kolodziejczak}, {McCauliff}, \& {Morris}}]{Stumpe2012}
{Stumpe}, M.~C., {Smith}, J.~C., {Van Cleve}, J.~E., {et~al.} 2012, \pasp, 124, 985

\bibitem[{{Sun} {et~al.}(2019){Sun}, {Ioannidis}, {Gu}, {Schmitt}, {Wang}, \& {Kouwenhoven}}]{Sun2019}
{Sun}, L., {Ioannidis}, P., {Gu}, S., {et~al.} 2019, \aap, 624, A15

\bibitem[{{Tabernero} {et~al.}(2022){Tabernero}, {Marfil}, {Montes}, \& {Gonz{\'a}lez Hern{\'a}ndez}}]{Tabernero2022}
{Tabernero}, H.~M., {Marfil}, E., {Montes}, D., \& {Gonz{\'a}lez Hern{\'a}ndez}, J.~I. 2022, \aap, 657, A66

\bibitem[{{Tayar} {et~al.}(2022){Tayar}, {Claytor}, {Huber}, \& {van Saders}}]{Tayar2022}
{Tayar}, J., {Claytor}, Z.~R., {Huber}, D., \& {van Saders}, J. 2022, \apj, 927, 31

\bibitem[{{Tokovinin}(2018)}]{Tokovinin2018}
{Tokovinin}, A. 2018, \pasp, 130, 035002

\bibitem[{{Twicken} {et~al.}(2018){Twicken}, {Catanzarite}, {Clarke}, {Girouard}, {Jenkins}, {Klaus}, {Li}, {McCauliff}, {Seader}, {Tenenbaum}, {Wohler}, {Bryson}, {Burke}, {Caldwell}, {Haas}, {Henze}, \& {Sanderfer}}]{Twicken2018}
{Twicken}, J.~D., {Catanzarite}, J.~H., {Clarke}, B.~D., {et~al.} 2018, \pasp, 130, 064502

\bibitem[{{Twicken} {et~al.}(2010){Twicken}, {Clarke}, {Bryson}, {Tenenbaum}, {Wu}, {Jenkins}, {Girouard}, \& {Klaus}}]{Twicken2010}
{Twicken}, J.~D., {Clarke}, B.~D., {Bryson}, S.~T., {et~al.} 2010, in \procspie, Vol. 7740, Software and Cyberinfrastructure for Astronomy, 774023

\bibitem[{{Vaughan} {et~al.}(1978){Vaughan}, {Preston}, \& {Wilson}}]{Vaughan1978}
{Vaughan}, A.~H., {Preston}, G.~W., \& {Wilson}, O.~C. 1978, \pasp, 90, 267

\bibitem[{{Wenger} {et~al.}(2000){Wenger}, {Ochsenbein}, {Egret}, {Dubois}, {Bonnarel}, {Borde}, {Genova}, {Jasniewicz}, {Lalo{\"e}}, {Lesteven}, \& {Monier}}]{Wenger2000}
{Wenger}, M., {Ochsenbein}, F., {Egret}, D., {et~al.} 2000, \aaps, 143, 9

\bibitem[{{Zechmeister} \& {K{\"u}rster}(2009)}]{Zechmeister2009}
{Zechmeister}, M. \& {K{\"u}rster}, M. 2009, \aap, 496, 577

\bibitem[{{Zeng} {et~al.}(2019){Zeng}, {Jacobsen}, {Sasselov}, {Petaev}, {Vanderburg}, {Lopez-Morales}, {Perez-Mercader}, {Mattsson}, {Li}, {Heising}, {Bonomo}, {Damasso}, {Berger}, {Cao}, {Levi}, \& {Wordsworth}}]{Zeng2019}
{Zeng}, L., {Jacobsen}, S.~B., {Sasselov}, D.~D., {et~al.} 2019, Proceedings of the National Academy of Science, 116, 9723

\bibitem[{{Zhao} {et~al.}(2022){Zhao}, {Fischer}, {Ford}, {Wise}, {Cretignier}, {Aigrain}, {Barragan}, {Bedell}, {Buchhave}, {Camacho}, {Cegla}, {Cisewski-Kehe}, {Collier Cameron}, {de Beurs}, {Dodson-Robinson}, {Dumusque}, {Faria}, {Gilbertson}, {Haley}, {Harrell}, {Hogg}, {Holzer}, {John}, {Klein}, {Lafarga}, {Lienhard}, {Maguire-Rajpaul}, {Mortier}, {Nicholson}, {Palumbo}, {Ramirez Delgado}, {Shallue}, {Vanderburg}, {Viana}, {Zhao}, {Zicher}, {Cabot}, {Henry}, {Roettenbacher}, {Brewer}, {Llama}, {Petersburg}, \& {Szymkowiak}}]{Zhao2022}
{Zhao}, L.~L., {Fischer}, D.~A., {Ford}, E.~B., {et~al.} 2022, \aj, 163, 171

\bibitem[{{Ziegler} {et~al.}(2020){Ziegler}, {Tokovinin}, {Brice{\~n}o}, {Mang}, {Law}, \& {Mann}}]{Ziegler2020}
{Ziegler}, C., {Tokovinin}, A., {Brice{\~n}o}, C., {et~al.} 2020, \aj, 159, 19

\end{thebibliography}

\appendix

\onecolumn

\section{TESS light curves}\label{ap:sap_lightcurves}

In this appendix we show the sector-by sector TESS SAP and PDCSAP light curves. 

\begin{figure*}[htb]
    \centering
    \includegraphics[width=1\textwidth]{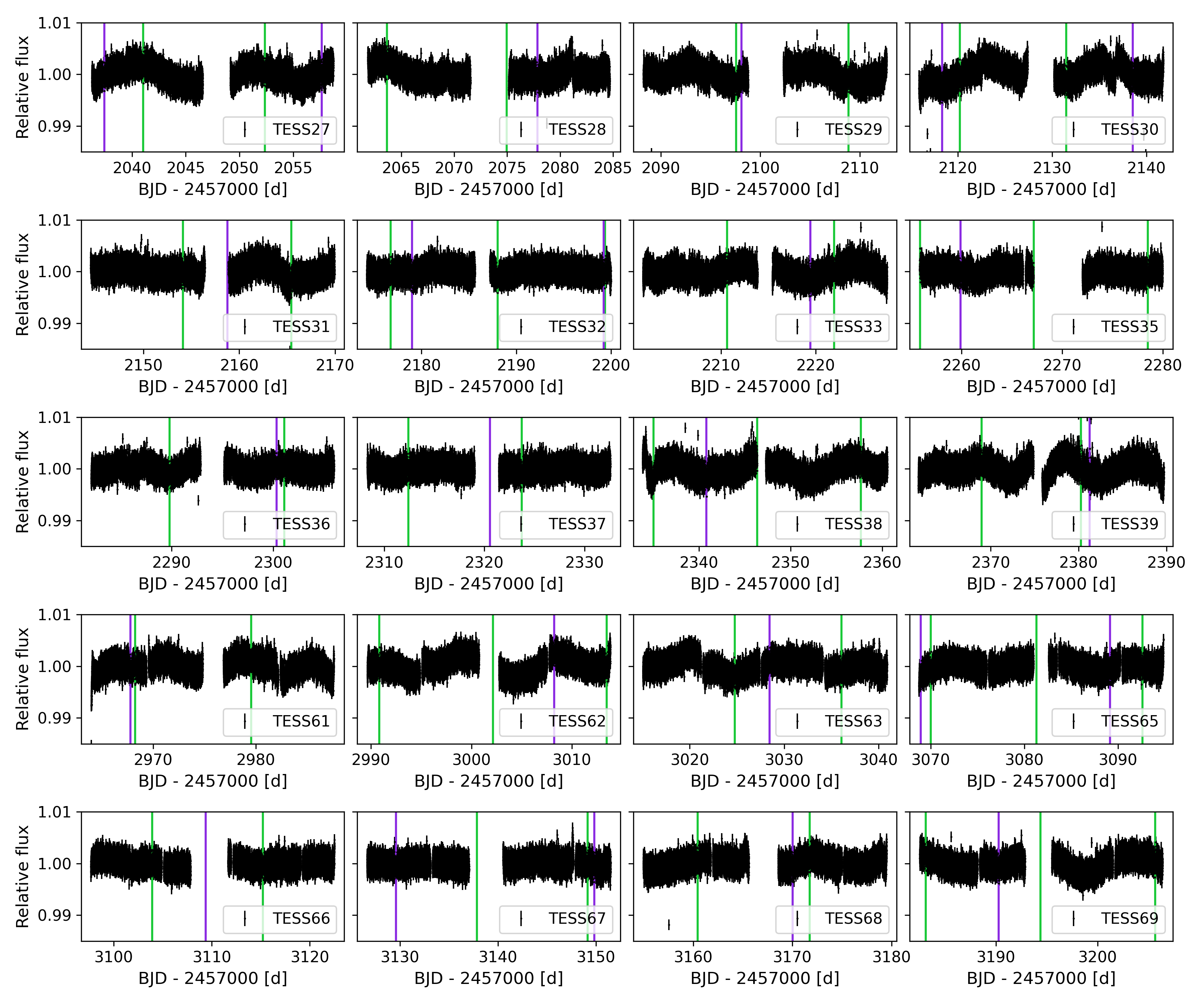}
    \caption{TESS PDCSAP light curves for TOI-2322. The transits of TOI-2322.02 and TOI-2322.01 are marked by vertical green and purple lines respectively. The first three rows show the light curves from the first extended mission, while the last two show the light curves from the second extended mission.}
    \label{fig:pdcsap-lightcurves}
\end{figure*}

\begin{figure*}[htb!]
    \centering
    \includegraphics[width=1\textwidth]{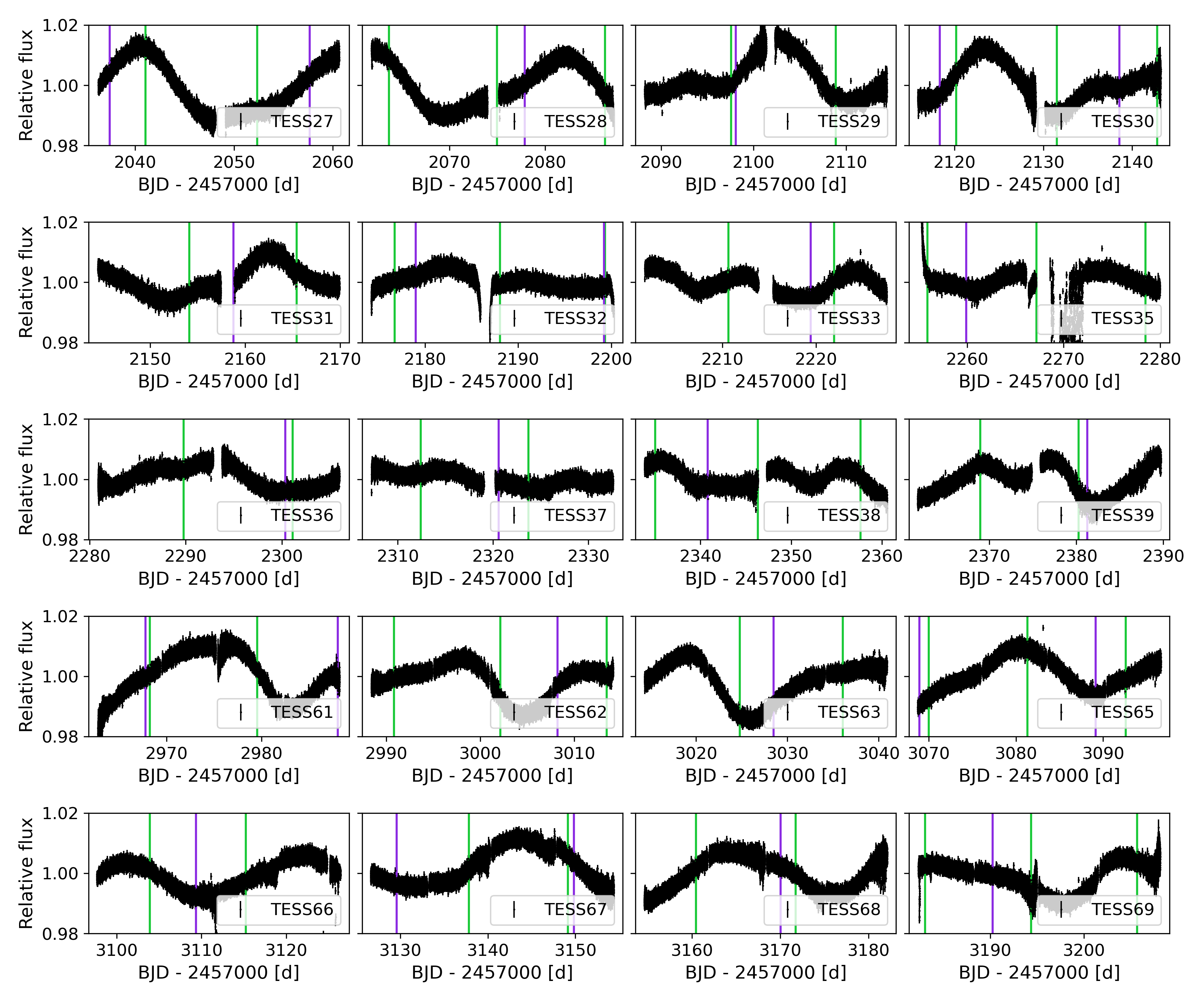}
    \caption{TESS SAP light curves for TOI-2322. The transits of TOI-2322.02 and TOI-2322.01 are marked by vertical green and purple lines respectively. The first three rows show the light curves from the first extended mission, while the last two show the light curves from the second extended mission.}
    \label{fig:sap-lightcurves}
\end{figure*}

\clearpage

\section{TESS pixel file plots}\label{ap:tpfplotter}

\begin{figure*}[hbt]
    \centering
    \includegraphics[width=.24\textwidth]{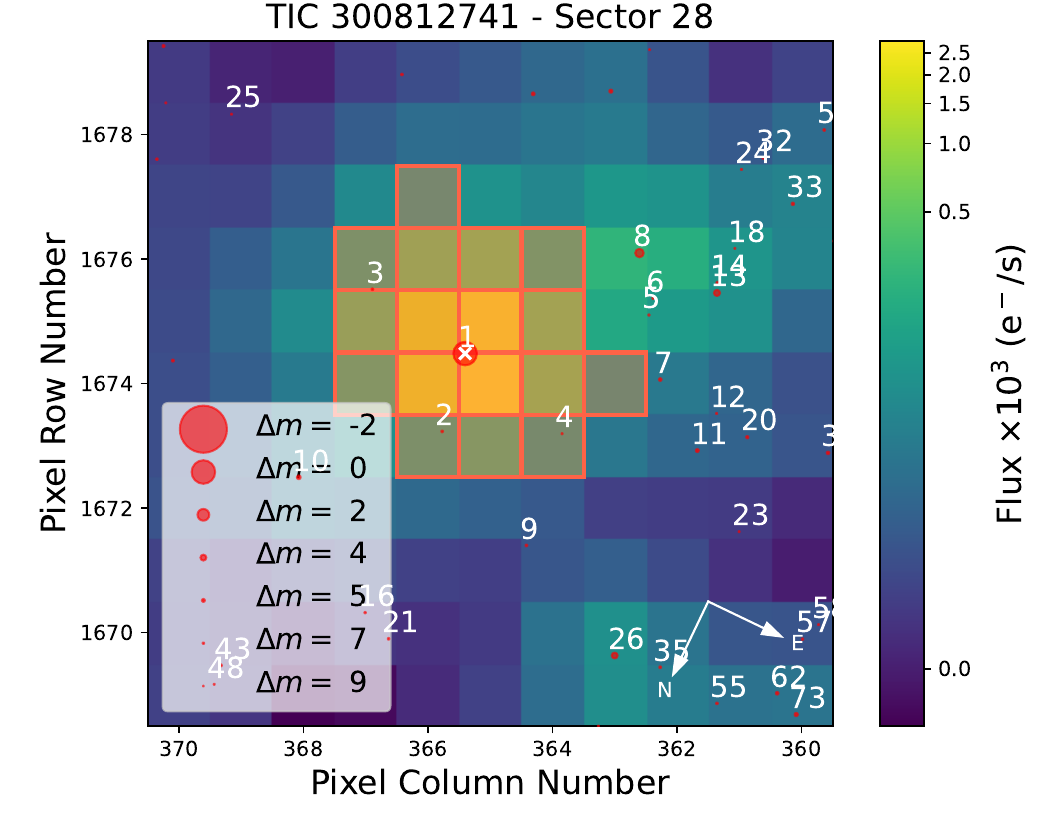}
    \includegraphics[width=.24\textwidth]{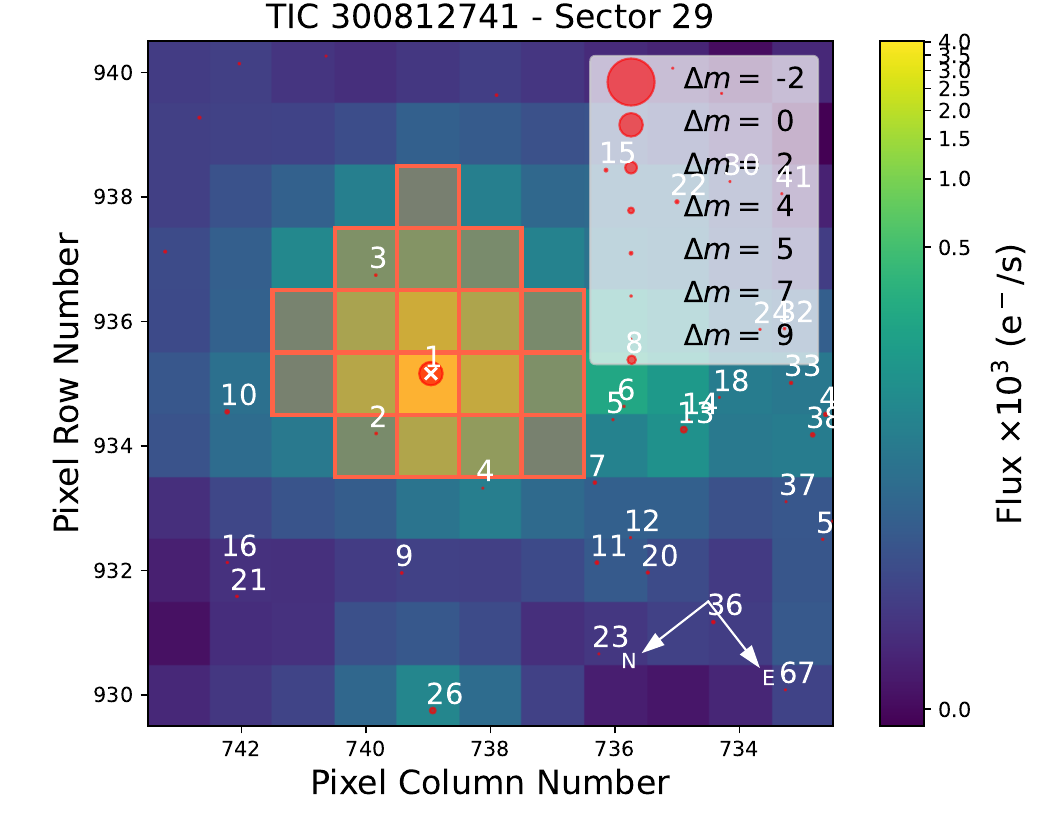}
    \includegraphics[width=.24\textwidth]{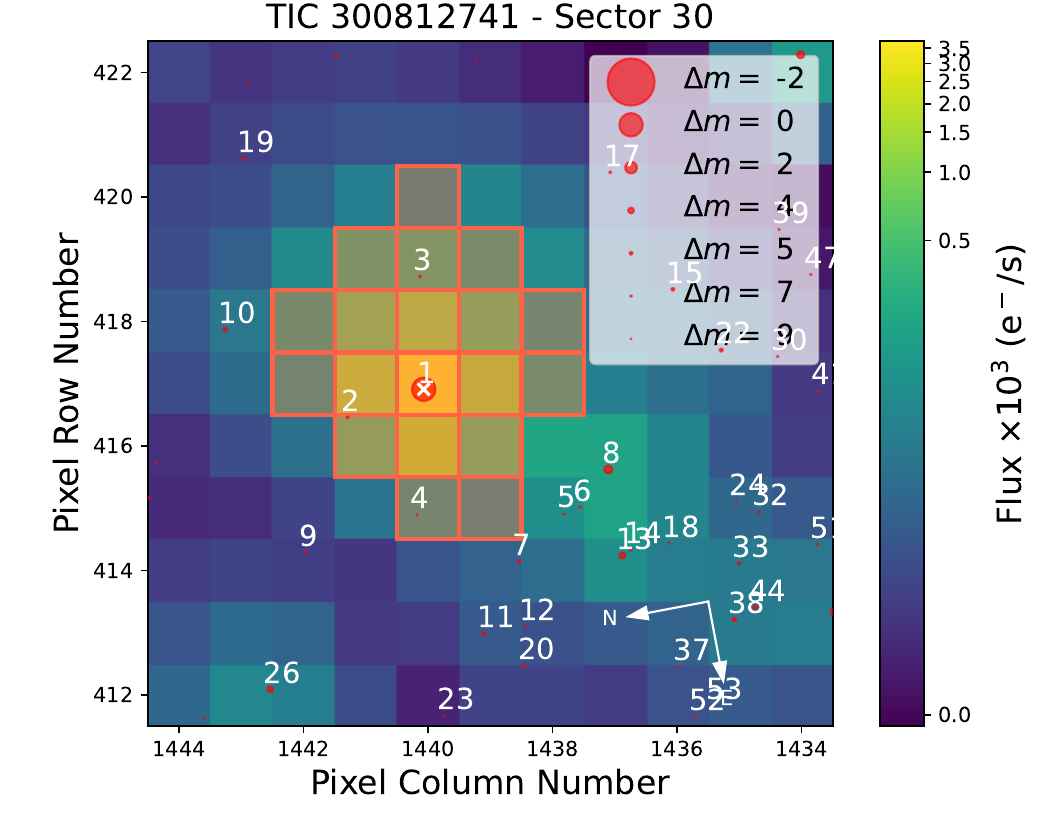}
    \includegraphics[width=.24\textwidth]{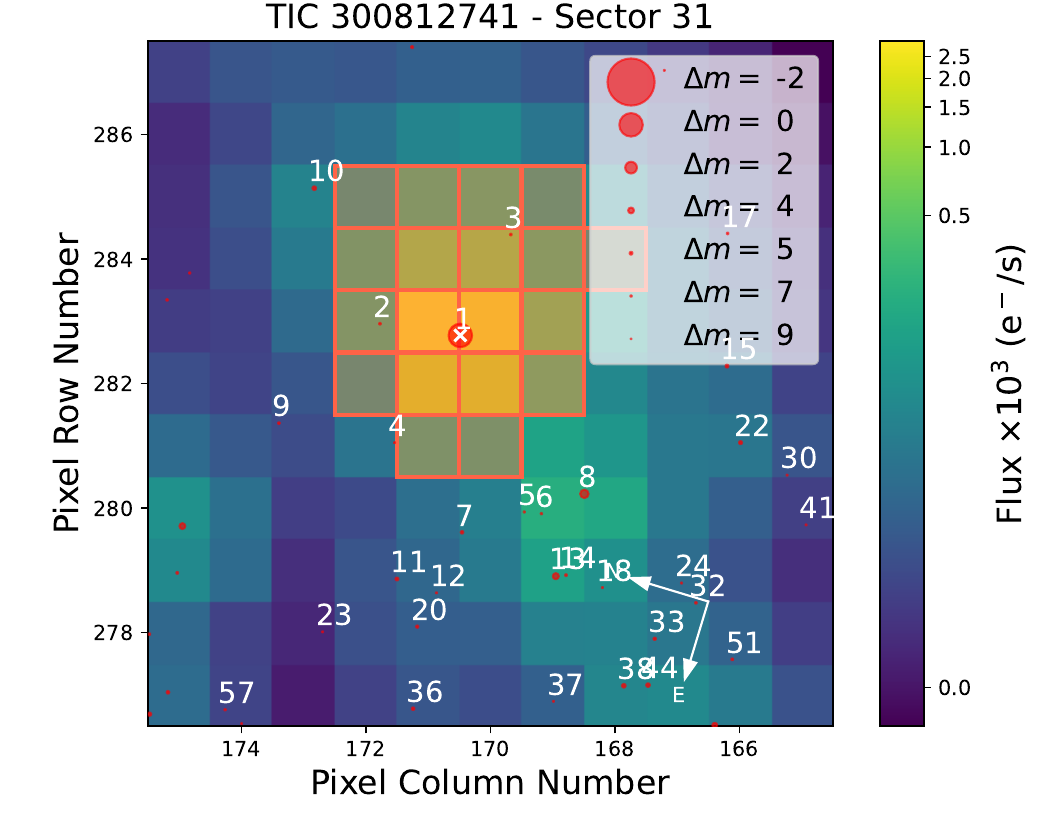}
    \includegraphics[width=.24\textwidth]{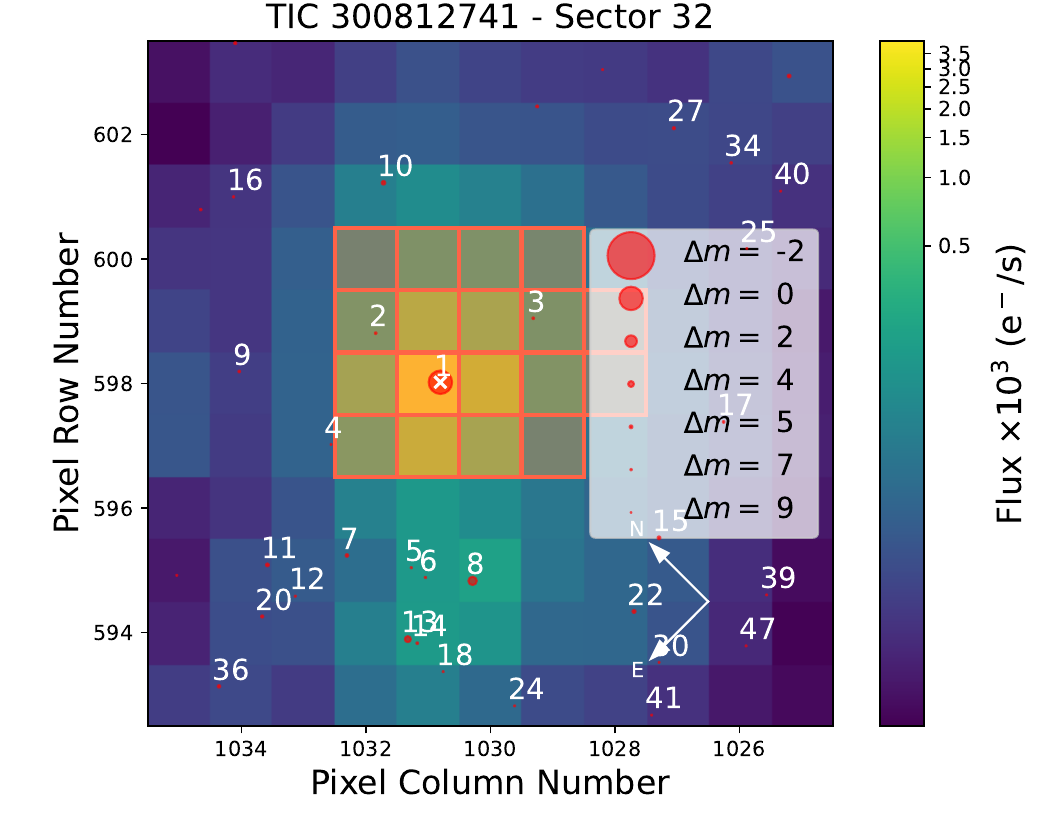}
    \includegraphics[width=.24\textwidth]{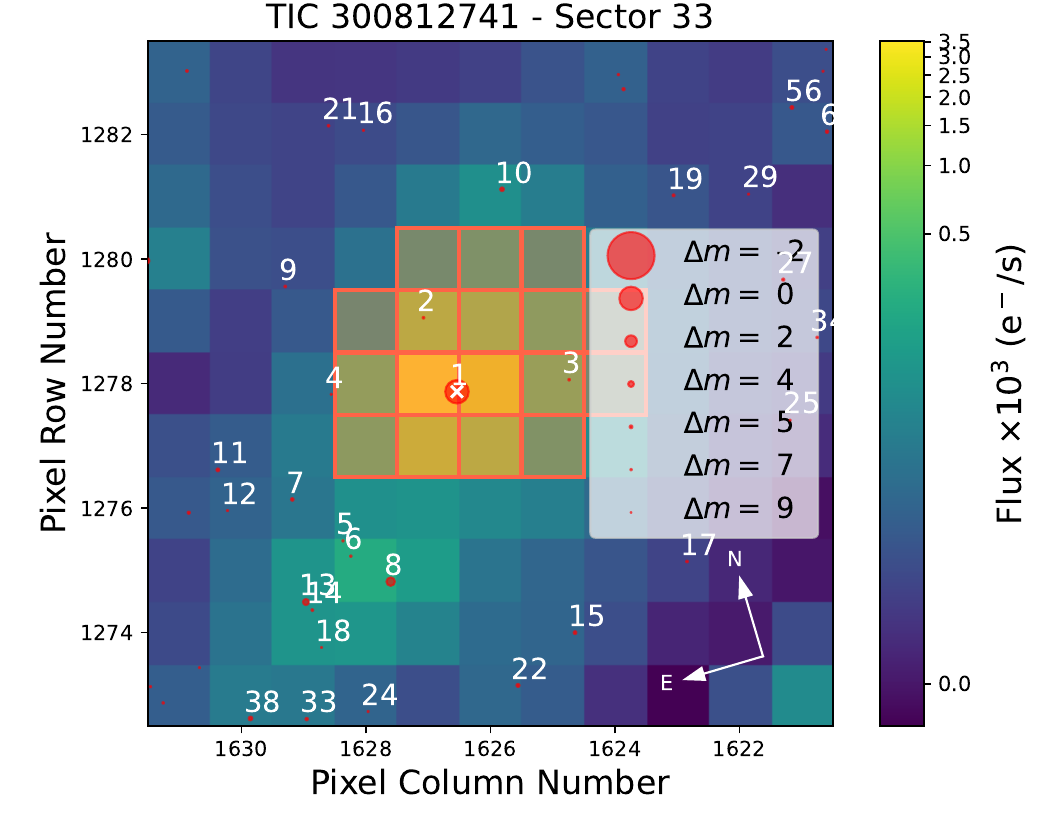}
    \includegraphics[width=.24\textwidth]{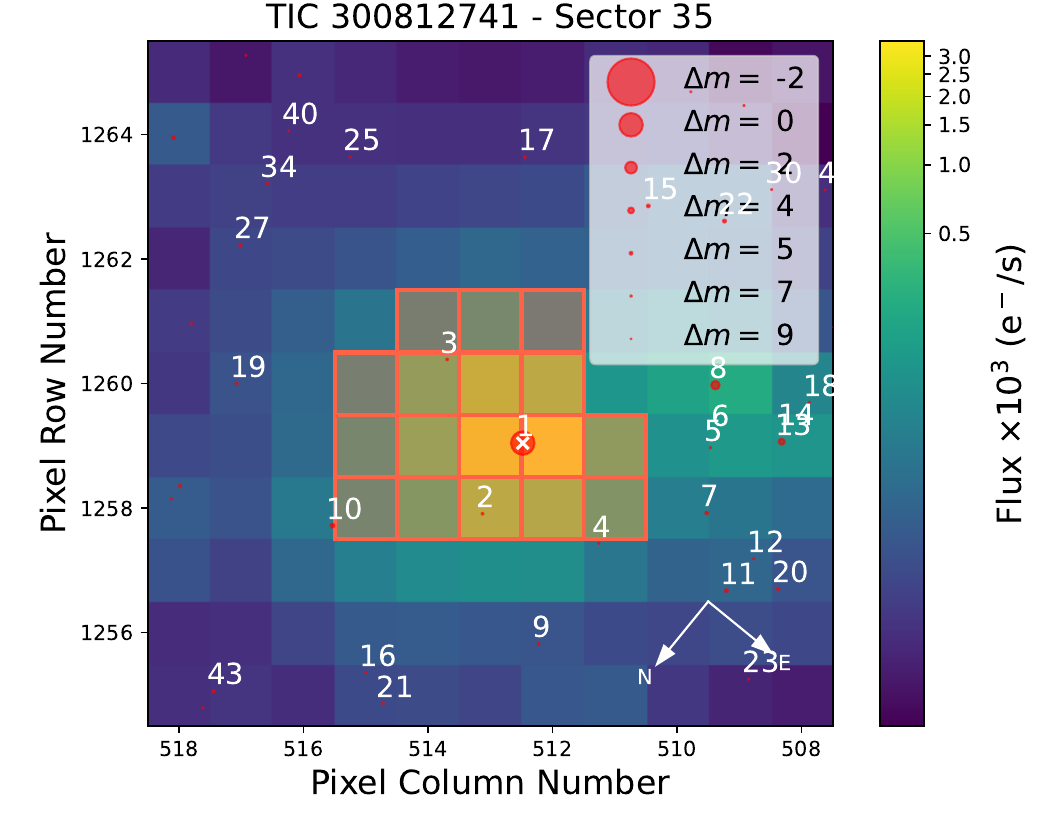}
    \includegraphics[width=.24\textwidth]{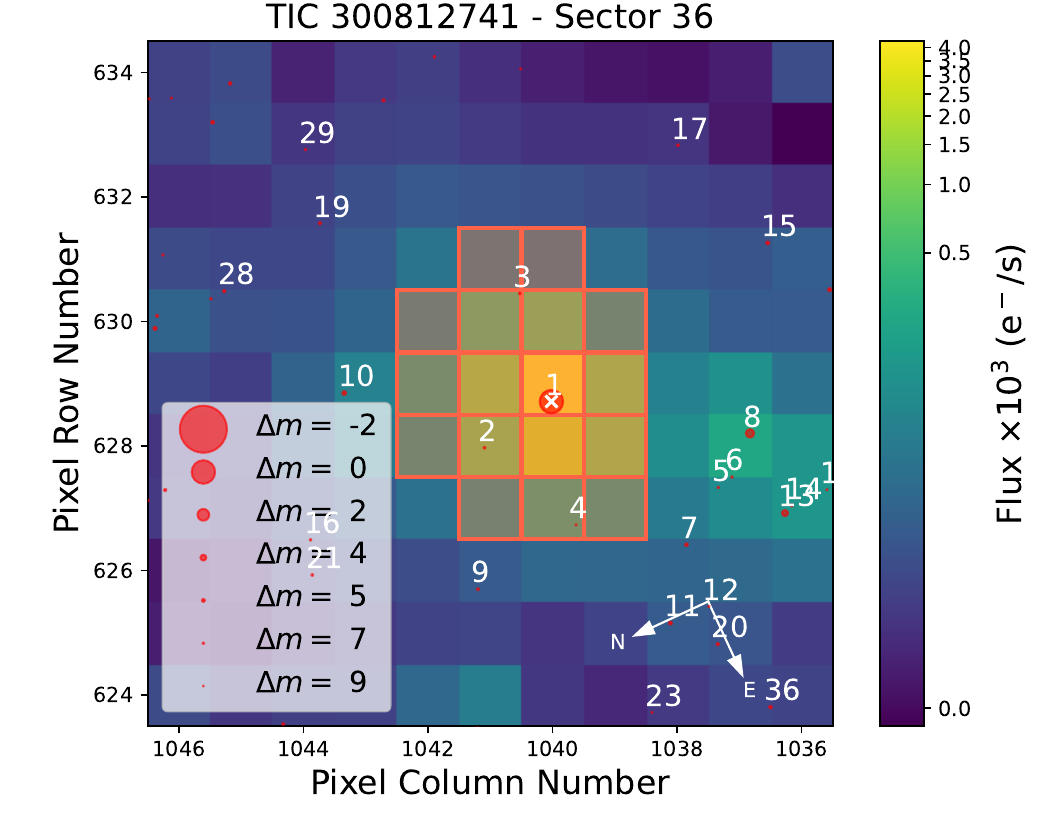}
    \includegraphics[width=.24\textwidth]{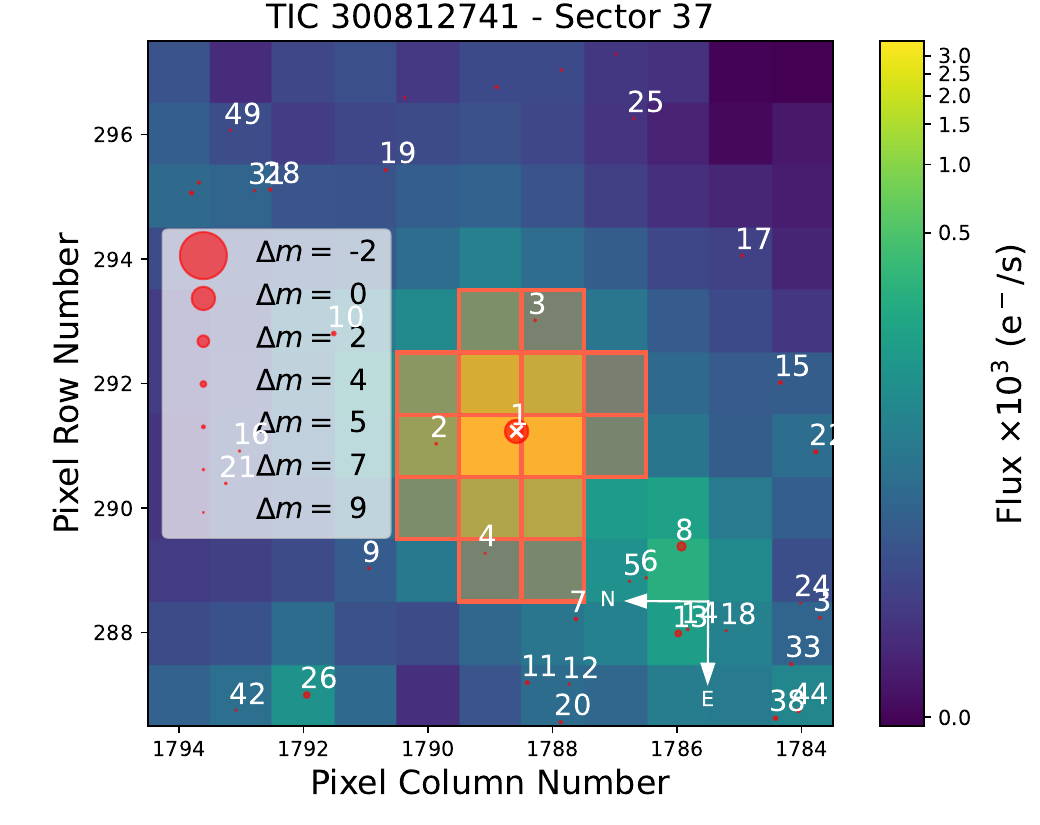}
    \includegraphics[width=.24\textwidth]{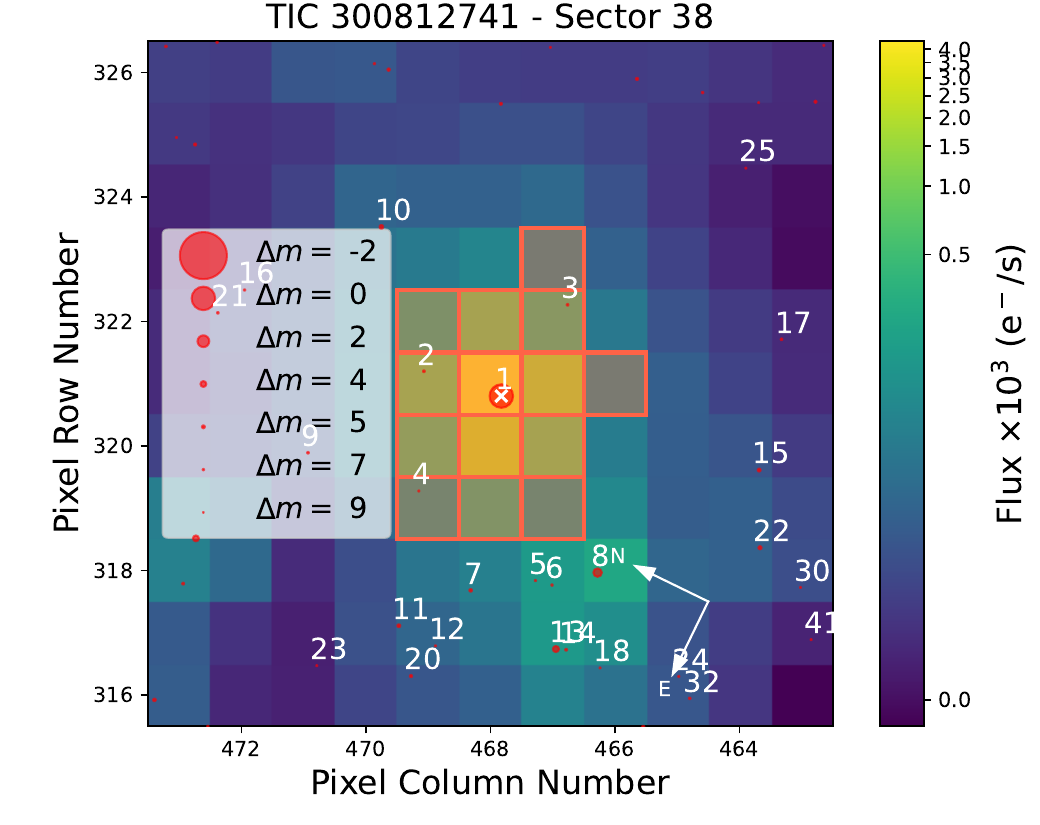}
    \includegraphics[width=.24\textwidth]{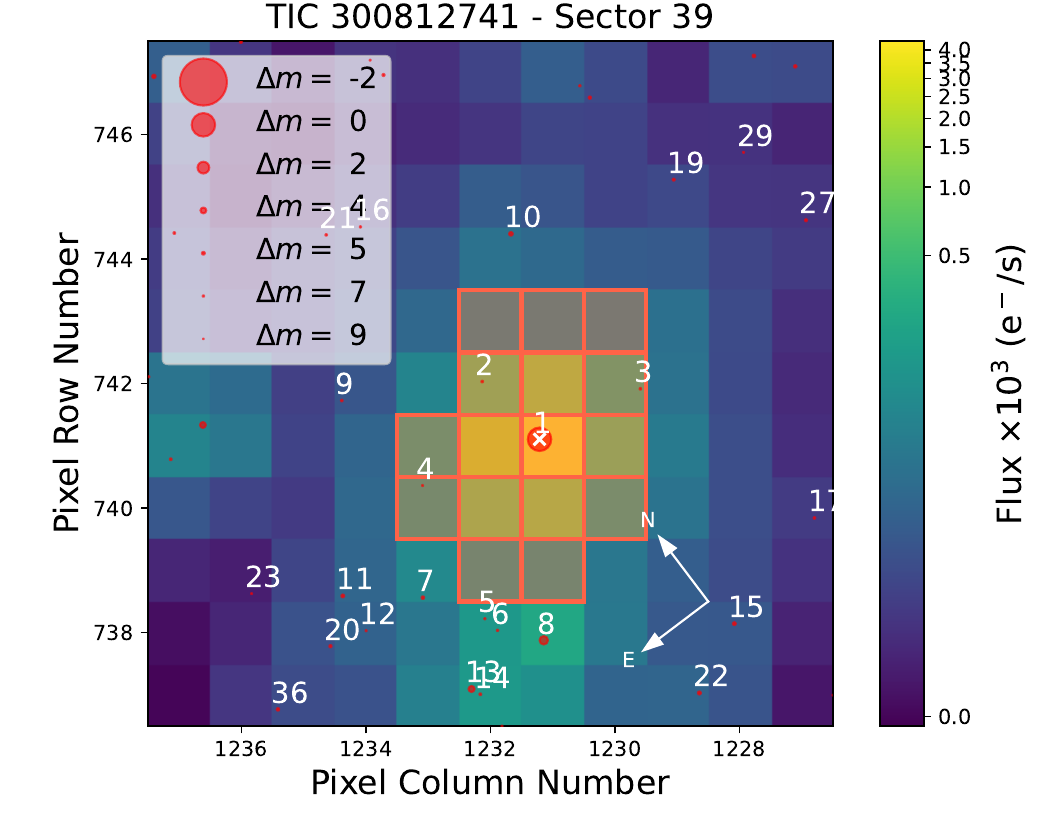}
    \caption{TESS target pixel files for sectors 28-33 and 35-39, observed during the first extended mission. The target star is labelled as 1 and marked by a white cross in each case. All sources from the Gaia DR3 catalogue down to a magnitude contrast of 9 are shown as red circles, with the size proportional to the contrast. The SPOC pipeline aperture is overplotted in shaded red squares.}
    \label{fig:tpfplotter-EM1}
\end{figure*}

\begin{figure*}[hbt]
    \centering
    \includegraphics[width=.24\textwidth]{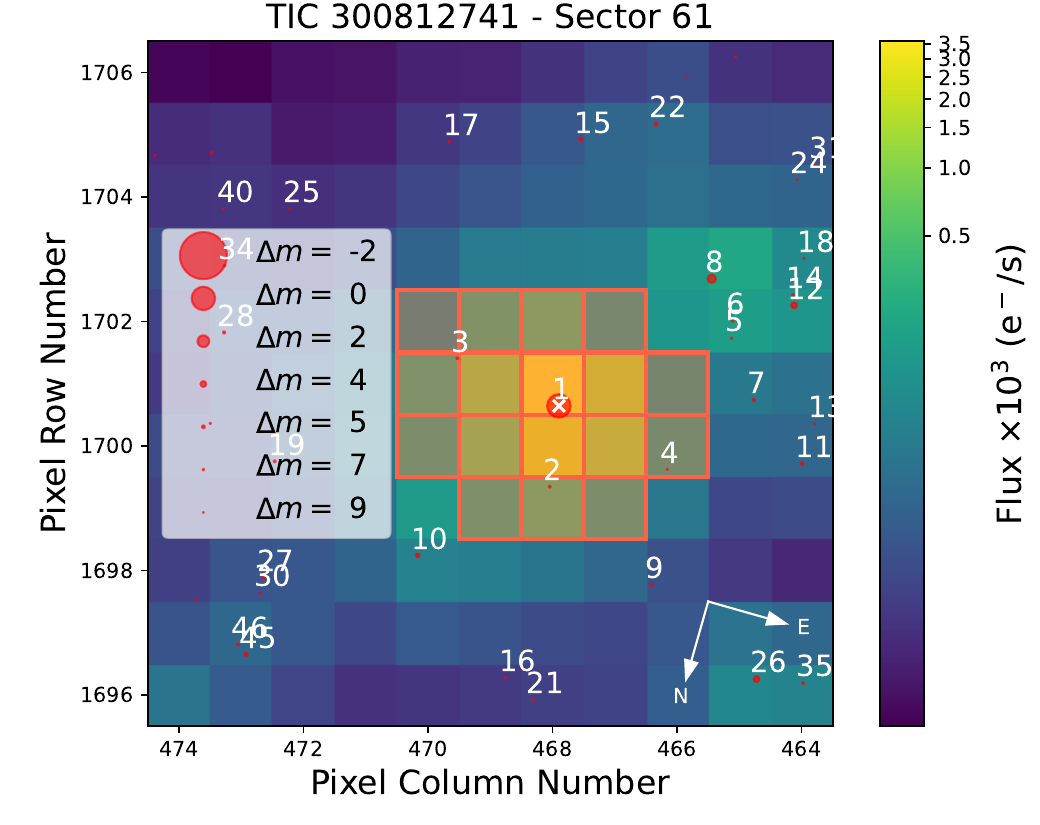}
    \includegraphics[width=.24\textwidth]{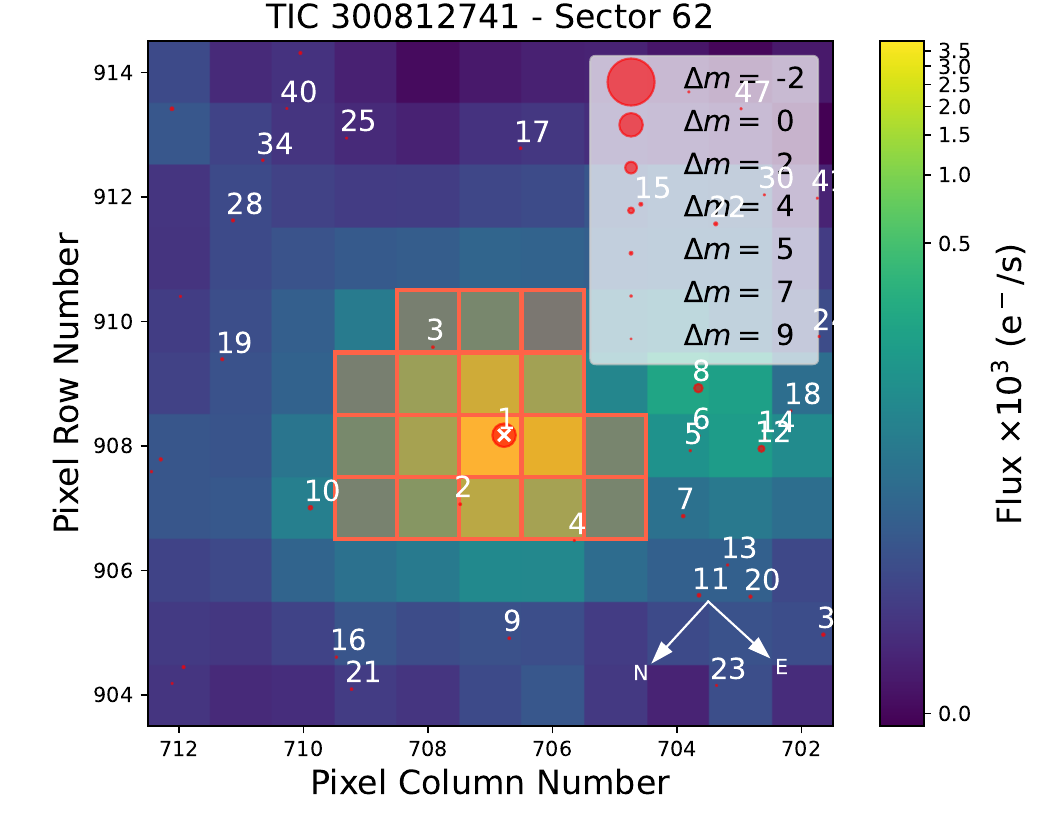}
    \includegraphics[width=.24\textwidth]{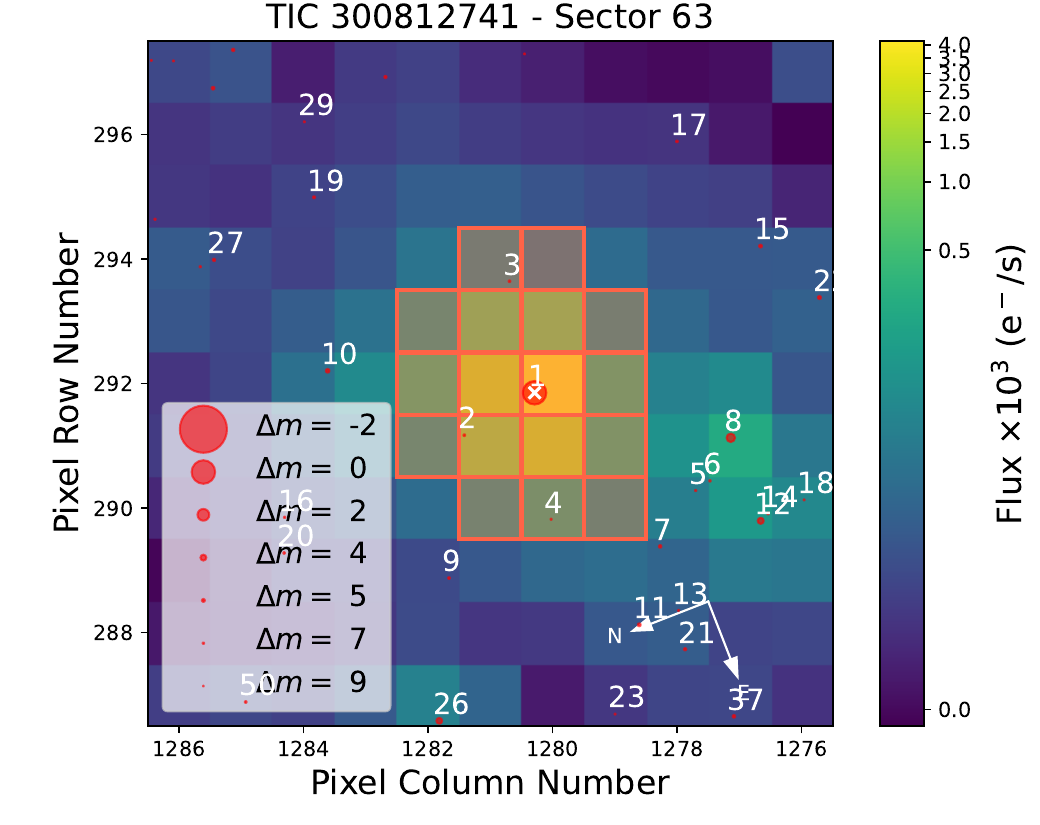}
    \includegraphics[width=.24\textwidth]{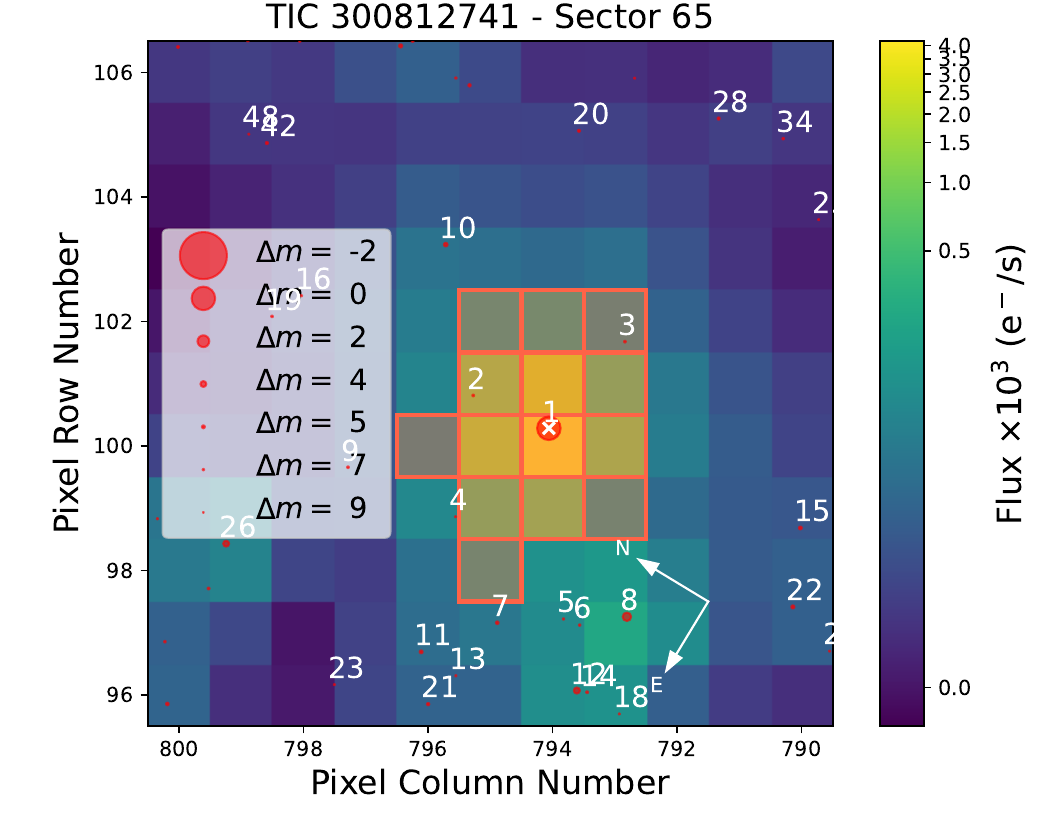}
    \includegraphics[width=.24\textwidth]{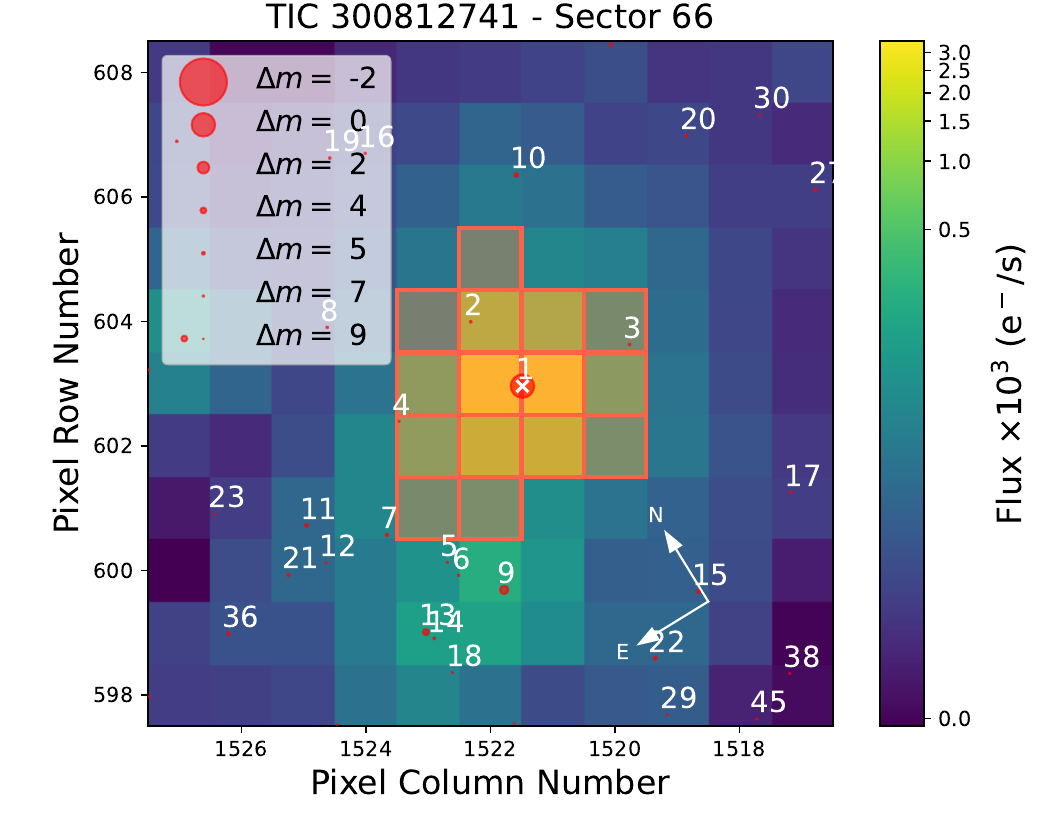}
    \includegraphics[width=.24\textwidth]{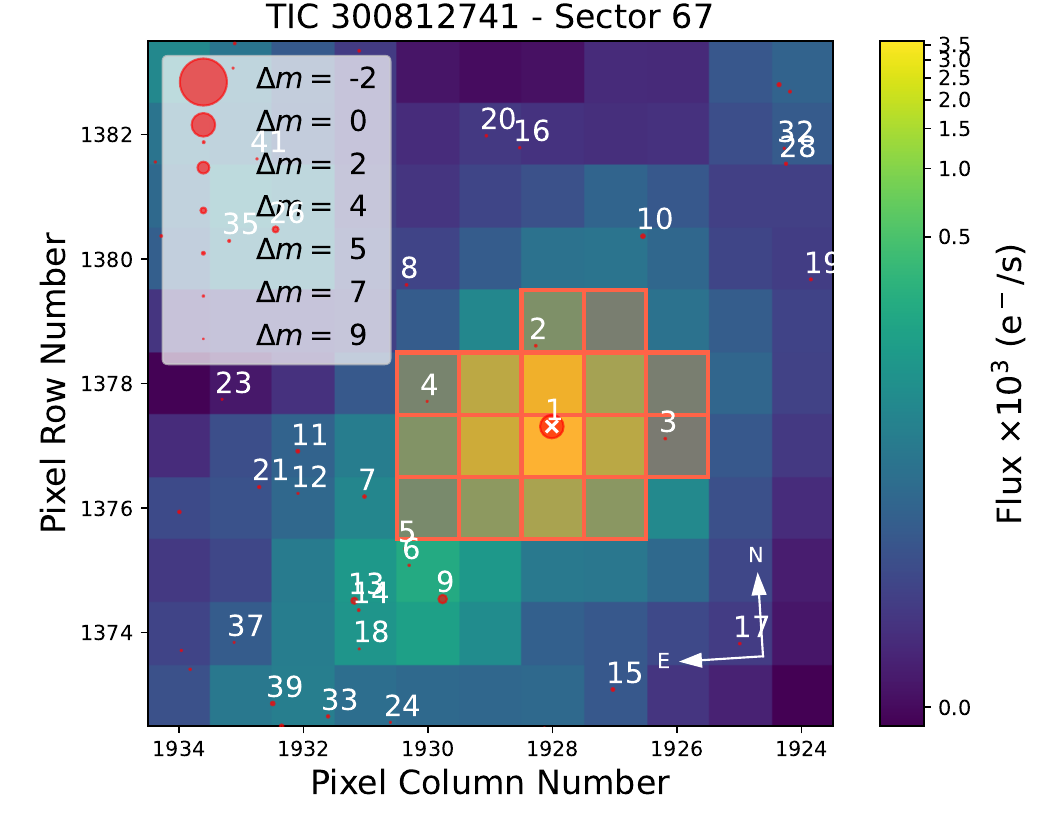}
    \includegraphics[width=.24\textwidth]{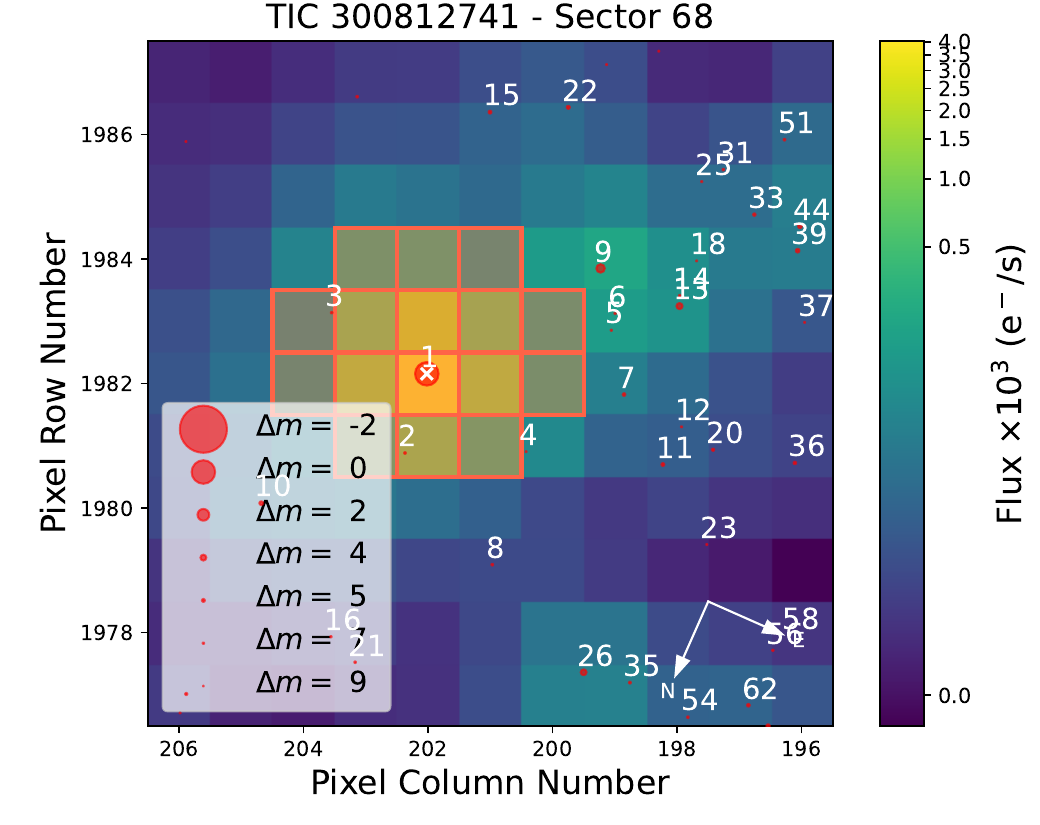}
    \includegraphics[width=.24\textwidth]{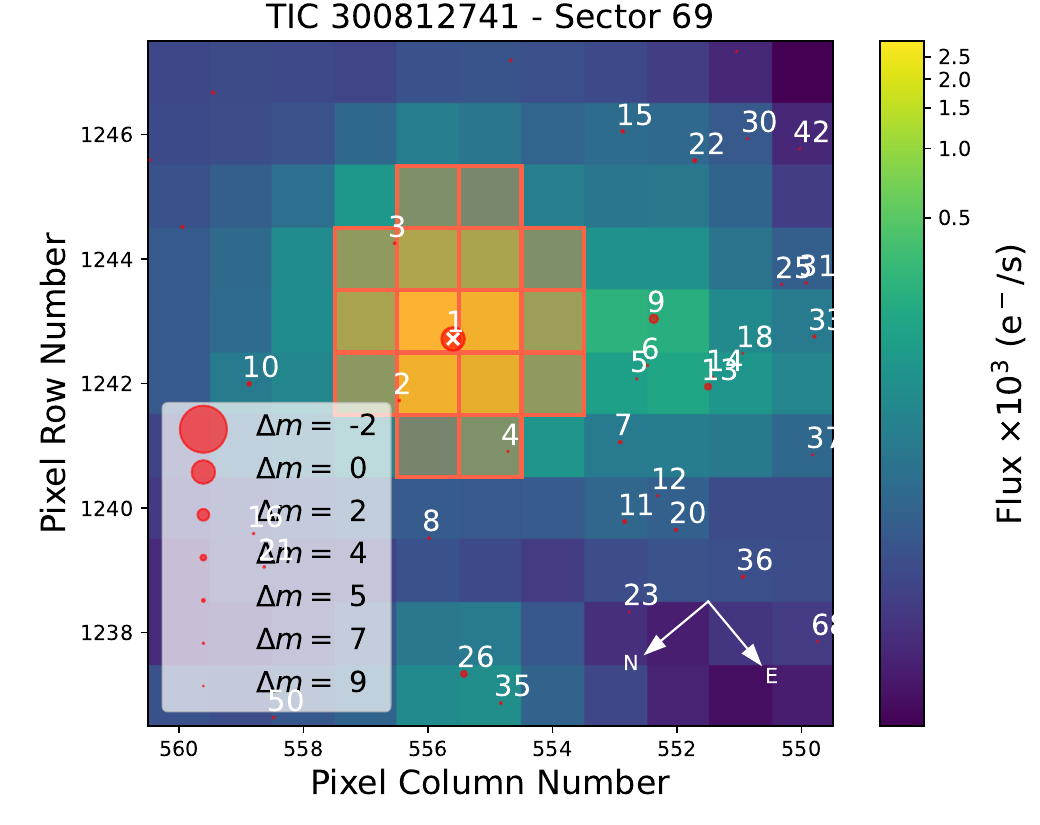}
    \caption{TESS target pixel files for sectors 61-63 and 65-69, observed during the second extended mission. The target star is labelled as 1 and marked by a white cross in each case. All sources from the Gaia DR3 catalogue down to a magnitude contrast of 9 are shown as red circles, with the size proportional to the contrast. The SPOC pipeline aperture is overplotted in shaded red squares.}
    \label{fig:tpfplotter-EM2}
\end{figure*}

In this appendix we show the TESS pixel file plots for all observed sectors save sector 27, which is shown in Figure \ref{fig:tpfplotter}. Those for the first extended mission are shown in Fig. \ref{fig:tpfplotter-EM1}, and those for the second extended mission in Fig. \ref{fig:tpfplotter-EM2}. 

\clearpage

\section{Radial velocities and activity indicators}\label{ap:RVs}

\begin{figure}[ht!]
    \centering
    \includegraphics[width=.43\textwidth]{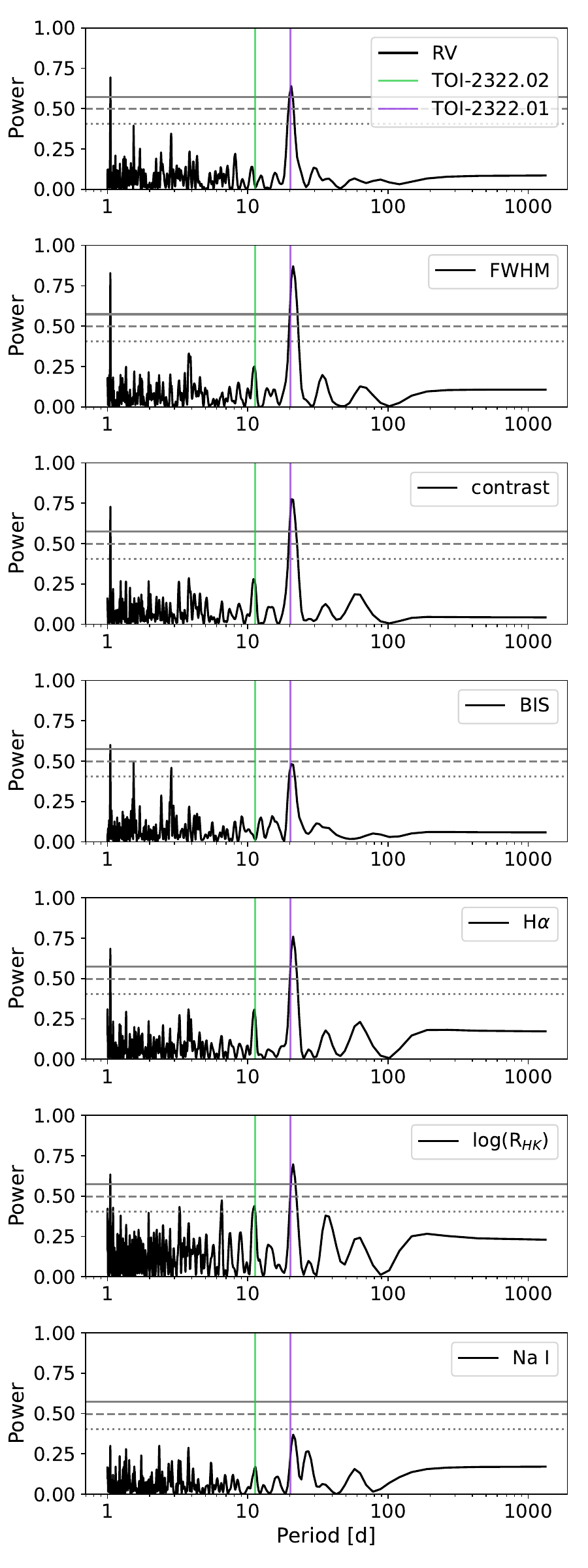}
    \caption{GLS periodograms of the ESPRESSO RVs (top) and activity indicators (second to bottom: CCF FWHM, CCF contrast, CCF bisector, $\mathrm{H_\alpha}$, $\mathrm{\log R'_{HK}}$, and Na I). The vertical green and purple lines indicate the periods of TOI-2322.02 and TOI-2322.01 respectively. The dotted, dashed, and solid horizontal grey lines indicate the 10\%, 1\%, and 0.1\% FAP levels respectively.}
    \label{fig:spectral_periodograms_ESPRESSO}
\end{figure}

\begin{figure}[ht!]
    \centering
    \includegraphics[width=.43\textwidth]{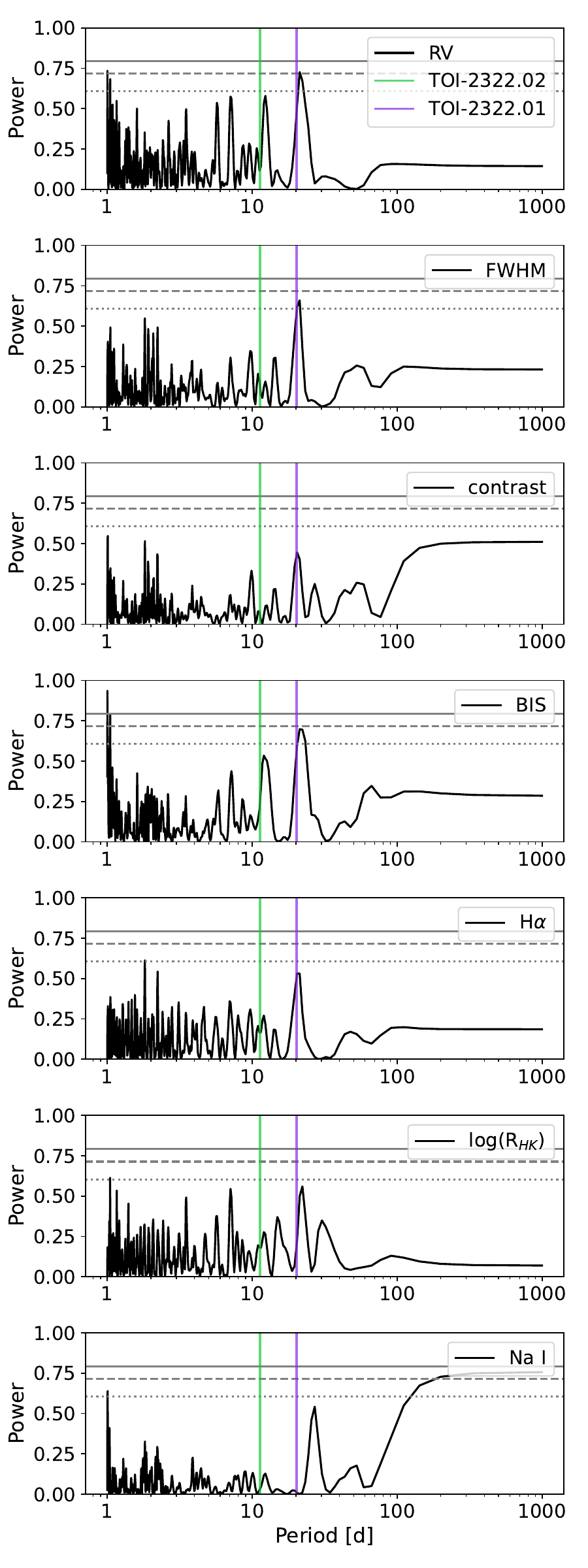}
    \caption{GLS periodograms of the HARPS RVs (top) and activity indicators (second to bottom: CCF FWHM, CCF contrast, CCF bisector, $\mathrm{H_\alpha}$, $\mathrm{\log R'_{HK}}$, and Na I). The vertical green and purple lines indicate the periods of TOI-2322.02 and TOI-2322.01 respectively. The dotted, dashed, and solid horizontal grey lines indicate the 10\%, 1\%, and 0.1\% FAP levels respectively.}
    \label{fig:spectral_periodograms_HARPS}
\end{figure}

In this appendix we present the ESPRESSO RVs and activity indicators (table \ref{tab:TOI-2322_RV_ESPRESSO_data}), and the HARPS RVs and activity indicators (table \ref{tab:TOI-2322_RV_HARPS_data}). Both datasets were obtained with the ESPRESSO DRS v3.2.5. Figures \ref{fig:spectral_periodograms_ESPRESSO} and \ref{fig:spectral_periodograms_HARPS} show the periodograms of the RVs and activity indicators for ESPRESSO and HARPS respectively.

\begin{table*}[pht]
\begin{center}
\caption{RVs and activity indicators obtained from the ESPRESSO spectra for TOI-2322.}
\label{tab:TOI-2322_RV_ESPRESSO_data}
\centering
\resizebox{\textwidth}{!}{%
\begin{tabular}{lllllllll}
\hline  \hline
BJD - 2457000 [d] & RV [m/s] & Bisector & FWHM & Contrast & H$_\alpha$ & $S_{MW}$ & log($R^\prime_{HK}$) & Na~I \\
\hline
2896.7231 & -4456.44 $\pm$ 1.14 & 7.46 $\pm$ 2.27 & 6088.43 $\pm$ 2.27 & 51.9708 $\pm$ 0.0194 & 0.440419 $\pm$ 0.000136 & 0.99331 $\pm$ 0.00212 & -4.53752 $\pm$ 0.00103 & 0.132340 $\pm$ 0.000087 \\ 
2900.6387 & -4484.75 $\pm$ 1.67 & 31.86 $\pm$ 3.34 & 6107.67 $\pm$ 3.34 & 52.0040 $\pm$ 0.0284 & 0.431776 $\pm$ 0.000190 & 0.95959 $\pm$ 0.00492 & -4.55422 $\pm$ 0.00248 & 0.135131 $\pm$ 0.000131 \\ 
2903.6299 & -4471.62 $\pm$ 1.31 & 24.89 $\pm$ 2.63 & 6067.59 $\pm$ 2.63 & 52.1076 $\pm$ 0.0226 & 0.426606 $\pm$ 0.000152 & 0.93151 $\pm$ 0.00311 & -4.56863 $\pm$ 0.00162 & 0.134131 $\pm$ 0.000102 \\ 
2906.6752 & -4494.35 $\pm$ 0.77 & 34.09 $\pm$ 1.54 & 6018.54 $\pm$ 1.54 & 52.3162 $\pm$ 0.0134 & 0.416491 $\pm$ 0.000089 & 0.91000 $\pm$ 0.00116 & -4.58001 $\pm$ 0.00062 & 0.125123 $\pm$ 0.000055 \\ 
2909.6583 & -4485.12 $\pm$ 1.25 & 23.53 $\pm$ 2.50 & 5998.87 $\pm$ 2.50 & 52.5688 $\pm$ 0.0219 & 0.398560 $\pm$ 0.000142 & 0.85464 $\pm$ 0.00320 & -4.61074 $\pm$ 0.00184 & 0.123655 $\pm$ 0.000096 \\ 
2911.8259 & -4490.60 $\pm$ 1.36 & 23.71 $\pm$ 2.73 & 5988.21 $\pm$ 2.73 & 52.6496 $\pm$ 0.0240 & 0.403844 $\pm$ 0.000162 & 0.79809 $\pm$ 0.00294 & -4.64455 $\pm$ 0.00182 & 0.121652 $\pm$ 0.000108 \\ 
2912.8149 & -4489.02 $\pm$ 0.97 & 20.69 $\pm$ 1.94 & 5975.60 $\pm$ 1.94 & 52.6778 $\pm$ 0.0171 & 0.401533 $\pm$ 0.000116 & 0.82588 $\pm$ 0.00160 & -4.62760 $\pm$ 0.00096 & 0.125955 $\pm$ 0.000074 \\ 
2913.6085 & -4480.98 $\pm$ 0.87 & 17.21 $\pm$ 1.73 & 5982.04 $\pm$ 1.73 & 52.5424 $\pm$ 0.0152 & 0.399456 $\pm$ 0.000099 & 0.84621 $\pm$ 0.00152 & -4.61562 $\pm$ 0.00088 & 0.126195 $\pm$ 0.000064 \\ 
2916.6026 & -4457.32 $\pm$ 2.78 & 7.70 $\pm$ 5.57 & 6066.41 $\pm$ 5.57 & 52.4248 $\pm$ 0.0481 & 0.421021 $\pm$ 0.000321 & 0.53181 $\pm$ 0.01632 & -4.85301 $\pm$ 0.01640 & 0.135948 $\pm$ 0.000233 \\ 
2919.6304 & -4480.06 $\pm$ 1.16 & 31.60 $\pm$ 2.32 & 6105.81 $\pm$ 2.32 & 51.6817 $\pm$ 0.0196 & 0.435507 $\pm$ 0.000133 & 0.96773 $\pm$ 0.00232 & -4.55013 $\pm$ 0.00116 & 0.130373 $\pm$ 0.000087 \\ 
2930.6015 & -4485.66 $\pm$ 1.36 & 13.22 $\pm$ 2.71 & 5992.58 $\pm$ 2.71 & 52.6482 $\pm$ 0.0238 & 0.401808 $\pm$ 0.000158 & 0.79527 $\pm$ 0.00350 & -4.64630 $\pm$ 0.00218 & 0.125527 $\pm$ 0.000107 \\ 
2932.6473 & -4491.02 $\pm$ 1.02 & 27.32 $\pm$ 2.05 & 5985.97 $\pm$ 2.05 & 52.6458 $\pm$ 0.0180 & 0.392557 $\pm$ 0.000120 & 0.82950 $\pm$ 0.00201 & -4.62544 $\pm$ 0.00120 & 0.121360 $\pm$ 0.000078 \\ 
2933.7073 & -4487.73 $\pm$ 1.64 & 32.06 $\pm$ 3.27 & 5973.16 $\pm$ 3.27 & 52.9763 $\pm$ 0.0290 & 0.394148 $\pm$ 0.000197 & 0.74847 $\pm$ 0.00453 & -4.67655 $\pm$ 0.00303 & 0.122081 $\pm$ 0.000135 \\ 
2936.5984 & -4468.60 $\pm$ 1.64 & 13.77 $\pm$ 3.28 & 6014.28 $\pm$ 3.28 & 52.6210 $\pm$ 0.0287 & 0.402086 $\pm$ 0.000192 & 0.79876 $\pm$ 0.00485 & -4.64414 $\pm$ 0.00301 & 0.122933 $\pm$ 0.000132 \\ 
2939.6139 & -4461.35 $\pm$ 1.56 & 16.47 $\pm$ 3.12 & 6087.35 $\pm$ 3.12 & 52.0702 $\pm$ 0.0267 & 0.447840 $\pm$ 0.000186 & 0.99645 $\pm$ 0.00393 & -4.53599 $\pm$ 0.00190 & 0.127557 $\pm$ 0.000125 \\ 
2942.6093 & -4487.02 $\pm$ 2.19 & 26.52 $\pm$ 4.39 & 6095.12 $\pm$ 4.39 & 52.2331 $\pm$ 0.0376 & 0.452663 $\pm$ 0.000260 & 0.84398 $\pm$ 0.00746 & -4.61691 $\pm$ 0.00436 & 0.124571 $\pm$ 0.000181 \\ 
2945.6178 & -4483.90 $\pm$ 0.98 & 37.45 $\pm$ 1.96 & 6059.43 $\pm$ 1.96 & 52.1260 $\pm$ 0.0169 & 0.426374 $\pm$ 0.000113 & 0.91886 $\pm$ 0.00186 & -4.57529 $\pm$ 0.00098 & 0.130655 $\pm$ 0.000074 \\ 
2948.5719 & -4497.79 $\pm$ 1.57 & 37.58 $\pm$ 3.14 & 6011.84 $\pm$ 3.14 & 52.4014 $\pm$ 0.0273 & 0.406711 $\pm$ 0.000183 & 0.89375 $\pm$ 0.00435 & -4.58881 $\pm$ 0.00238 & 0.126581 $\pm$ 0.000126 \\ 
2952.5809 & -4485.46 $\pm$ 1.14 & 19.99 $\pm$ 2.27 & 5995.43 $\pm$ 2.27 & 52.4469 $\pm$ 0.0199 & 0.410393 $\pm$ 0.000133 & 0.84549 $\pm$ 0.00241 & -4.61603 $\pm$ 0.00140 & 0.129901 $\pm$ 0.000089 \\ 
2955.7866 & -4487.72 $\pm$ 0.76 & 24.64 $\pm$ 1.53 & 5975.02 $\pm$ 1.53 & 52.6395 $\pm$ 0.0135 & 0.408030 $\pm$ 0.000090 & 0.89232 $\pm$ 0.00112 & -4.58959 $\pm$ 0.00061 & 0.123307 $\pm$ 0.000056 \\ 
2958.6395 & -4464.21 $\pm$ 0.99 & 6.69 $\pm$ 1.99 & 6017.47 $\pm$ 1.99 & 52.4785 $\pm$ 0.0173 & 0.428976 $\pm$ 0.000119 & 0.91478 $\pm$ 0.00175 & -4.57746 $\pm$ 0.00093 & 0.122745 $\pm$ 0.000075 \\ 
2961.5837 & -4466.76 $\pm$ 0.80 & 14.61 $\pm$ 1.60 & 6114.08 $\pm$ 1.60 & 51.7892 $\pm$ 0.0135 & 0.454696 $\pm$ 0.000093 & 1.05018 $\pm$ 0.00120 & -4.51073 $\pm$ 0.00055 & 0.125967 $\pm$ 0.000057 \\ 
2965.5862 & -4487.94 $\pm$ 1.20 & 33.05 $\pm$ 2.40 & 6086.26 $\pm$ 2.40 & 52.0957 $\pm$ 0.0205 & 0.430229 $\pm$ 0.000136 & 0.95897 $\pm$ 0.00275 & -4.55453 $\pm$ 0.00139 & 0.129625 $\pm$ 0.000092 \\ 
2969.5465 & -4494.25 $\pm$ 1.45 & 31.53 $\pm$ 2.90 & 6038.23 $\pm$ 2.90 & 52.4286 $\pm$ 0.0252 & 0.415885 $\pm$ 0.000166 & 0.87615 $\pm$ 0.00400 & -4.59854 $\pm$ 0.00224 & 0.126349 $\pm$ 0.000115 \\ 
2972.5899 & -4494.24 $\pm$ 0.99 & 26.23 $\pm$ 1.98 & 5990.16 $\pm$ 1.98 & 52.6469 $\pm$ 0.0174 & 0.406624 $\pm$ 0.000116 & 0.90916 $\pm$ 0.00179 & -4.58047 $\pm$ 0.00096 & 0.125067 $\pm$ 0.000076 \\ 
2975.7163 & -4483.99 $\pm$ 1.39 & 21.34 $\pm$ 2.78 & 6007.28 $\pm$ 2.78 & 52.6968 $\pm$ 0.0244 & 0.391655 $\pm$ 0.000163 & 0.82401 $\pm$ 0.00344 & -4.62872 $\pm$ 0.00206 & 0.122536 $\pm$ 0.000110 \\ 
2978.6078 & -4472.58 $\pm$ 1.21 & 18.70 $\pm$ 2.43 & 6004.29 $\pm$ 2.43 & 52.5720 $\pm$ 0.0212 & 0.410131 $\pm$ 0.000144 & 0.86974 $\pm$ 0.00250 & -4.60214 $\pm$ 0.00141 & 0.122195 $\pm$ 0.000095 \\ 
2988.6988 & -4490.25 $\pm$ 1.17 & 26.94 $\pm$ 2.34 & 6043.35 $\pm$ 2.34 & 52.2963 $\pm$ 0.0202 & 0.411561 $\pm$ 0.000134 & 0.88566 $\pm$ 0.00249 & -4.59326 $\pm$ 0.00138 & 0.120534 $\pm$ 0.000088 \\ 
2989.5559 & -4487.34 $\pm$ 1.45 & 29.03 $\pm$ 2.91 & 6034.20 $\pm$ 2.91 & 52.5438 $\pm$ 0.0253 & 0.409517 $\pm$ 0.000170 & 0.79282 $\pm$ 0.00377 & -4.64784 $\pm$ 0.00236 & 0.124585 $\pm$ 0.000114 \\ 
2991.5927 & -4485.68 $\pm$ 0.97 & 26.19 $\pm$ 1.94 & 6010.49 $\pm$ 1.94 & 52.4987 $\pm$ 0.0169 & 0.419979 $\pm$ 0.000115 & 0.90649 $\pm$ 0.00171 & -4.58190 $\pm$ 0.00092 & 0.122435 $\pm$ 0.000073 \\ 
3006.5205 & -4490.11 $\pm$ 0.97 & 34.95 $\pm$ 1.94 & 6082.46 $\pm$ 1.94 & 51.9292 $\pm$ 0.0166 & 0.475018 $\pm$ 0.000114 & 1.15287 $\pm$ 0.00173 & -4.46618 $\pm$ 0.00071 & 0.133772 $\pm$ 0.000072 \\ 
3028.6191 & -4494.54 $\pm$ 1.23 & 34.53 $\pm$ 2.45 & 6079.58 $\pm$ 2.45 & 52.2226 $\pm$ 0.0211 & 0.439905 $\pm$ 0.000144 & 0.97865 $\pm$ 0.00298 & -4.54470 $\pm$ 0.00147 & 0.128704 $\pm$ 0.000094 \\ 
3029.5506 & -4495.45 $\pm$ 1.22 & 23.47 $\pm$ 2.45 & 6079.19 $\pm$ 2.45 & 52.2290 $\pm$ 0.0210 & 0.436224 $\pm$ 0.000140 & 0.97221 $\pm$ 0.00294 & -4.54789 $\pm$ 0.00146 & 0.129442 $\pm$ 0.000093 \\ 
\hline
\end{tabular}
}
\end{center}
\end{table*}

\begin{table*}[pht]
\begin{center}
\caption{RVs and activity indicators obtained from the HARPS spectra for TOI-2322.}
\label{tab:TOI-2322_RV_HARPS_data}
\centering
\resizebox{\textwidth}{!}{%
\begin{tabular}{lllllllll}
\hline  \hline
BJD - 2457000 [d] & RV [m/s] & Bisector & FWHM & Contrast & H$_\alpha$ & $S_{MW}$ & log($R^\prime_{HK}$) & Na~I \\
\hline
2860.8335 & -4466.79 $\pm$ 2.81 & 36.85 $\pm$ 5.62 & 6165.86 $\pm$ 5.62 & 50.8388 $\pm$ 0.0463 & 0.441509 $\pm$ 0.000510 & 0.85147 $\pm$ 0.00474 & -4.58581 $\pm$ 0.00244 & 0.158163 $\pm$ 0.000339 \\ 
2862.7714 & -4462.11 $\pm$ 2.72 & 38.47 $\pm$ 5.43 & 6167.05 $\pm$ 5.43 & 50.5155 $\pm$ 0.0445 & 0.440464 $\pm$ 0.000472 & 0.80226 $\pm$ 0.00466 & -4.61187 $\pm$ 0.00254 & 0.156998 $\pm$ 0.000311 \\ 
2864.8016 & -4465.80 $\pm$ 2.93 & 47.55 $\pm$ 5.87 & 6137.64 $\pm$ 5.87 & 50.7243 $\pm$ 0.0485 & 0.440076 $\pm$ 0.000552 & 0.79864 $\pm$ 0.00510 & -4.61385 $\pm$ 0.00280 & 0.165289 $\pm$ 0.000360 \\ 
2869.8431 & -4485.22 $\pm$ 2.29 & 38.44 $\pm$ 4.58 & 6083.69 $\pm$ 4.58 & 51.3918 $\pm$ 0.0387 & 0.410138 $\pm$ 0.000406 & 0.84681 $\pm$ 0.00377 & -4.58821 $\pm$ 0.00195 & 0.157926 $\pm$ 0.000266 \\ 
2871.8360 & -4476.61 $\pm$ 3.01 & 25.75 $\pm$ 6.01 & 6093.36 $\pm$ 6.01 & 51.2554 $\pm$ 0.0506 & 0.440862 $\pm$ 0.000558 & 0.80841 $\pm$ 0.00559 & -4.60853 $\pm$ 0.00303 & 0.162772 $\pm$ 0.000364 \\ 
2873.8396 & -4452.15 $\pm$ 1.38 & 24.94 $\pm$ 2.77 & 6114.67 $\pm$ 2.77 & 51.3776 $\pm$ 0.0233 & 0.424323 $\pm$ 0.000231 & 0.93434 $\pm$ 0.00182 & -4.54517 $\pm$ 0.00085 & 0.142391 $\pm$ 0.000137 \\ 
2885.7305 & -4479.59 $\pm$ 2.09 & 35.91 $\pm$ 4.18 & 6111.33 $\pm$ 4.18 & 51.3151 $\pm$ 0.0351 & 0.417831 $\pm$ 0.000352 & 0.84715 $\pm$ 0.00344 & -4.58804 $\pm$ 0.00178 & 0.146104 $\pm$ 0.000229 \\ 
2887.7843 & -4479.84 $\pm$ 1.58 & 28.75 $\pm$ 3.16 & 6090.28 $\pm$ 3.16 & 51.4143 $\pm$ 0.0267 & 0.418298 $\pm$ 0.000256 & 0.89443 $\pm$ 0.00229 & -4.56427 $\pm$ 0.00112 & 0.143317 $\pm$ 0.000159 \\ 
2902.7992 & -4465.53 $\pm$ 1.33 & 33.18 $\pm$ 2.66 & 6130.45 $\pm$ 2.66 & 51.2167 $\pm$ 0.0222 & 0.435249 $\pm$ 0.000216 & 0.93884 $\pm$ 0.00171 & -4.54307 $\pm$ 0.00080 & 0.138246 $\pm$ 0.000127 \\ 
2904.7922 & -4480.00 $\pm$ 1.54 & 45.54 $\pm$ 3.09 & 6121.50 $\pm$ 3.09 & 51.3117 $\pm$ 0.0259 & 0.426998 $\pm$ 0.000252 & 0.90518 $\pm$ 0.00217 & -4.55904 $\pm$ 0.00105 & 0.138516 $\pm$ 0.000155 \\ 
2906.7984 & -4492.03 $\pm$ 1.53 & 41.46 $\pm$ 3.06 & 6098.06 $\pm$ 3.06 & 51.4801 $\pm$ 0.0259 & 0.406201 $\pm$ 0.000251 & 0.88474 $\pm$ 0.00215 & -4.56903 $\pm$ 0.00106 & 0.135137 $\pm$ 0.000153 \\ 
2928.7615 & -4492.95 $\pm$ 1.29 & 35.46 $\pm$ 2.59 & 6071.76 $\pm$ 2.59 & 51.6583 $\pm$ 0.0220 & 0.403908 $\pm$ 0.000198 & 0.85151 $\pm$ 0.00187 & -4.58579 $\pm$ 0.00096 & 0.134610 $\pm$ 0.000112 \\ 
2932.7581 & -4490.73 $\pm$ 1.91 & 34.61 $\pm$ 3.82 & 6052.59 $\pm$ 3.82 & 51.7615 $\pm$ 0.0327 & 0.396961 $\pm$ 0.000330 & 0.80182 $\pm$ 0.00326 & -4.61211 $\pm$ 0.00178 & 0.136015 $\pm$ 0.000194 \\ 
2936.6596 & -4470.81 $\pm$ 1.40 & 15.40 $\pm$ 2.81 & 6073.45 $\pm$ 2.81 & 51.6830 $\pm$ 0.0239 & 0.418589 $\pm$ 0.000233 & 0.91184 $\pm$ 0.00211 & -4.55583 $\pm$ 0.00101 & 0.137730 $\pm$ 0.000128 \\ 
2937.6710 & -4455.27 $\pm$ 1.38 & 14.75 $\pm$ 2.76 & 6100.71 $\pm$ 2.76 & 51.5397 $\pm$ 0.0233 & 0.419973 $\pm$ 0.000220 & 0.90730 $\pm$ 0.00209 & -4.55802 $\pm$ 0.00101 & 0.136168 $\pm$ 0.000122 \\ 
2955.8094 & -4486.97 $\pm$ 1.82 & 36.56 $\pm$ 3.64 & 6065.62 $\pm$ 3.64 & 51.6588 $\pm$ 0.0310 & 0.412909 $\pm$ 0.000297 & 0.80237 $\pm$ 0.00362 & -4.61181 $\pm$ 0.00198 & 0.139859 $\pm$ 0.000167 \\ 
2956.7686 & -4484.80 $\pm$ 1.64 & 27.84 $\pm$ 3.28 & 6077.40 $\pm$ 3.28 & 51.6854 $\pm$ 0.0279 & 0.404511 $\pm$ 0.000264 & 0.81212 $\pm$ 0.00304 & -4.60652 $\pm$ 0.00164 & 0.136068 $\pm$ 0.000147 \\ 
2958.7585 & -4462.15 $\pm$ 1.83 & 18.49 $\pm$ 3.66 & 6097.62 $\pm$ 3.66 & 51.5486 $\pm$ 0.0310 & 0.452940 $\pm$ 0.000314 & 0.92749 $\pm$ 0.00357 & -4.54839 $\pm$ 0.00168 & 0.141573 $\pm$ 0.000171 \\ 
2960.6718 & -4456.63 $\pm$ 1.74 & 10.57 $\pm$ 3.47 & 6164.63 $\pm$ 3.47 & 51.1586 $\pm$ 0.0288 & 0.440002 $\pm$ 0.000300 & 0.91439 $\pm$ 0.00297 & -4.55461 $\pm$ 0.00142 & 0.141533 $\pm$ 0.000163 \\ 
\hline
\end{tabular}
}
\end{center}
\end{table*}

\twocolumn 

\clearpage
\twocolumn 
\section{\texttt{juliet} fit to the ESPRESSO and TESS data}\label{ap:juliet}

In this Appendix, we present the detailed \texttt{juliet} fit from Section \ref{s:juliet}. \texttt{juliet} enables the joint fitting of RVs and transit through the \texttt{radvel} package \citep{Fulton2018} and the \texttt{batman} package \citep{Kreidberg2015} respectively, and incorporates Gaussian processes (GPs) via the \texttt{celerite} package \citep{Foreman-Mackey2017}. Importance nested sampling, using the \texttt{dynesty} package \citep{Speagle2020}, is employed to explore the parameter space. By default, \texttt{juliet} adopts random walk sampling with 500 live points.

For joint RV and transit fits, \texttt{juliet} takes as parameters the stellar density $\rho$; the period $\mathrm{P_{i}}$, time of mid-transit $\mathrm{t_{0,i}}$, RV semi-amplitude $\mathrm{K_{i}}$, planet-to-star radius ratio $\mathrm{p_{i}}$, impact parameter $\mathrm{b_{i}}$, and either eccentricity $\mathrm{e_{i}}$ and angle of periastron $\mathrm{\omega_{i}}$ or a derived parametrisation such as $\mathrm{\sqrt{e_{i}} \sin \omega_{i}, \sqrt{e_{i}} \cos \omega_{i}}$ for each planet $\mathrm{i}$; the systemic radial velocity $\mathrm{\mu_{instrument}}$ and jitter $\mathrm{\sigma_{w,instrument}}$ for each RV instrument; and the dilution factor $\mathrm{m_{dilution,instrument}}$, flux offset $\mathrm{m_{flux,instrument}}$, jitter $\mathrm{\sigma_{w,instrument}}$, and limb-darkening parameters $\mathrm{q_{1,instrument}}$ and $\mathrm{q_{2,instrument}}$ for each transit instrument. For the limb-darkening parameters, we used the quadratic law parametrisation of \cite{Kipping2013}. We take separate dilution factor, flux offset, and jitter for each TESS sector, but joint limb-darkening parameters.

As noted in Section \ref{s:juliet}, we use GPs to model the stellar activity. For the RVs, we used the quasi-periodic kernel (called exp-sine-squared kernel in \texttt{juliet}) of \cite{Haywood2014}, fitted simultaneously with the planet(s). To constrain the priors of this GP, we first fitted a GP to the FWHM activity indicator time series, which has been shown to be a good tracer of stellar activity in ESPRESSO data \citep{LilloBox2020,Lavie2023,Castro2023}, and which shows the highest power at $\mathrm{P_{rot}\sim 21 \,d}$. The priors and posteriors for this fit are shown in Table \ref{tab:fwhm-gp}. Subsequently, we used the resulting parameters as priors on the radial velocity GP. We use truncated normal priors for $\mathrm{\sigma_{GP,rv}}$, $\mathrm{\alpha_{GP,rv}}$, and $\mathrm{\Gamma_{GP,rv}}$, centred on the FWHM fit posteriors and taking the larger error bar as the width of the normal prior, truncating at 0 to avoid convergence issues. We also apply a scaling factor to the $\mathrm{\sigma_{GP,FWHM}}$ posterior, computed as the ratio of the standard deviation of the two time series, $\mathrm{{\sigma_{RV}}/{\sigma_{FWHM}}}$. For $\mathrm{Prot_{GP,FWHM}}$ the values are better constrained, so we can use a normal prior on $\mathrm{Prot_{GP,rv}}$. 

\begin{table}[pht] 
\begin{center} 
\caption{Prior and posterior planetary parameter distributions obtained with \texttt{juliet} for the FWHM GP model. } 
\label{tab:fwhm-gp} 
\centering 
\begin{tabular}{lll} 
\hline  \hline 
Parameter & Prior\tablefootmark{$\star$} & Posterior \\ 
\hline 
$\mathrm{\mu_{FWHM}}$ \dotfill & $\mathcal{U}(5973.8,6115.3)$ & \fwhmmufwhm \\
$\mathrm{\sigma_{w,FWHM}}$ \dotfill & $\mathcal{J}(0.001,100)$ & \fwhmsigmawfwhm \\
$\mathrm{\sigma_{GP,FWHM}}$ \dotfill & $\mathcal{U}(0.001,1000)$ & \fwhmGPsigmafwhm \\
$\mathrm{\alpha_{GP,FWHM}}$ \dotfill & $\mathcal{U}(0.001,1000)$ & \fwhmGPalphafwhm \\
$\mathrm{\Gamma_{GP,FWHM}}$ \dotfill & $\mathcal{U}(0.001,1000)$ & \fwhmGPGammafwhm \\
$\mathrm{Prot_{GP,FWHM}}$ \dotfill & $\mathcal{N}(20,5)$ & \fwhmGPProtfwhm \\
\hline 
\end{tabular} 
\end{center} 
\tablefoot{\tablefoottext{$\star$}{$\mathcal{U}(a,b)$ indicates a uniform distribution between a and b; $\mathcal{J}(a, b)$ a Jeffreys or log-uniform distribution between a and b; $\mathcal{N}(a,b)$ a normal distribution with mean $a$ and standard deviation $b$.}}
\end{table} 

For the photometry, meanwhile, we apply a two-step process on a sector-by-sector basis to detrend the in-transit data, which is used in the final fit. The full process is described in Section \ref{s:juliet}.

\begin{figure*}[htb!]
    \centering
    \includegraphics[width=.95\textwidth]{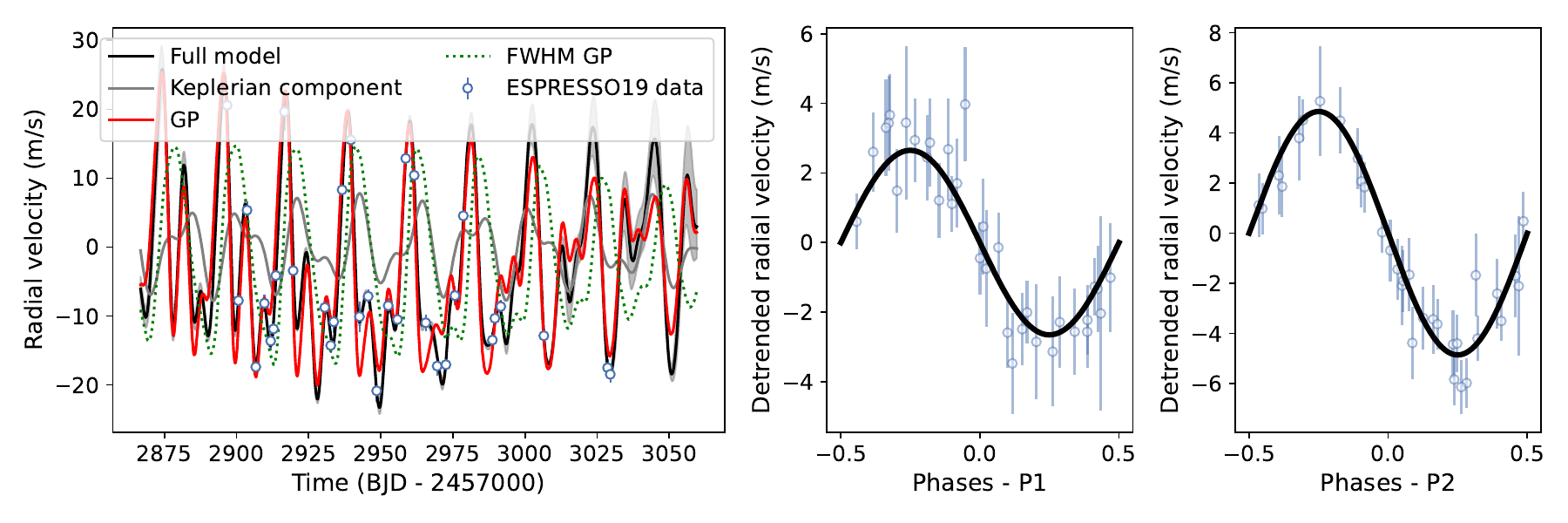}
    \caption{Left: ESPRESSO RVs (blue dots), model components (GP: red, Keplerian: grey), and median model (black) for the \texttt{juliet} $2c$ model. The error bars show the RV errors and jitter added in quadrature. The instrumental systemic velocity has been subtracted. The GP fit to the FWHM is also shown (dotted green line) for comparison. Centre and right: phase-folded RVs and median Keplerian model for the inner candidate planet TOI-2322.02 (centre) and outer candidate planet TOI-2322.01 (right).}
    \label{fig:RVs_2plcirc_juliet}
\end{figure*}

\begin{figure*}[htb!]
    \centering
    \includegraphics[width=.95\textwidth]{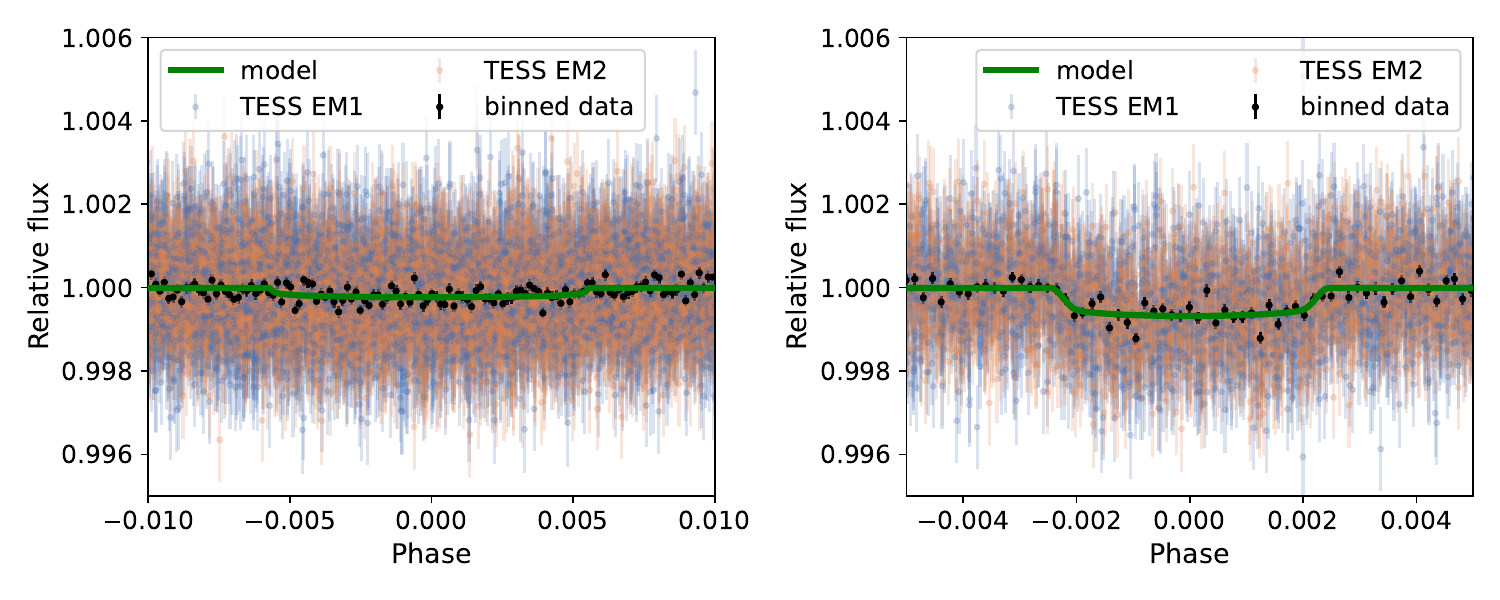}
    \caption{Stacked phase-folded PDCSAP TESS data with the median model (green line), for the inner candidate TOI-2322.02 (left) and outer candidate TOI-2322.01 (right). The points show the data from the TESS first extended mission (blue), second extended mission (orange), and binned data (black).}
    \label{fig:TESS_stacked_2plcirc_juliet}
\end{figure*}

The Bayesian log-evidence comparison strongly favours a two-planet model over either of the single-planet models, with $\Delta \log Z_{2c-in} \approx 119$ and $\Delta \log Z_{2c-out} \approx 21$. The free-eccentricity two-planet model is not significantly favoured over the circular one, with $\Delta \log Z_{2e-2c} \approx 0.2$. We also tested the same four models without GP detrending, finding that the addition of the GP is favoured in all cases, with $\Delta \log Z_{GP-noGP} \approx 19-24$. The priors and posteriors for the favoured $2c$ model are given in Table \ref{tab:TOI-2322_2cqp_priors_posteriors_juliet}. Figure \ref{fig:RVs_2plcirc_juliet} shows the RVs with the fitted model and phase-folded RVs for both planets, and Figure \ref{fig:TESS_stacked_2plcirc_juliet} shows the phase-folded TESS photometry. Since we do not reach a $3\sigma$ measurement of the semi-amplitudes, we compute only upper limits on the planetary masses. 

\begin{table}[pht] 
\begin{center} 
\caption{Prior and posterior planetary parameter distributions obtained with \texttt{juliet} for the $2c$ model. \textit{Top}: Fitted parameters. \textit{Bottom}: derived orbital parameters and physical parameters.} 
\label{tab:TOI-2322_2cqp_priors_posteriors_juliet} 
\centering 
\resizebox{\columnwidth}{!}{%
\begin{tabular}{lll} 
\hline  \hline 
Parameter & Prior & Posterior \\ 
\hline 
$\mathrm{\mu_{ESPRESSO19}}$ \dotfill [$\mathrm{m \, s^{-1}}$] & $\mathcal{U}(-4497.9,-4456.5)$ & \jmuESPRESSOnineteen \\
$\mathrm{\sigma_{w,ESPRESSO19}}$ \dotfill [$\mathrm{m \, s^{-1}}$] & $\mathcal{J}(0.001,10)$ & \jsigmawESPRESSOnineteen \\
$\mathrm{P_{b}}$ \dotfill [d] & $\mathcal{N}(11.307,0.001)$ & \jPpone \\
$\mathrm{t_{0,b}}$ \dotfill [BJD] & $\mathcal{N}(2460160.45,0.01)$ & \jtzeropone \\
$\mathrm{K_{b}}$ \dotfill [$\mathrm{m \, s^{-1}}$] & $\mathcal{U}(0,10)$ & \jKpone \\
$\mathrm{\rho_{}}$ \dotfill [$\mathrm{kg \, m^{-3}}$] & $\mathcal{J}(100,10000)$ & \jrho \\
$\mathrm{p_{b}}$ \dotfill & $\mathcal{U}(0.005,0.05)$ & \jppone \\
$\mathrm{b_{b}}$ \dotfill & $\mathcal{U}(0,1)$ & \jbpone \\
$\mathrm{P_{c}}$ \dotfill [d] & $\mathcal{N}(20.225,0.001)$ & \jPptwo \\
$\mathrm{t_{0,c}}$ \dotfill [BJD] & $\mathcal{N}(2459037.42,0.01)$ & \jtzeroptwo \\
$\mathrm{K_{c}}$ \dotfill [$\mathrm{m \, s^{-1}}$] & $\mathcal{U}(0,10)$ & \jKptwo \\
$\mathrm{p_{c}}$ \dotfill & $\mathcal{U}(0.005,0.05)$ & \jpptwo \\
$\mathrm{b_{c}}$ \dotfill & $\mathcal{U}(0,1)$ & \jbptwo \\
$\mathrm{q_{1,TESS}}$ \dotfill & $\mathcal{U}(0,1)$ & \jqoneTESS \\
$\mathrm{q_{2,TESS}}$ \dotfill & $\mathcal{U}(0,1)$ & \jqtwoTESS \\
$\mathrm{\sigma_{GP,rv}}$ \dotfill & $\mathcal{TN}(11.2,16.2,0,1000)$ & \jGPsigmarv \\
$\mathrm{\alpha_{GP,rv}}$ \dotfill & $\mathcal{TN}(0.00011,0.00065,0,1000)$ & \jGPalpharv \\
$\mathrm{\Gamma_{GP,rv}}$ \dotfill & $\mathcal{TN}(2.9,3.4,0.0,1000.0)$ & \jGPGammarv \\
$\mathrm{Prot_{GP,rv}}$ \dotfill & $\mathcal{N}(21.44,0.41)$ & \jGPProtrv \\
\hline 
$\mathrm{e_{b}}$ \dotfill & $\mathrm{fixed}$ & \jeccpone \\
$\mathrm{\omega_{b}}$ \dotfill & $\mathrm{fixed}$ & \jomegapone \\
$\mathrm{a_{b}}$ \dotfill [au] & $-$ & \toitwothreetwotwoab \\
$\mathrm{i_{b}}$ \dotfill [$\degr$] & $-$ & \toitwothreetwotwoincb \\
$\mathrm{T_{14,b}}$ \dotfill [h] & $-$ & \toitwothreetwotwoTdurb \\
$\mathrm{M_{b}}$ \dotfill [$\mathrm{M_\oplus}$] & $-$ & \toitwothreetwotwomassb  \\
$\mathrm{R_{b}}$ \dotfill [$\mathrm{R_\oplus}$] & $-$ & \toitwothreetwotworadb \\
$\mathrm{\rho_{b}}$ \dotfill [$\mathrm{g \, cm^{-3}}$] & $-$ & \toitwothreetwotworhoplb \\
$\mathrm{T_{eq,b}}$ \dotfill [K] & $-$ & \toitwothreetwotwoteqb \\
$\mathrm{S_{ins,b}}$ \dotfill [$\mathrm{S_\oplus}$] & $-$ & \toitwothreetwotwoSb \\
$\mathrm{e_{c}}$ \dotfill & $\mathrm{fixed}$ & \jeccptwo \\
$\mathrm{\omega_{c}}$ \dotfill & $\mathrm{fixed}$ & \jomegaptwo \\
$\mathrm{a_{c}}$ \dotfill [au] & $-$ & \toitwothreetwotwoac \\
$\mathrm{i_{c}}$ \dotfill [$\degr$] & $-$ & \toitwothreetwotwoincc \\
$\mathrm{T_{14,c}}$ \dotfill [h] & $-$ & \toitwothreetwotwoTdurc \\
$\mathrm{M_{c}}$ \dotfill [$\mathrm{M_\oplus}$] & $-$ & \toitwothreetwotwomassc  \\
$\mathrm{R_{c}}$ \dotfill [$\mathrm{R_\oplus}$] & $-$ & \toitwothreetwotworadc \\
$\mathrm{\rho_{c}}$ \dotfill [$\mathrm{g \, cm{-3}}$] & $-$ & \toitwothreetwotworhoplc \\
$\mathrm{T_{eq,c}}$ \dotfill [K] & $-$ & \toitwothreetwotwoteqc \\
$\mathrm{S_{ins,c}}$ \dotfill [$\mathrm{S_\oplus}$] & $-$ & \toitwothreetwotwoSc \\
$\mathrm{\log Z}$ \dotfill  & $-$ & $75594.1 \pm 0.4$ \\
\hline 
\end{tabular} 
} 
\end{center} 
\end{table} 

\begin{table}[pht] 
\begin{center} 
\caption{Prior and posterior planetary parameter distributions obtained with \texttt{juliet} for the sector-by-sector detrending of the TESS photometry from the first extended mission.} 
\label{tab:TESS_GPS_EM1} 
\centering 
\resizebox{\columnwidth}{!}{%
\begin{tabular}{lll} 
\hline  \hline 
Parameter & Prior & Posterior \\ 
\hline 
$\mathrm{m_{flux,TESS27}}$ \dotfill & $\mathcal{N}(0.0,0.1)$ & \TESSdetrmfluxTESStwentyseven \\
$\mathrm{\sigma_{w,TESS27}}$ \dotfill & $\mathcal{J}(0.1,1000)$ & \TESSdetrsigmawTESStwentyseven \\
$\mathrm{\sigma_{GP,TESS27}}$ \dotfill & $\mathcal{J}(1\times10^{-6},1000000)$ & \TESSdetrGPsigmaTESStwentyseven \\
$\mathrm{\rho_{GP,TESS27}}$ \dotfill & $\mathcal{J}(0.001,1000)$ & \TESSdetrGPrhoTESStwentyseven \\
$\mathrm{m_{flux,TESS28}}$ \dotfill & $\mathcal{N}(0.0,0.1)$ & \TESSdetrmfluxTESStwentyeight \\
$\mathrm{\sigma_{w,TESS28}}$ \dotfill & $\mathcal{J}(0.1,1000)$ & \TESSdetrsigmawTESStwentyeight \\
$\mathrm{\sigma_{GP,TESS28}}$ \dotfill & $\mathcal{J}(1\times10^{-6},1000000)$ & \TESSdetrGPsigmaTESStwentyeight \\
$\mathrm{\rho_{GP,TESS28}}$ \dotfill & $\mathcal{J}(0.001,1000)$ & \TESSdetrGPrhoTESStwentyeight \\
$\mathrm{m_{flux,TESS29}}$ \dotfill & $\mathcal{N}(0.0,0.1)$ & \TESSdetrmfluxTESStwentynine \\
$\mathrm{\sigma_{w,TESS29}}$ \dotfill & $\mathcal{J}(0.1,1000)$ & \TESSdetrsigmawTESStwentynine \\
$\mathrm{\sigma_{GP,TESS29}}$ \dotfill & $\mathcal{J}(1\times10^{-6},1000000)$ & \TESSdetrGPsigmaTESStwentynine \\
$\mathrm{\rho_{GP,TESS29}}$ \dotfill & $\mathcal{J}(0.001,1000)$ & \TESSdetrGPrhoTESStwentynine \\
$\mathrm{m_{flux,TESS30}}$ \dotfill & $\mathcal{N}(0.0,0.1)$ & \TESSdetrmfluxTESSthirty \\
$\mathrm{\sigma_{w,TESS30}}$ \dotfill & $\mathcal{J}(0.1,1000)$ & \TESSdetrsigmawTESSthirty \\
$\mathrm{\sigma_{GP,TESS30}}$ \dotfill & $\mathcal{J}(1\times10^{-6},1000000)$ & \TESSdetrGPsigmaTESSthirty \\
$\mathrm{\rho_{GP,TESS30}}$ \dotfill & $\mathcal{J}(0.001,1000)$ & \TESSdetrGPrhoTESSthirty \\
$\mathrm{m_{flux,TESS31}}$ \dotfill & $\mathcal{N}(0.0,0.1)$ & \TESSdetrmfluxTESSthirtyone \\
$\mathrm{\sigma_{w,TESS31}}$ \dotfill & $\mathcal{J}(0.1,1000)$ & \TESSdetrsigmawTESSthirtyone \\
$\mathrm{\sigma_{GP,TESS31}}$ \dotfill & $\mathcal{J}(1\times10^{-6},1000000)$ & \TESSdetrGPsigmaTESSthirtyone \\
$\mathrm{\rho_{GP,TESS31}}$ \dotfill & $\mathcal{J}(0.001,1000)$ & \TESSdetrGPrhoTESSthirtyone \\
$\mathrm{m_{flux,TESS32}}$ \dotfill & $\mathcal{N}(0.0,0.1)$ & \TESSdetrmfluxTESSthirtytwo \\
$\mathrm{\sigma_{w,TESS32}}$ \dotfill & $\mathcal{J}(0.1,1000)$ & \TESSdetrsigmawTESSthirtytwo \\
$\mathrm{\sigma_{GP,TESS32}}$ \dotfill & $\mathcal{J}(1\times10^{-6},1000000)$ & \TESSdetrGPsigmaTESSthirtytwo \\
$\mathrm{\rho_{GP,TESS32}}$ \dotfill & $\mathcal{J}(0.001,1000)$ & \TESSdetrGPrhoTESSthirtytwo \\
$\mathrm{m_{flux,TESS33}}$ \dotfill & $\mathcal{N}(0.0,0.1)$ & \TESSdetrmfluxTESSthirtythree \\
$\mathrm{\sigma_{w,TESS33}}$ \dotfill & $\mathcal{J}(0.1,1000)$ & \TESSdetrsigmawTESSthirtythree \\
$\mathrm{\sigma_{GP,TESS33}}$ \dotfill & $\mathcal{J}(1\times10^{-6},1000000)$ & \TESSdetrGPsigmaTESSthirtythree \\
$\mathrm{\rho_{GP,TESS33}}$ \dotfill & $\mathcal{J}(0.001,1000)$ & \TESSdetrGPrhoTESSthirtythree \\
$\mathrm{m_{flux,TESS35}}$ \dotfill & $\mathcal{N}(0.0,0.1)$ & \TESSdetrmfluxTESSthirtyfive \\
$\mathrm{\sigma_{w,TESS35}}$ \dotfill & $\mathcal{J}(0.1,1000)$ & \TESSdetrsigmawTESSthirtyfive \\
$\mathrm{\sigma_{GP,TESS35}}$ \dotfill & $\mathcal{J}(1\times10^{-6},1000000)$ & \TESSdetrGPsigmaTESSthirtyfive \\
$\mathrm{\rho_{GP,TESS35}}$ \dotfill & $\mathcal{J}(0.001,1000)$ & \TESSdetrGPrhoTESSthirtyfive \\
$\mathrm{m_{flux,TESS36}}$ \dotfill & $\mathcal{N}(0.0,0.1)$ & \TESSdetrmfluxTESSthirtysix \\
$\mathrm{\sigma_{w,TESS36}}$ \dotfill & $\mathcal{J}(0.1,1000)$ & \TESSdetrsigmawTESSthirtysix \\
$\mathrm{\sigma_{GP,TESS36}}$ \dotfill & $\mathcal{J}(1\times10^{-6},1000000)$ & \TESSdetrGPsigmaTESSthirtysix \\
$\mathrm{\rho_{GP,TESS36}}$ \dotfill & $\mathcal{J}(0.001,1000)$ & \TESSdetrGPrhoTESSthirtysix \\
$\mathrm{m_{flux,TESS37}}$ \dotfill & $\mathcal{N}(0.0,0.1)$ & \TESSdetrmfluxTESSthirtyseven \\
$\mathrm{\sigma_{w,TESS37}}$ \dotfill & $\mathcal{J}(0.1,1000)$ & \TESSdetrsigmawTESSthirtyseven \\
$\mathrm{\sigma_{GP,TESS37}}$ \dotfill & $\mathcal{J}(1\times10^{-6},1000000)$ & \TESSdetrGPsigmaTESSthirtyseven \\
$\mathrm{\rho_{GP,TESS37}}$ \dotfill & $\mathcal{J}(0.001,1000)$ & \TESSdetrGPrhoTESSthirtyseven \\
$\mathrm{m_{flux,TESS38}}$ \dotfill & $\mathcal{N}(0.0,0.1)$ & \TESSdetrmfluxTESSthirtyeight \\
$\mathrm{\sigma_{w,TESS38}}$ \dotfill & $\mathcal{J}(0.1,1000)$ & \TESSdetrsigmawTESSthirtyeight \\
$\mathrm{\sigma_{GP,TESS38}}$ \dotfill & $\mathcal{J}(1\times10^{-6},1000000)$ & \TESSdetrGPsigmaTESSthirtyeight \\
$\mathrm{\rho_{GP,TESS38}}$ \dotfill & $\mathcal{J}(0.001,1000)$ & \TESSdetrGPrhoTESSthirtyeight \\
$\mathrm{m_{flux,TESS39}}$ \dotfill & $\mathcal{N}(0.0,0.1)$ & \TESSdetrmfluxTESSthirtynine \\
$\mathrm{\sigma_{w,TESS39}}$ \dotfill & $\mathcal{J}(0.1,1000)$ & \TESSdetrsigmawTESSthirtynine \\
$\mathrm{\sigma_{GP,TESS39}}$ \dotfill & $\mathcal{J}(1\times10^{-6},1000000)$ & \TESSdetrGPsigmaTESSthirtynine \\
$\mathrm{\rho_{GP,TESS39}}$ \dotfill & $\mathcal{J}(0.001,1000)$ & \TESSdetrGPrhoTESSthirtynine \\
\hline 
\end{tabular} 
} 
\end{center} 
\end{table} 

\begin{table}[pht] 
\begin{center} 
\caption{Prior and posterior planetary parameter distributions obtained with \texttt{juliet} for the sector-by-sector detrending of the TESS photometry from the second extended mission.} 
\label{tab:TESS_GPS_EM2} 
\centering 
\resizebox{\columnwidth}{!}{%
\begin{tabular}{lll} 
\hline  \hline 
Parameter & Prior & Posterior \\ 
\hline 
$\mathrm{m_{flux,TESS61}}$ \dotfill & $\mathcal{N}(0.0,0.1)$ & \TESSdetrmfluxTESSsixtyone \\
$\mathrm{\sigma_{w,TESS61}}$ \dotfill & $\mathcal{J}(0.1,1000)$ & \TESSdetrsigmawTESSsixtyone \\
$\mathrm{\sigma_{GP,TESS61}}$ \dotfill & $\mathcal{J}(1\times10^{-6},1000000)$ & \TESSdetrGPsigmaTESSsixtyone \\
$\mathrm{\rho_{GP,TESS61}}$ \dotfill & $\mathcal{J}(0.001,1000)$ & \TESSdetrGPrhoTESSsixtyone \\
$\mathrm{m_{flux,TESS62}}$ \dotfill & $\mathcal{N}(0.0,0.1)$ & \TESSdetrmfluxTESSsixtytwo \\
$\mathrm{\sigma_{w,TESS62}}$ \dotfill & $\mathcal{J}(0.1,1000)$ & \TESSdetrsigmawTESSsixtytwo \\
$\mathrm{\sigma_{GP,TESS62}}$ \dotfill & $\mathcal{J}(1\times10^{-6},1000000)$ & \TESSdetrGPsigmaTESSsixtytwo \\
$\mathrm{\rho_{GP,TESS62}}$ \dotfill & $\mathcal{J}(0.001,1000)$ & \TESSdetrGPrhoTESSsixtytwo \\
$\mathrm{m_{flux,TESS63}}$ \dotfill & $\mathcal{N}(0.0,0.1)$ & \TESSdetrmfluxTESSsixtythree \\
$\mathrm{\sigma_{w,TESS63}}$ \dotfill & $\mathcal{J}(0.1,1000)$ & \TESSdetrsigmawTESSsixtythree \\
$\mathrm{\sigma_{GP,TESS63}}$ \dotfill & $\mathcal{J}(1\times10^{-6},1000000)$ & \TESSdetrGPsigmaTESSsixtythree \\
$\mathrm{\rho_{GP,TESS63}}$ \dotfill & $\mathcal{J}(0.001,1000)$ & \TESSdetrGPrhoTESSsixtythree \\
$\mathrm{m_{flux,TESS65}}$ \dotfill & $\mathcal{N}(0.0,0.1)$ & \TESSdetrmfluxTESSsixtyfive \\
$\mathrm{\sigma_{w,TESS65}}$ \dotfill & $\mathcal{J}(0.1,1000)$ & \TESSdetrsigmawTESSsixtyfive \\
$\mathrm{\sigma_{GP,TESS65}}$ \dotfill & $\mathcal{J}(1\times10^{-6},1000000)$ & \TESSdetrGPsigmaTESSsixtyfive \\
$\mathrm{\rho_{GP,TESS65}}$ \dotfill & $\mathcal{J}(0.001,1000)$ & \TESSdetrGPrhoTESSsixtyfive \\
$\mathrm{m_{flux,TESS66}}$ \dotfill & $\mathcal{N}(0.0,0.1)$ & \TESSdetrmfluxTESSsixtysix \\
$\mathrm{\sigma_{w,TESS66}}$ \dotfill & $\mathcal{J}(0.1,1000)$ & \TESSdetrsigmawTESSsixtysix \\
$\mathrm{\sigma_{GP,TESS66}}$ \dotfill & $\mathcal{J}(1\times10^{-6},1000000)$ & \TESSdetrGPsigmaTESSsixtysix \\
$\mathrm{\rho_{GP,TESS66}}$ \dotfill & $\mathcal{J}(0.001,1000)$ & \TESSdetrGPrhoTESSsixtysix \\
$\mathrm{m_{flux,TESS67}}$ \dotfill & $\mathcal{N}(0.0,0.1)$ & \TESSdetrmfluxTESSsixtyseven \\
$\mathrm{\sigma_{w,TESS67}}$ \dotfill & $\mathcal{J}(0.1,1000)$ & \TESSdetrsigmawTESSsixtyseven \\
$\mathrm{\sigma_{GP,TESS67}}$ \dotfill & $\mathcal{J}(1\times10^{-6},1000000)$ & \TESSdetrGPsigmaTESSsixtyseven \\
$\mathrm{\rho_{GP,TESS67}}$ \dotfill & $\mathcal{J}(0.001,1000)$ & \TESSdetrGPrhoTESSsixtyseven \\
$\mathrm{m_{flux,TESS68}}$ \dotfill & $\mathcal{N}(0.0,0.1)$ & \TESSdetrmfluxTESSsixtyeight \\
$\mathrm{\sigma_{w,TESS68}}$ \dotfill & $\mathcal{J}(0.1,1000)$ & \TESSdetrsigmawTESSsixtyeight \\
$\mathrm{\sigma_{GP,TESS68}}$ \dotfill & $\mathcal{J}(1\times10^{-6},1000000)$ & \TESSdetrGPsigmaTESSsixtyeight \\
$\mathrm{\rho_{GP,TESS68}}$ \dotfill & $\mathcal{J}(0.001,1000)$ & \TESSdetrGPrhoTESSsixtyeight \\
$\mathrm{m_{flux,TESS69}}$ \dotfill & $\mathcal{N}(0.0,0.1)$ & \TESSdetrmfluxTESSsixtynine \\
$\mathrm{\sigma_{w,TESS69}}$ \dotfill & $\mathcal{J}(0.1,1000)$ & \TESSdetrsigmawTESSsixtynine \\
$\mathrm{\sigma_{GP,TESS69}}$ \dotfill & $\mathcal{J}(1\times10^{-6},1000000)$ & \TESSdetrGPsigmaTESSsixtynine \\
$\mathrm{\rho_{GP,TESS69}}$ \dotfill & $\mathcal{J}(0.001,1000)$ & \TESSdetrGPrhoTESSsixtynine \\
\hline 
\end{tabular} 
} 
\end{center} 
\end{table} 

\clearpage
\onecolumn
\section{S+LEAF kernel models}\label{ap:s+leaf}

In this Appendix, we show the fitted models using the four S+LEAF kernels described in \ref{s:kernels}.

\begin{figure*}[htb]
    \centering
    \includegraphics[width=0.45\linewidth]{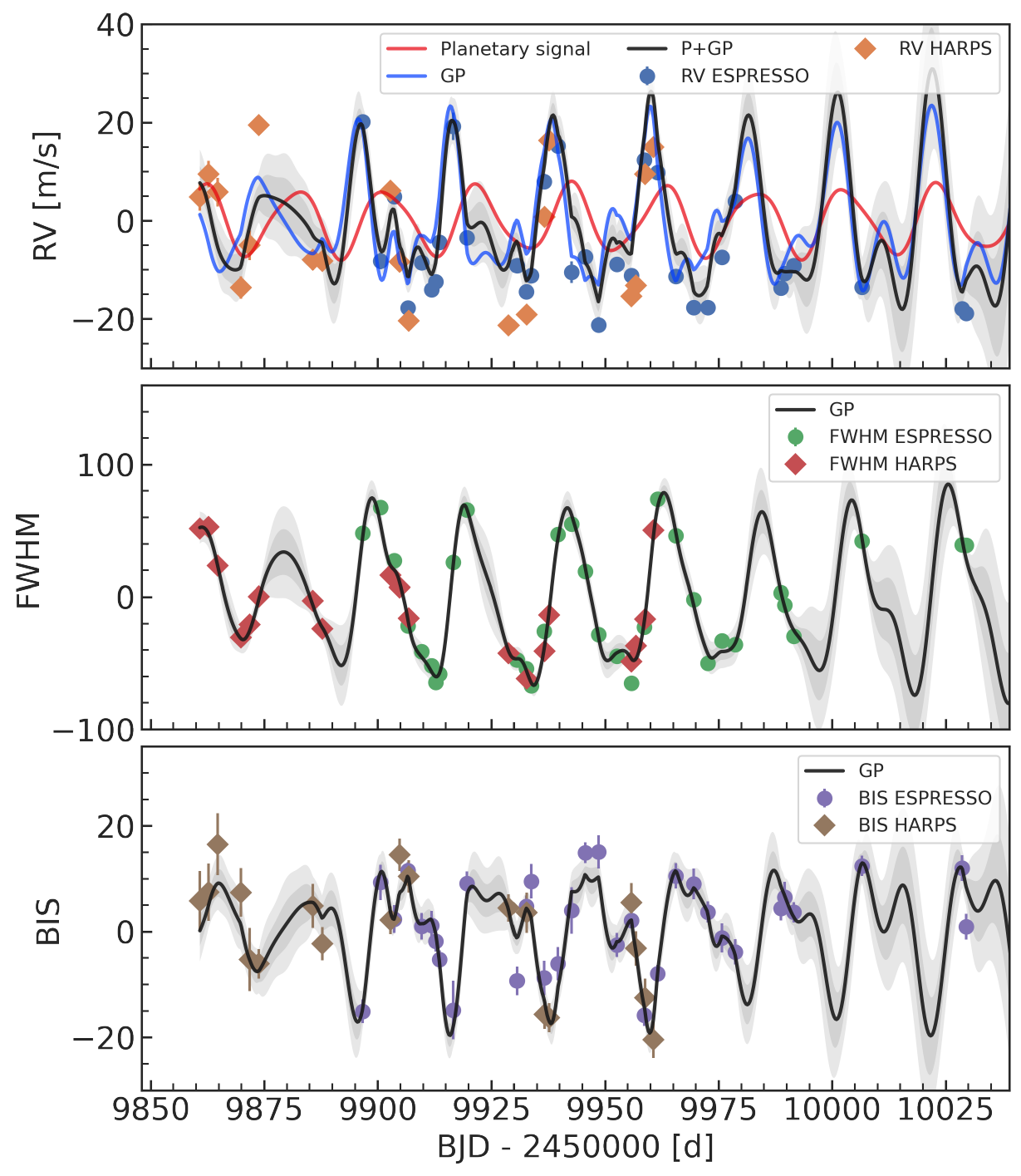}
    \includegraphics[width=0.45\linewidth]{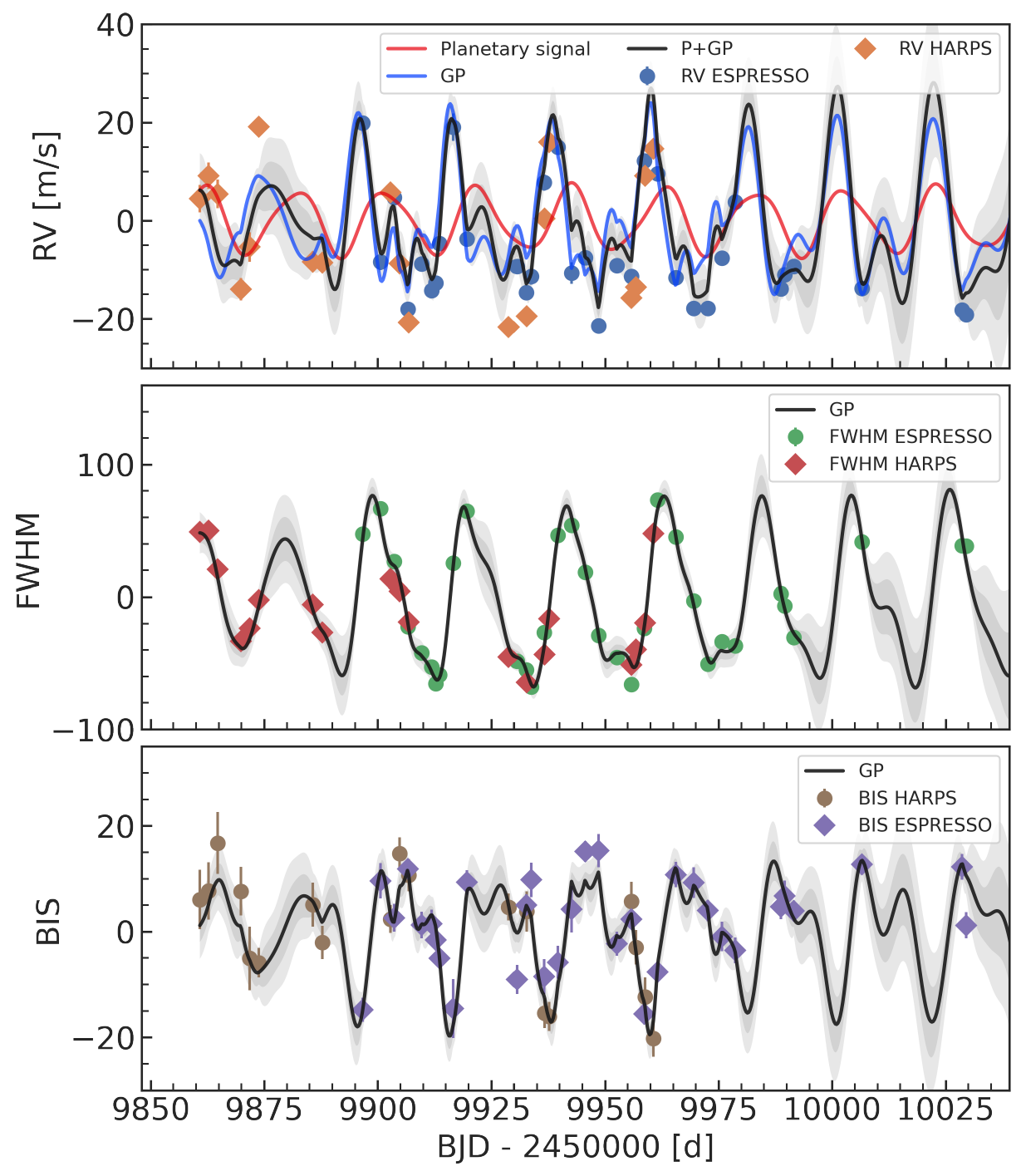}
    \includegraphics[width=0.45\linewidth]{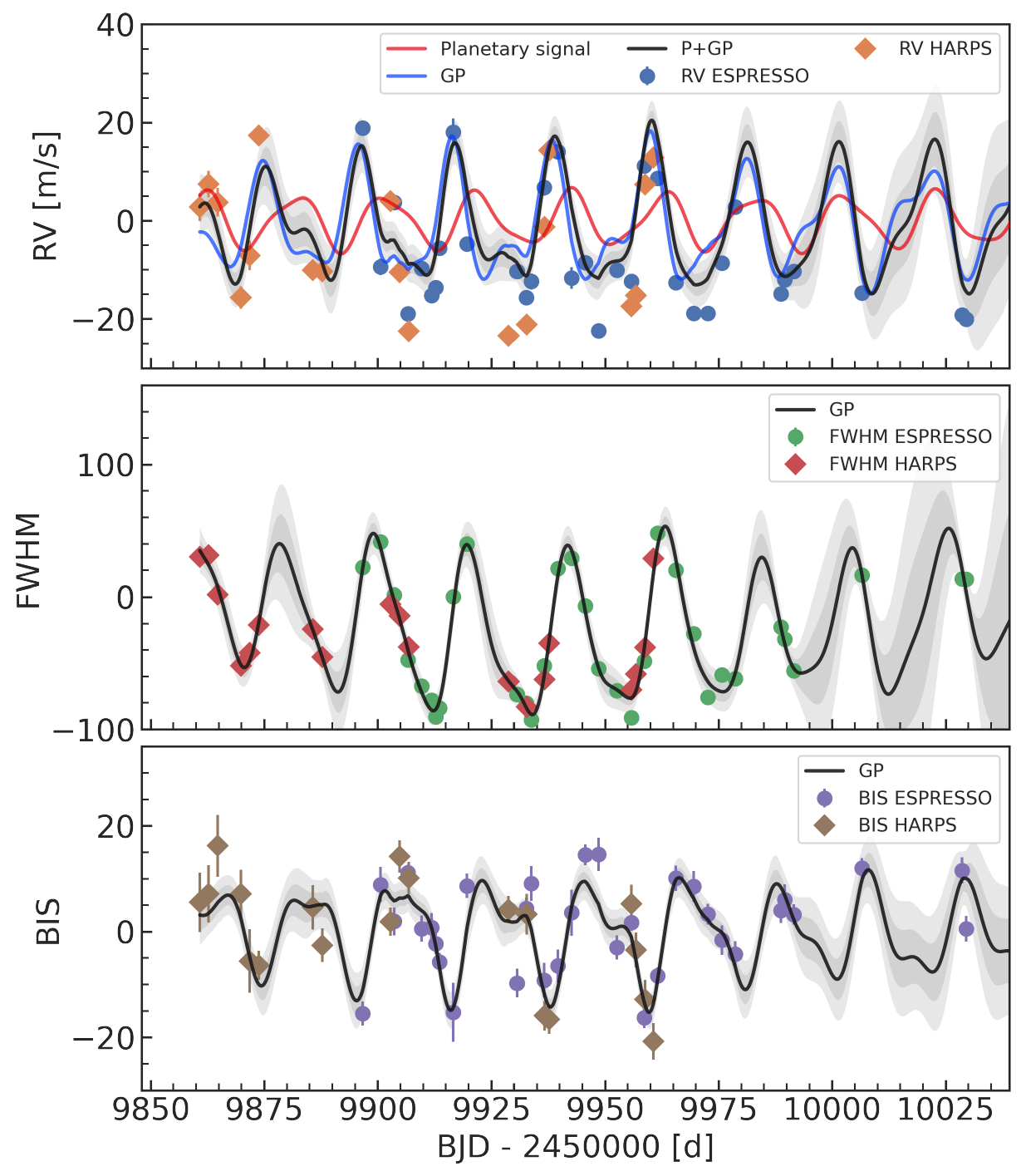}
    \includegraphics[width=0.45\linewidth]{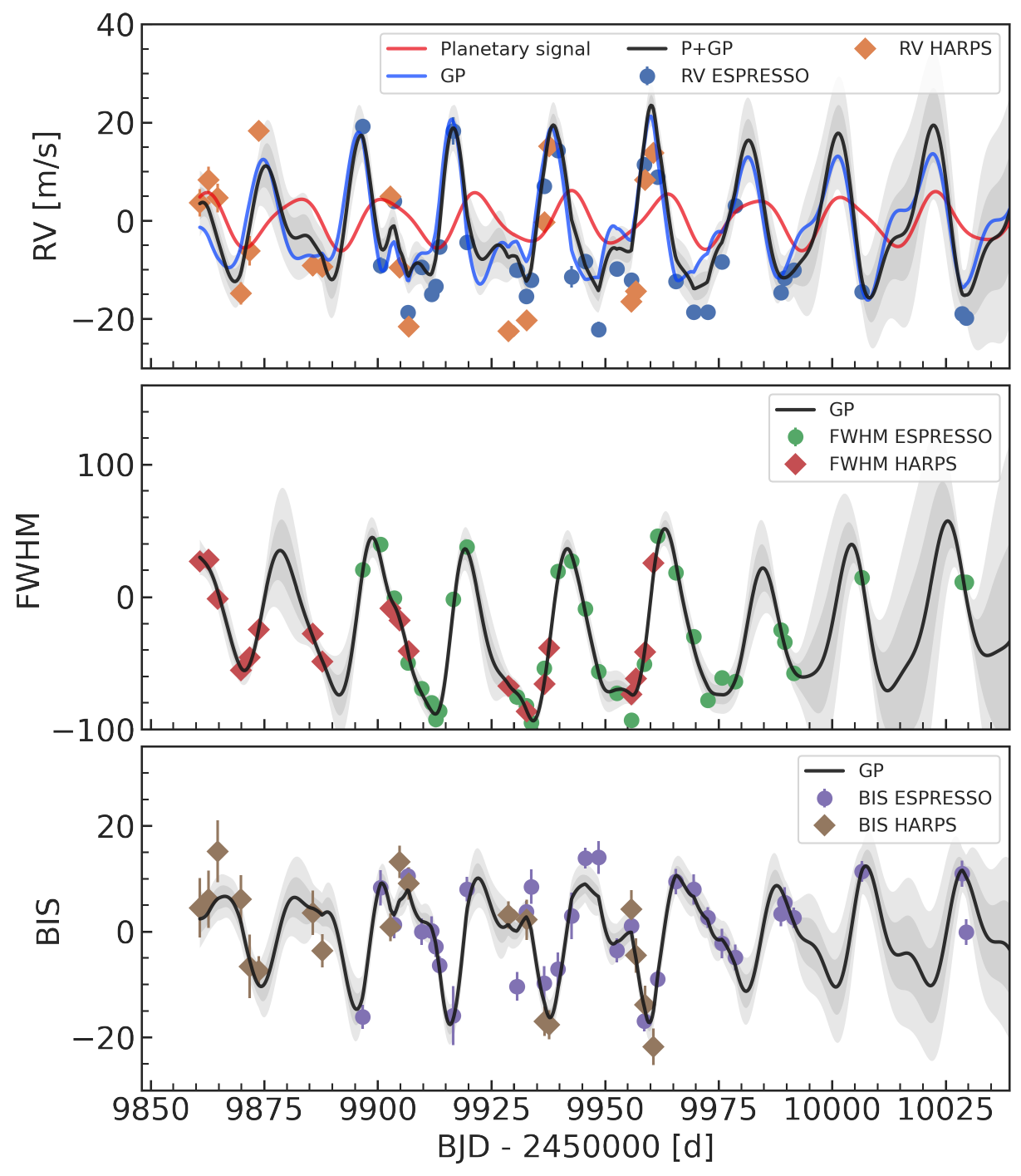}
    \caption{Fitted models using the S+LEAF kernels. Top to bottom and left to right: 2$\times$SHO kernel, MEP kernel, ESP kernel, and ESP-4 kernel. Within each panel, the subplots are analogous to those in Fig. \ref{fig:RVs_FWHM_BIS_pyaneti}.}
\end{figure*}

\end{document}